%% file: LRDM_long.tex
\documentclass[12pt,a4paper,prd,superscriptaddress,nofootinbib,tightenlines,preprintnumbers]{revtex4}
\pdfoutput=1
\usepackage{amstext,amssymb,amsmath}
\usepackage{graphicx}
\usepackage{dcolumn}
\usepackage{xspace}
\usepackage{braket}
\usepackage{color}
\usepackage{units}
\usepackage{slashed} 
\usepackage[english]{babel}
\usepackage{multirow}
\usepackage{soul}
\usepackage{float}
\usepackage[plainpages=false,hyperfootnotes=false]{hyperref}

\newcommand{\del}{\partial}
\newcommand{\dd}{\mathrm{d}}

\def \P {\mathcal{P}} 
\def \C {\mathcal{C}} 

\def \L {\mathcal{L}} 

\def \epsilon {\varepsilon} 

\def \vec#1{{\boldsymbol{#1}}} 
\newcommand{\matrixx}[1]{\begin{pmatrix} #1 \end{pmatrix}} 
\newcommand{\BR}{\mathrm{BR}}
\newcommand{\tr}{\mathrm{tr}}
\newcommand{\hc}{\ensuremath{\text{h.c.}}}

\allowdisplaybreaks

\begin{document}

\title{Phenomenology of left--right symmetric dark matter}

\preprint{ULB-TH/15-22}

\author{Camilo Garcia-Cely}
\email{Camilo.Alfredo.Garcia.Cely@ulb.ac.be}
\affiliation{Service de Physique Th\'eorique, Universit\'e Libre de Bruxelles, Boulevard du Triomphe, CP225, 1050 Brussels, Belgium}

\author{Julian \surname{Heeck}}
\email{Julian.Heeck@ulb.ac.be}
\affiliation{Service de Physique Th\'eorique, Universit\'e Libre de Bruxelles, Boulevard du Triomphe, CP225, 1050 Brussels, Belgium}

\hypersetup{
    pdftitle={Phenomenology of left--right symmetric dark matter},
    pdfauthor={Camilo Garcia-Cely, Julian Heeck}
}

\begin{abstract}

We present a detailed study of dark matter phenomenology in low-scale left--right symmetric models. Stability of new fermion or scalar multiplets is ensured by an accidental matter parity that survives the spontaneous symmetry breaking of the gauge group by scalar triplets. The relic abundance of these particles is set by gauge interactions and gives rise to dark matter candidates with masses above the electroweak scale. Dark matter annihilations are thus modified by the Sommerfeld effect, not only in the early Universe, but also today, for instance, in the Center of the Galaxy. Majorana candidates -- triplet, quintuplet, bi-doublet, and bi-triplet -- bring only one new parameter to the model, their mass, and are hence highly testable at colliders and through astrophysical observations. Scalar candidates -- doublet and 7-plet, the latter being only stable at the renormalizable level -- have additional scalar--scalar interactions that give rise to rich phenomenology.
The particles under discussion share many features with the well-known candidates wino, Higgsino, inert doublet scalar, sneutrino, and Minimal Dark Matter. In particular, they all predict a large gamma-ray flux from dark matter annihilations, which can be searched for with Cherenkov telescopes.
We furthermore discuss models with unequal left--right gauge couplings, $g_R \neq g_L$, taking the recent experimental hints for a charged gauge boson with \unit[2]{TeV} mass as a benchmark point. In this case, the dark matter mass is determined by the observed relic density.

\end{abstract}


\maketitle
\newpage
\tableofcontents
\newpage


\section{Introduction}
\label{sec:introduction}

The Standard Model (SM) gives a highly satisfactory account of the forces and interactions between known particles. 
Its shortcomings are however severe when it comes to the issue of neutrino masses and the existence of dark matter (DM).
At least the former finds a natural solution in left--right (LR) symmetric extensions of the electroweak gauge group $SU(2)_L \times U(1)_{Y} \to SU(2)_L \times SU(2)_R \times U(1)_{B-L}$~\cite{Pati:1974yy,Mohapatra:1974gc,Senjanovic:1975rk,Senjanovic:1978ev}, in which small Majorana neutrino masses can arise via the type-I and type-II seesaw mechanism~\cite{Mohapatra:1979ia,Mohapatra:1980yp}. A key feature is the introduction of right-handed neutrinos as imposed by the gauge group, rather than ad hoc. In addition, LR models explain the obscure parity violation at low energies via spontaneous symmetry breaking.

Not resolved in LR models is, however, the issue of DM. While one of the right-handed neutrinos can be tuned 
to be in the keV mass range relevant for long-lived warm DM, its gauge interactions typically overproduce them and require 
a non-standard production/dilution mechanism~\cite{Bezrukov:2009th,Nemevsek:2012cd}. Assuming this production 
mechanism to be in place, the (unstable) keV neutrino can give rise to testable signatures~\cite{Nemevsek:2012cd,Barry:2014ika}.
Less fine-tuned DM, e.g.~a cold thermal relic, typically referred to as a weakly interacting massive particle (WIMP), requires the addition of a new particle to LR models together with a stabilizing symmetry.
A new framework for stable cold DM along these lines was recently brought forward in Ref.~\cite{Heeck:2015qra}, employing 
the fact that the LR gauge group is actually broken down to the non-trivial $\mathbb{Z}_2^{B-L}$ 
by the scalar triplets.  New LR-symmetric fermion triplets $(\vec{3},\vec{1},0)\oplus(\vec{1},\vec{3},0)$ or quintuplets $(\vec{5},\vec{1},0)\oplus (\vec{1},\vec{5},0)$, for example, are then absolutely 
stable, contain Minimal Dark Matter (MDM) as a component,\footnote{The idea of MDM~\cite{Cirelli:2005uq,Cirelli:2007xd,Cirelli:2009uv,Cirelli:2014dsa,Cirelli:2015bda,Garcia-Cely:2015dda} is to introduce multiplets to the SM that are of large enough $SU(2)_L$ dimension to forbid renormalizable couplings that could lead to decay, the prime example being a chiral fermion quintuplet without hypercharge.} and only bring with them one additional parameter: the mass of the multiplet (degenerate due to LR exchange symmetry).

The $\mathbb{Z}_2^{B-L}$ is well-known as \emph{matter parity} $(-1)^{3 (B-L)}$ in the literature and has already been employed as a 
stabilizing symmetry for DM in grand unified $SO(10)$ theories~\cite{Kadastik:2009dj,Kadastik:2009cu,Frigerio:2009wf,
Mambrini:2015vna,Nagata:2015dma,Arbelaez:2015ila} and general supersymmetric models (where it is known as $R$ parity).
In this work, we will expand on the idea of Ref.~\cite{Heeck:2015qra} and discuss a variety of DM candidates and their 
signatures within low-scale non-supersymmetric LR models. The plan of the paper is as follows. 
In Sec.~\ref{sec:left-right_models} we give an overview of LR models and introduce the relevant formulae.
In Sec.~\ref{sec:stability} we list new multiplets that can lead to DM candidates and describe qualitative 
features as well as estimate Landau poles. The following sections describe in more detail the phenomenology 
of the simplest DM candidates, namely (Majorana) fermions in Sec.~\ref{sec:majorana} and (real) scalars in Sec.~\ref{sec:real}.
Sec.~\ref{sec:diboson} is dedicated to LR models that might explain the recent $\unit[2]{TeV}$ excesses seen in various channels in ATLAS and CMS, in particular those that require $g_R < g_L$.
Finally, we conclude in Sec.~\ref{sec:conclusion}.
Several appendices provide details that would interrupt the flow of the main text. In Appendix~\ref{app:gauge_boson_mixing} we give the gauge boson mixing formulae for the case $g_L\neq g_R$ as well as gauge boson partial decay widths.
Appendix~\ref{app:real_reps} reviews real representations of $SU(2)$ in the context of field theory; Appendix~\ref{app:mass_splitting} gives the formulae for the radiative mass splitting of $SU(2)$ multiplets. In Appendix~\ref{sec:SEinGC} we discuss the Sommerfeld effect in the context of  indirect detection DM searches and, finally, in Appendix~\ref{sec:SU2LL} we describe the $SU(2)_L$-symmetric limit for the calculation of the relic density.

\section{Left--right models}
\label{sec:left-right_models}

Let us briefly introduce left--right symmetric models and provide the relevant notation and formulae, following Ref.~\cite{Duka:1999uc}.
Under the left--right gauge group 
$SU(2)_L\times SU(2)_R\times U(1)_{B-L}$ -- omitting the $SU(3)_C$ color factor for simplicity -- 
the usual fermion content of the SM, plus right-handed neutrinos~$\nu_R$, falls into the following 
representations
\begin{align}
 \ell_L &= \matrixx{\nu_L \\ e_L} \sim (\vec{2},\vec{1},-1)\,, &
 \ell_R &= \matrixx{\nu_R \\ e_R} \sim (\vec{1},\vec{2},-1)\,, \label{eq:LRleptons}\\
 q_L &= \matrixx{u_L \\ d_L} \sim (\vec{2},\vec{1}, 1/3)\,, &
 q_R &= \matrixx{u_R \\ d_R} \sim (\vec{1},\vec{2}, 1/3)\,,
\label{eq:LRquarks}
\end{align}
suppressing flavor indices.
Fermion masses are provided by the vacuum expectation value (VEV) of a scalar 
bi-doublet $H$ 
\begin{align}
 H = \matrixx{ h_1^0 & h_1^+ \\ h^-_2 & h^0_2} \sim (\vec{2},\overline{\vec{2}}, 0) \,,
\label{eq:bidoublet_higgs}
\end{align}
via the Yukawa couplings $y_f\overline{f}_L H f_R$ and $\tilde y_f\overline{f}_L \tilde{H} f_R$, 
with $f=q,\ell$. The conjugate field $\tilde{H} \equiv \sigma_2 H^* \sigma_2$ transforms again as a bi-doublet, $\sigma_2$ denoting the antisymmetric Pauli matrix.

Additional scalars beyond the bi-doublet are necessary to break the gauge group $SU(2)_L \times SU(2)_R \times U(1)_{B-L}$ down 
to $U(1)_\text{EM}$. Since the generator of electric charge is given by 
\begin{align}
Q = T^3_L + T^3_R + \tfrac12 (B-L) \,,
\end{align}
where $T^3_X$ denotes the diagonal generator of $SU(2)_X$, we need a scalar that carries $B-L$ charge.
A common choice is to introduce two scalar triplets $\Delta_{L,R}$~\cite{Mohapatra:1979ia,Mohapatra:1980yp}
\begin{align}
\Delta_L = \matrixx{\delta_L^+/\sqrt{2} & \delta^{++}_L \\ \delta^0_L & -\delta_L^+/\sqrt{2} } \sim (\vec{3},\vec{1},2)\,, &&
\Delta_R = \matrixx{\delta_R^+/\sqrt{2} & \delta^{++}_R \\ \delta^0_R & -\delta_R^+/\sqrt{2} }\sim (\vec{1},\vec{3},2)\,.
\label{eq:triplet_higgs}
\end{align}
A non-zero VEV of the neutral component of $\Delta_R$, $\langle \delta_R^0 \rangle \equiv 
v_R/\sqrt{2}$, breaks $SU(2)_R \times U(1)_{B-L}\to U(1)_Y$ at a scale above TeV and furthermore generates large Majorana masses 
for the right-handed neutrinos -- leading to seesaw neutrino masses for the active neutrinos -- and 
masses for the new gauge bosons $W_R$ and $Z_R$. The bi-doublet VEV
\begin{align}
\langle H \rangle &= \matrixx{ \kappa_1/\sqrt{2} & 0 \\ 0 & \kappa_2/\sqrt{2}},
\end{align}
with $\kappa \equiv \sqrt{\kappa_1^2 + \kappa_2^2}\simeq\unit[246]{GeV}$,
induces a mixing between left- and right-handed gauge bosons and gives the dominant masses to $W_L$, $Z_L$, 
and all charged SM fermions. We will assume $\kappa_{1,2}$ to be real in the following for simplicity and define a new angle $\beta$ via $\tan\beta \equiv \kappa_1/\kappa_2$. 
The VEV of the left-handed triplet $\langle \delta_L^0 \rangle \equiv v_L/\sqrt{2}$ is typically given 
by a seesaw relation of the form $v_L \propto \kappa^2/v_R$~\cite{Deshpande:1990ip} 
and hence small, in accordance with constraints from the $\rho$ parameter ($v_L$ breaks the custodial 
symmetry relation $\rho = M_{W}^2/(M_{Z}^2 c_W^2) = 1$). Even if small, say $v_L <\unit{GeV}$, or 
even $v_L <\unit{eV}$, it can give potentially important contributions to Majorana neutrinos masses 
(type-II seesaw mechanism). Fine tuning in the scalar potential can be reduced by choosing $v_L = 0$~\cite{Deshpande:1990ip,Dekens:2014ina}, and we will henceforth ignore $v_L$ in our discussion. 

The gauge couplings of $SU(2)_L$, $SU(2)_R$, and $U(1)_{B-L}$ will be denoted by $g_L$, $g_R$, and $g_{BL}$, respectively.
A further ingredient of LR models with $g_L = g_R$ is an additional discrete left--right exchange symmetry, 
corresponding either to generalized parity $\P$
\begin{align}
\ell_L \stackrel{\P}{\longleftrightarrow} \ell_R \,, &&
q_L \stackrel{\P}{\longleftrightarrow} q_R \,, &&
H \stackrel{\P}{\longleftrightarrow} H^\dagger \,, &&
\Delta_L \stackrel{\P}{\longleftrightarrow} \Delta_R \,, 
\end{align}
or generalized charge conjugation $\C$
\begin{align}
\ell_L \stackrel{\C}{\longleftrightarrow} \ell_R^c \,, &&
q_L \stackrel{\C}{\longleftrightarrow} q_R^c \,, &&
H \stackrel{\C}{\longleftrightarrow} H^T \,, &&
\Delta_L \stackrel{\C}{\longleftrightarrow} \Delta_R^* \,,
\end{align}
which also act on the gauge bosons~\cite{Duka:1999uc} and are obviously broken by the triplet VEVs. Imposing either $\P$ or $\C$ gives the gauge coupling relation $g_R = g_L $ at high scales, but different constraints on the Yukawa 
coupling matrices, namely $y_f = y_f^\dagger$ for $\P$ and $y_f = y_f^T$ for $\C$ (same for $\tilde y_f$)~\cite{Dekens:2014ina}.
For most of our paper we will assume 
$\P$ symmetry and $g_R = g_L \equiv g_2$, but comment on deviations from this 
when appropriate (see Sec.~\ref{sec:diboson} and Appendix~\ref{app:gauge_boson_mixing}).

Denoting the $B-L$ gauge boson by $B_\mu$ and the $SU(2)_X$ gauge bosons by $W_X^a$, with $W_X^\pm \equiv (W^1_X \mp i W^2_X)/\sqrt{2}$, the mixing matrices of the charged and neutral gauge bosons can be parametrized as~\cite{Duka:1999uc}
\begin{align}
\matrixx{W_L^+\\W_R^+} &= \matrixx{\cos\xi & \sin\xi\\-\sin\xi & \cos\xi} \matrixx{W_1^+\\W_2^+} ,\\
\begin{split}
\matrixx{W_{L}^3 \\ W_{R}^3 \\ B} &= \matrixx{c_W c_\phi & c_W s_\phi & s_W \\ -s_W s_M c_\phi  - 
c_M s_\phi & -s_W s_M s_\phi +c_M c_\phi &  c_W s_M\\ -s_W c_M c_\phi + s_M s_\phi & -s_W c_M s_\phi  - s_M c_\phi & c_W c_M} \matrixx{Z_1\\Z_2\\A} ,
\end{split}
\end{align}
where $W_{1,2}^+$ denote the charged mass eigenstates and $Z_{1,2}$ the massive neutral ones, $A$ being the massless photon. 
Here, $s_W \equiv \sin \theta_W$ is the sine of the weak mixing angle, $s_\phi\equiv \sin\phi$ the sine of the neutral mixing angle, and $s_M \equiv \sin \theta_M = \tan \theta_W$ for $g_R = g_L$ (see Appendix~\ref{app:gauge_boson_mixing} for formulae in the general case $g_L\neq g_R$).
For later convenience, we also note
\begin{align}
W_L^3 - W_R^3 \simeq Z_1/c_W - c_M Z_2\,, &&
W_L^3 + W_R^3 \simeq \cos(2\theta_W) Z_1/c_W +c_M Z_2 + 2 s_W A\,,
\end{align}
neglecting $\phi$.
In the phenomenologically relevant limit $\kappa \ll v_R$, one finds the masses
\begin{align}
M_{W_1} &\simeq \frac{g_2}{2} \kappa \,, &
M_{W_2} &\simeq \frac{g_2}{\sqrt{2}} v_R \,, \\
M_{Z_1} &\simeq \frac{g_2}{2 c_W} \kappa\,, &
M_{Z_2} &\simeq \frac{g_2 c_W}{\sqrt{\cos 2\theta_W}} v_R\,,
\end{align}
and the suppressed mixing angles
\begin{align}
\xi \simeq -\frac{M_{W_1}^2}{M_{W_2}^2} \sin 2\beta \,, &&
\phi \simeq - \frac{M_{Z_1}^2}{M_{Z_2}^2} \sqrt{\cos 2\theta_W}\,.
\end{align}

Note that the ratio $M_{Z_2}/M_{W_2}$ crucially depends on the way $SU(2)_R\times U(1)_{B-L}$ is broken to the hypercharge group $U(1)_Y$. 
If broken by scalar doublets $\sim (\vec{1},\vec{2},1)$, one finds $M_{Z_2}/M_{W_2} \simeq 
1.2$~\cite{Mohapatra:1974gc,Senjanovic:1975rk,Senjanovic:1978ev}, no discrete $U(1)_{B-L}$ subgroup survives, 
and neutrinos are typically Dirac particles.\footnote{Even though no stabilizing symmetry exists, one can, of course, still have dimension-four-stable MDM~\cite{Ko:2015uma}.}
If broken by triplets $\sim (\vec{1},\vec{3},2)$ -- which 
is the case discussed in this article -- one finds $M_{Z_2}/M_{W_2} \simeq 1.7$~\cite{Mohapatra:1979ia,Mohapatra:1980yp}, 
a $\mathbb{Z}_2^{B-L}$ subgroup~\cite{Heeck:2015qra}, and Majorana neutrinos. If broken by quintuplets $\sim (\vec{1},\vec{5},4)$, one finds $M_{Z_2}/M_{W_2} \simeq 2.4$~\cite{Heeck:2015pia}, a $\mathbb{Z}_4^{B-L}$ subgroup, and Dirac neutrinos (but with lepton number violating interactions~\cite{Heeck:2013rpa,Heeck:2013vha}).
The mass ratio thus increases for larger representations.
These discrete choices aside, the ratio $M_{Z_2}/M_{W_2}$ can vary from these benchmark points in models with $g_R\neq g_L$~\cite{Chang:1983fu} (see Appendix~\ref{app:gauge_boson_mixing}), 
and one can even have $M_{Z_2}/M_{W_2} < 1$ if $SU(2)_R\times U(1)_{B-L}$ is broken in two steps at different scales, shifting focus from $W'$ to $Z'$ searches at colliders~\cite{Patra:2015bga}. 

Let us briefly mention experimental constraints on the gauge boson parameters relevant to our following discussion.
A popular direct search channel for the right-handed charged gauge boson $W_2$ is given by its decay into leptons and heavy neutrinos, $W_2\to\ell N$, followed by the decay $N\to \ell' j j$~\cite{Keung:1983uu}. Due to the Majorana nature of $N$, the leptons $\ell$ and $\ell'$ can have the same charge.
This $pp \to \ell \ell jj$ process gives constraints up to $M_{W_2} < \unit[3]{TeV}$~\cite{Khachatryan:2014dka} if one of the heavy neutrinos is lighter than $W_2$, but no limit for $M_{W_2} < M_N$. These limitations are avoided to some degree in purely hadronic low-energy processes, such as the $K_L$--$K_S$ mass difference~\cite{Beall:1981ze}, which exclude $M_{W_2} < \unit[2.9]{TeV}\,(g_R/g_L)$ at $95\%$~C.L.~for the $\C$ case and $M_{W_2} < \unit[3.2]{TeV}\,(g_R/g_L)$ if $\P$ is employed~\cite{Bertolini:2014sua}, which is the limit we use in the following.\footnote{Such low values for $M_{W_2}$ also imply $\tan\beta = \mathcal{O}(1)$ in the case of $\P$ parity~\cite{Bertolini:2014sua}.}
It is important to note that the low-energy limits arise from an off-shell $W_2$ and hence do not depend on the $W_2$ width or branching ratios, as compared to on-shell searches by ATLAS and CMS. 
The relation $M_{Z_2}/M_{W_2} \simeq 1.7$ puts the neutral gauge boson $Z_2$ out of experimental reach for now, but will ultimately be an important discriminator of different models should new gauge boson(s) be found.
Recent excesses seen at ATLAS~\cite{Aad:2015owa} and CMS~\cite{Khachatryan:2014dka,Khachatryan:2014hpa,Khachatryan:2014gha} 
experiments point towards a $\unit[2]{TeV}$ mass of $W_2$, which can be consistently accommodated in LR models with $g_L \neq g_R$ (see Sec.~\ref{sec:diboson}).

\section{Dark matter stability}
\label{sec:stability}

As shown in Ref.~\cite{Heeck:2015qra}, the introduction of new multiplets to LR models can give 
rise to DM candidates without the need for ad hoc global stabilizing symmetries. These new particles can either be accidentally stable in the MDM spirit 
because the high $SU(2)$ dimensionality forbids renormalizable couplings that could 
lead to decay~\cite{Cirelli:2005uq}, or \emph{exactly} stable due to their quantum numbers 
under the unbroken $\mathbb{Z}_2\subset U(1)_{B-L}$ subgroup that remains by breaking the LR 
gauge group via the scalar triplets $\Delta_{L,R}$. 
Since all fermions from Eqs.~\eqref{eq:LRleptons}--\eqref{eq:LRquarks} are odd under 
this $\mathbb{Z}_2$ and all bosons (scalars and vectors) are even, new fermion (boson) 
multiplets are exactly stable if they carry even (odd) $B-L$ charge~\cite{Heeck:2015pia}.

More specifically, one should consider the generator $X\equiv 3 (B-L)$ instead of $B-L$ 
in order to have integer $U(1)'$ charges at the quark level. Quarks then carry charge 
$X(q) = 1$, leptons $X(\ell) = -3$, and the scalars have $X(H)=0$ and $X(\Delta) = -6$. 
Since the $U(1)_X$ is broken by the VEV of $\Delta$ by six units, a $\mathbb{Z}_6 \cong 
\mathbb{Z}_2\times \mathbb{Z}_3$ subgroup remains unbroken, under which quarks transform 
as $q \to e^{i\pi/3} q$ and leptons as $\ell\to -\ell$ (see for example Ref.~\cite{Batell:2010bp} 
for a discussion of such discrete gauge symmetries~\cite{Krauss:1988zc}); the scalars transform 
trivially under the~$\mathbb{Z}_6$. At hadron level, the baryons transform as $N\to - N$ under the $\mathbb{Z}_6$, which is why we say that 
$B-L$ is broken to a $\mathbb{Z}_2$ subgroup under which all SM fermions (bosons) are 
odd (even). 
In more mathematical terms, the subgroup $\mathbb{Z}_3 \subset \mathbb{Z}_6$ is actually 
nothing but the \emph{center} subgroup of $SU(3)_C$, so only the $\mathbb{Z}_2$ remains 
as a global symmetry~\cite{Martin:1992mq}, and can be identified with matter parity $(-1)^X$.
It is then clear that a new fermion (boson) with even (odd) $B-L$ charge is 
stable, as long as it does not 
obtain a VEV (in the scalar case).

\subsection{Fermions}

Let us consider the introduction of new chiral fermions $\Psi$ first, where we restrict 
ourselves to colorless representations for simplicity. Allowing for a parity exchange symmetry, 
$\Psi_L \stackrel{\P}{\longleftrightarrow} \Psi_R$, the chiral fermion representations 
$\Psi_L\oplus \Psi_R$ with a stable component due to matter parity are given by
\begin{align}
(\vec{n_1},\vec{n_2},2 k )\oplus (\vec{n_2},\vec{n_1}, 2 k) \,, \quad n_{1,2}\in 
\mathbb{N},\, k\in \mathbb{Z} \,.
\label{eq:fermion_reps}
\end{align}
$n_1$ and $n_2$ must both be either odd or even, i.e.~$(n_1+n_2)/2\in \mathbb{N}$, in order to obtain integer electric charges for the components and cancel Witten's 
anomaly~\cite{Witten:1982fp}. For $n_1 = n_2$ this can be reduced to one chiral 
bi-multiplet
\begin{align}
(\vec{n},\vec{n},2 k )\,,  \quad n\in \mathbb{N},\, k\in \mathbb{Z} \,.
\label{eq:fermion_bi-multiplet}
\end{align}
The requirement for a neutral component gives additional constraints on $n_{1,2}$ 
and $k$, as does cancellation of triangle anomalies~\cite{Choi:1992am} and sufficiently 
high Landau poles of the gauge couplings~\cite{Lindner:1996tf} (to be discussed below 
in Sec.~\ref{sec:landau_poles}).
A bare Majorana mass term can only be written down for $k=0$, so the multiplets with $k\neq 0$ should be introduced as non-chiral Dirac fermions (which also solves the problem of triangle anomalies, e.g.~$SU(2)_X$--$SU(2)_X$--$U(1)_{B-L}$ 
(with $X=L,R$) and $U(1)_{B-L}^3$).

The most transparent case is given by $k=0$, where one can have a stable neutral 
Majorana fermion from the real chiral representations
\begin{align}
(\vec{2j_1+1},\vec{2j_2+1},0 )\oplus (\vec{2j_2+1},\vec{2j_1+1},0) && \mathrm{or} 
&& (\vec{2j+1},\vec{2j+1},0 )\,,
\label{eq:majorana_reps}
\end{align}
with $j,j_{1,2}\in \mathbb{N}$.
We will discuss the simplest examples, namely the triplet $(\vec{3},\vec{1},0 )\oplus (\vec{1},
\vec{3},0)$, the quintuplet $(\vec{5},\vec{1},0 )\oplus (\vec{1},\vec{5},0)$, and the bi-triplet $(\vec{3},\vec{3},0 )$ 
in Sec.~\ref{sec:majorana}. Stable neutral \emph{Dirac fermions} can also be obtained for $k=0$, 
namely when $n_{1,2}$ are even, e.g.~$(\vec{2},\vec{2},0 )$, $(\vec{4},\vec{4},0 )$, 
or $(\vec{2},\vec{4},0 )\oplus (\vec{4},\vec{2},0)$. In these cases, the electrically 
neutral component carries hypercharge and thus couples to the light $Z_1$ boson, which 
is typically at odds with constraints from direct detection experiments if a thermal freeze-out abundance is assumed.
At one-loop level, and for non-vanishing $W_L^-$--$W_R^-$ mixing, the neutral Dirac fermion can however split into two quasi-degenerate Majorana fermions; for a mass splitting above $\unit[\mathcal{O}(100)]{keV}$, the direct-detection bounds are then circumvented~\cite{TuckerSmith:2001hy}. This will be discussed in Sec.~\ref{sec:bidoublet} for the bi-doublet $(\vec{2},\vec{2},0 )$.

\subsection{Scalars}

A new VEV-less scalar multiplet is stable if it has an odd $B-L$ charge. This, of course, eliminates the conceptually simpler possibility of a real (self-conjugate) scalar. Let us therefore consider first \emph{real} scalars, i.e.~with vanishing $B-L$ charge, that might be stable at the renormalizable level in the MDM sense. This leads us back to the assignments from the Majorana fermions from Eq.~\eqref{eq:majorana_reps}:\footnote{The scalar representation $(\vec{\mathrm{even}},\vec{\mathrm{even}},0)$ can only have a \emph{complex} neutral scalar at tree level (with coupling to $Z_1$), and will not be discussed here. Higher order mass splittings into \emph{real} scalars can again alleviate direct-detection constraints~\cite{TuckerSmith:2001hy}.}
\begin{align}
(\vec{2j_1+1},\vec{2j_2+1},0 )\oplus (\vec{2j_2+1},\vec{2j_1+1},0) && \mathrm{or} && (\vec{2j+1},\vec{2j+1},0 )\,,
\label{eq:real_scalar_rep}
\end{align}
with $j,j_{1,2}\in \mathbb{N}$.
In order to discuss stability, we note that $\tr (H \sigma_j H^\dagger)$ transforms as $(\vec{1},\vec{3},0)$, $\tr (H^\dagger \sigma_j H)\sim (\vec{3},\vec{1},0)$, and $\tr (H^\dagger \sigma_j H \sigma_i) \sim (\vec{3},\vec{3},0)$. Similarly, two $\Delta$ triplets can be coupled to quintuplets, $(\vec{5},\vec{1},0)$ or $(\vec{1},\vec{5},0)$, using the product decomposition
\begin{align}
\vec{3}\otimes \vec{3} = \vec{1}\oplus\vec{3}\oplus\vec{5}\,.
\end{align}
Real scalars with $j_1$ and $j_2 \leq 2$ are hence unstable at the renormalizable level, while those with $j_1>2$ or $j_{2}>2$ -- the simplest being the 7-plet $(\vec{7},\vec{1},0 )\oplus (\vec{1},\vec{7},0)$ -- are only unstable through dimension-five operators of the form $\phi H^4$, $\phi \Delta^4$ or $\phi^3 H^2$ (in very abstract notation). Similar to the often discussed MDM case~\cite{Cirelli:2005uq,DiLuzio:2015oha}, we can still consider the 7-plet $(\vec{7},\vec{1},0 )\oplus (\vec{1},\vec{7},0)$ by arguing that these dangerous dimension-five operators are absent or highly suppressed. We will study this example in more detail in Sec.~\ref{sec:7-plet}.
Bi-multiplets with $j > 1$, e.g.~the bi-quintuplet $(\vec{5},\vec{5},0)$, are also stable at the renormalizable level, but will decay through dimension-five operators.

Lastly, let us consider scalar multiplets that are \emph{exactly} stable due to matter parity. These stable scalars with integer electric charge reside in the representations
\begin{align}
(\vec{2 j_1},\vec{2 j_2+1},2 k -1 )\oplus (\vec{2 j_2+1},\vec{2 j_1}, 2 k-1) \,,
\label{eq:complex_scalar_rep}
\end{align}
where $j_{1,2}\in \mathbb{N}$ and $k\in \mathbb{Z}$.
The simplest example has the same gauge quantum numbers of the leptons and is hence reminiscent of sleptons in supersymmetric models,
\begin{align}
(\vec{2},\vec{1}, -1 )\oplus (\vec{1},\vec{2},-1) \,.
\end{align}
Since the neutral components generally mix by coupling to $H$, we obtain single-component DM.
The VEV of $\Delta_R$ will split the masses of the complex neutral fields and thus lead to a \emph{real} scalar DM candidate.  The hypercharge-neutral right-handed (sneutrino-like) component is then the prime DM candidate to evade direct-detection bounds, further discussed in Sec.~\ref{sec:inert_doublet}.

\subsection{Landau poles}
\label{sec:landau_poles}

The addition of higher $SU(2)_{L,R}$ representations can severely modify the running of the corresponding gauge couplings $g_{L,R}$ and even lead to a Landau pole $\Lambda_\text{LP}$. Landau poles are commonly banished to far above the Planck scale $M_\mathrm{Pl}$ in the hopes that quantum gravity will solve the issue; another solution to gauge-coupling Landau poles is the unification into a sufficiently large non-abelian gauge group (without a Landau pole) at scales below $\Lambda_\text{LP}$. The prospects of unification with the addition of our DM multiplets will be discussed in a separate publication (see also Refs.~\cite{Lindner:1996tf,Kadastik:2009dj,Kadastik:2009cu,Frigerio:2009wf,Mambrini:2015vna,Dev:2015pga,Nagata:2015dma}). Here, we simply estimate $\Lambda_\text{LP}$ to obtain a feeling for possible \emph{upper} bounds on the $SU(2)$ dimension of our new fields. We are only concerned with the running of $g_L = g_R \equiv g_2$, because this will be affected most strongly by our multiplets (compared to $g_{BL}$).
The discussion is qualitatively similar to the MDM case~\cite{Cirelli:2005uq,DiLuzio:2015oha}.\footnote{As shown recently in Ref.~\cite{
Hamada:2015bra}, scalar multiplets introduced to the SM suffer from Landau poles in their quartic 
interactions with the SM doublet. We expect the same behavior in our LR theory.}

Defining the $SU(2)_L$ fine-structure coupling $\alpha_2 \equiv g_2^2/4\pi$, one finds the standard analytic one-loop solution for the renormalization-group running from a scale $\lambda$ to $\Lambda >\lambda$:
\begin{align}
\frac{1}{\alpha_2 (\Lambda)} = \frac{1}{\alpha_2 (\lambda)} - \frac{b_2}{2\pi}\log \left(\frac{\Lambda}{\lambda}\right) ,
\end{align}
and hence a Landau pole -- $\alpha_2^{-1} (\Lambda_\text{LP}) =0$ -- at the scale
\begin{align}
\Lambda_\text{LP} \simeq \lambda \exp \left[ \frac{2\pi}{b_2} \alpha_2^{-1} (\lambda) \right] ,
\end{align}
if $b_2 >0$.
The relevant one-loop coefficient $b_2$ for $SU(2)_L$ is given by~\cite{Gross:1973ju,Jones:1981we,Lindner:1996tf}
\begin{align}
b_2 = -\frac{22}{3} + \frac{2}{3} \sum_{f} c_2 (f)\, d_{SU(2)_R} (f)\, d_{SU(3)_C} (f)+ \frac{1}{3} \sum_{s} c_2 (s)\, d_{SU(2)_R} (s)\, d_{SU(3)_C} (s) \,,
\end{align}
where $d_X(f)$ is the dimension of the chiral fermion $f$ under the gauge group factor $X$, i.e.~$f\sim \vec{d_X}$ under group $X$ in our notation, and
\begin{align}
c_2(f) \equiv \frac{1}{12} \,d_{SU(2)_L} (f) \left[d_{SU(2)_L}^2 (f)-1 \right]
\end{align}
its index under $SU(2)_L$, e.g.~$c_2(\vec{2}) = \frac12$, $c_2(\vec{3})=2$, $c_2 (\vec{5})=10$; the same formulae hold for the complex scalar $s$. One finds $b_2 =-7/3$ with the standard LR particle content given in Sec.~\ref{sec:left-right_models}, whereas our new $SU(2)_L\times SU(2)_R$ representations give
\begin{align}
\Delta b_2 [(\vec{n_1},\vec{n_2})\oplus (\vec{n_2},\vec{n_1})] &= \frac{c}{72} n_1 n_2 (n_1^2+n_2^2-2) \,,\\
\Delta b_2 [(\vec{n},\vec{n})] &= \frac{c}{72} n^2 (n^2-1)\,,
\end{align}
omitting the irrelevant $B-L$ charge,
where $c=1$ for real scalars, $c=2$ for complex scalars, $c=4$ for chiral fermions, and $c=8$ for Dirac fermions.
 
Let us assume an LR breaking scale $\lambda$ and all new particle masses at $\unit[5]{TeV}$, so that $\alpha_2^{-1} (\lambda) \simeq 31.5$. The order-of-magnitude condition $\Lambda_\text{LP} \gtrsim M_\mathrm{Pl}$ then gives $\Delta b_2 \lesssim 7.9$, which means that  real, complex, Majorana, and Dirac bi-multiplet $(\vec{n},\vec{n})$ must satisfy $n \leq 5$, $ n\leq 4$, $n\leq3$, and $n\leq3$, respectively.
For the other large representation under study here, $(\vec{n},\vec{1})\oplus (\vec{1},\vec{n})$,  real, complex, Majorana, and Dirac must satisfy $n\leq 8$, $n\leq 6$, $n\leq 5$, and $n\leq 4$, respectively. (This accidentally coincides with the MDM upper limits~\cite{Cirelli:2005uq}.) Increasing the LR breaking scale or multiplet masses to $\lambda \gg \unit{TeV}$ will shift the Landau pole to higher values, so our upper limits on $n$ are conservative.
None of the multiplets studied in the following will hence induce gauge-coupling Landau poles below the Planck scale.

Besides Landau poles, demanding vacuum stability can provide additional constraints on coupling constants. Relevant here are however only the quartic couplings in the scalar potential, discussed in Refs.~\cite{Chakrabortty:2013zja,Chakrabortty:2013mha}. Correspondingly, our purely gauge-coupled new fermion multiplets will have no effect on vacuum stability, while the new scalars could have an effect depending on their quartic couplings. A discussion of these scalar--scalar couplings is beyond the scope of this article, seeing as they also severely modify the DM phenomenology.

\section{Fermionic dark matter}
\label{sec:majorana}

Here we discuss Majorana LR DM, including a more thorough discussion of the candidates of Ref.~\cite{Heeck:2015qra}. Specifically, we discuss the triplet $(\vec{3},\vec{1},0)\oplus (\vec{1},\vec{3},0)$, the quintuplet $(\vec{5},\vec{1},0)\oplus (\vec{1},\vec{5},0)$, the bi-doublet $(\vec{2},\vec{2},0)$, and the bi-triplet $(\vec{3},\vec{3},0)$.

For the numerical study of our models we modified the LR model implementation for \texttt{FeynRules}~\cite{Christensen:2008py,Alloul:2013bka} of Ref.~\cite{Roitgrund:2014zka} by including our new particles, which we then export to \texttt{CalcHEP}~\cite{Belyaev:2012qa} and \texttt{FeynArts}~\cite{Hahn:2000kx}.\footnote{We corrected a missing factor of 2 in the definition of the mixing angle $\xi$ in the version 1.1.5 of Ref.~\cite{Roitgrund:2014zka}.}

\subsection{Multiplets \texorpdfstring{$(\vec{2 n +1},\vec{1},0)\oplus (\vec{1},\vec{2 n +1},0)$}{Multiplets (2n+1,1,0)+(1,2n+1,0)}}
\label{sec:MLRDM}

We start our discussion with the simplest fermionic DM candidates -- triplet and quintuplet -- already studied in Ref.~\cite{Heeck:2015qra}.
The Lagrangian for the chiral multiplets $\phi_L\oplus \phi_R \sim (\vec{2 n +1},\vec{1},0) 
\oplus (\vec{1},\vec{2 n +1},0)$ is given by
\begin{align}
\L_\phi \ = \sum_{X=L,R}\left[\ i \overline{\phi}_X \slashed{D} P_X \phi_X - \dfrac{M}{2} 
\left(\overline{\phi}_X^c P_X \phi_X + \hc\right) \right] ,
\label{eq:MFDM}
\end{align}
$P_{R,L} \equiv (1\pm \gamma_5)/2$ being the usual chiral projection operators.
A brief review of the \emph{real} $SU(2)$ representations $\vec{2n+1}$ can be found in Appendix~\ref{app:real_reps}.
The key feature is to note that the charged components are Dirac fermions $\Psi^Q_X \equiv \phi^Q_X +(-1)^Q \left(\phi^{-Q}_X\right)^c$, $Q=1,\dots,n$ being their electric charge, while the neutral ones are Majorana $\Psi^0_X \equiv \phi^0_X + 
(\phi^0_X)^c$. 
The interactions of these charged and neutral components with gauge bosons are
\begin{align}
\begin{split}
\L_\phi &\supset \sum_{X=L,R}\left[ g_X \sum_{m=1}^n \left(  m \overline{\Psi}^m_X \slashed{W}_{X}^3 \Psi^m_X  \right) 
+ \frac{g_X}{\sqrt{2}} \left( \sum_{m=0}^{n-1} c_{n,m} \overline{\Psi}_X^{m+1} \slashed{W}^+_X \Psi^m_X 
+\hc\right) \right],
\end{split}
\end{align}
with $c_{n,m} \equiv \sqrt{(n+m+1)(n-m)}$. 
We stress that all axial-vector couplings cancel out in this basis, even though we introduced \emph{chiral} fields. The same will hold true for all multiplets discussed in this article.
The underlying reason for this is the following: we constructed our fermion multiplets in such a way that they are permitted a mass even in the unbroken (parity invariant) $SU(2)_L\times SU(2)_R\times U(1)_{B-L}$ phase. 
Massive fermions can only be coupled to massless gauge bosons by means of vector interactions, because axial currents are anomalous (not conserved), so it is no surprise to find that our mass eigenstates are only coupled vectorially.
 We will however continue to label the particles with subscripts $L$ and $R$, which must not be confused with chiral projections.

\begin{figure}[t]
\includegraphics[width=0.47\textwidth]{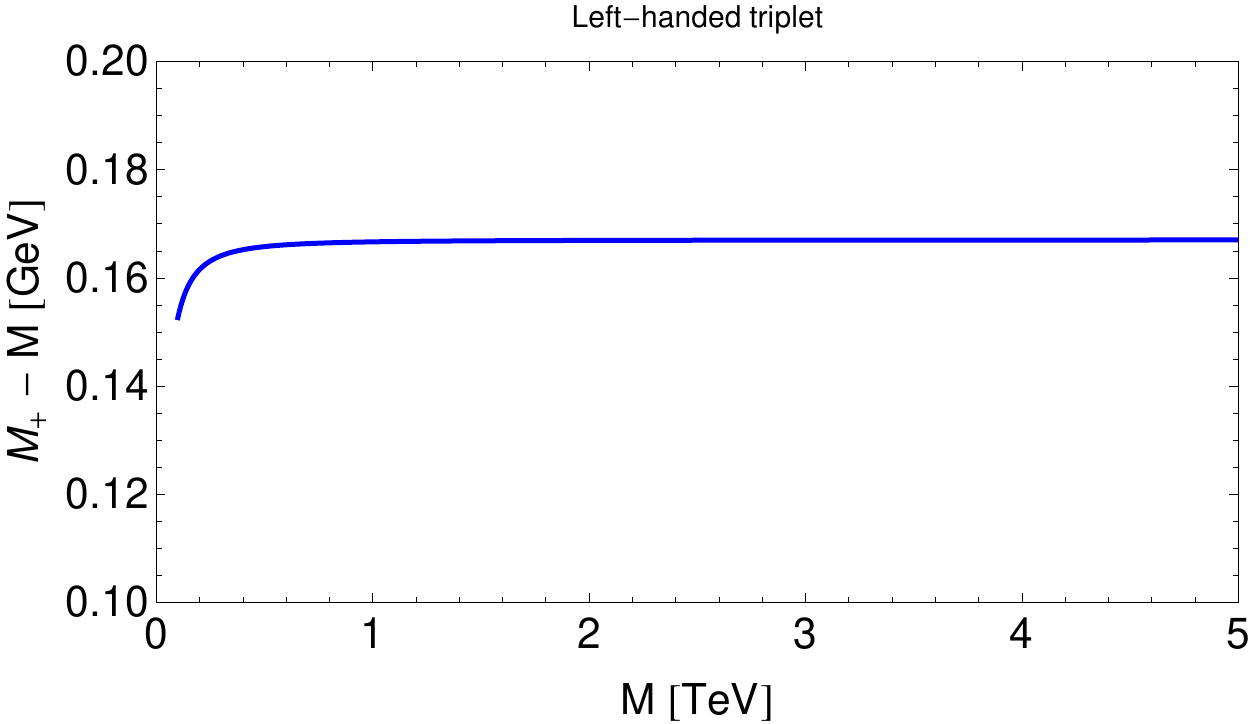}\hspace{2ex}
\includegraphics[width=0.47\textwidth]{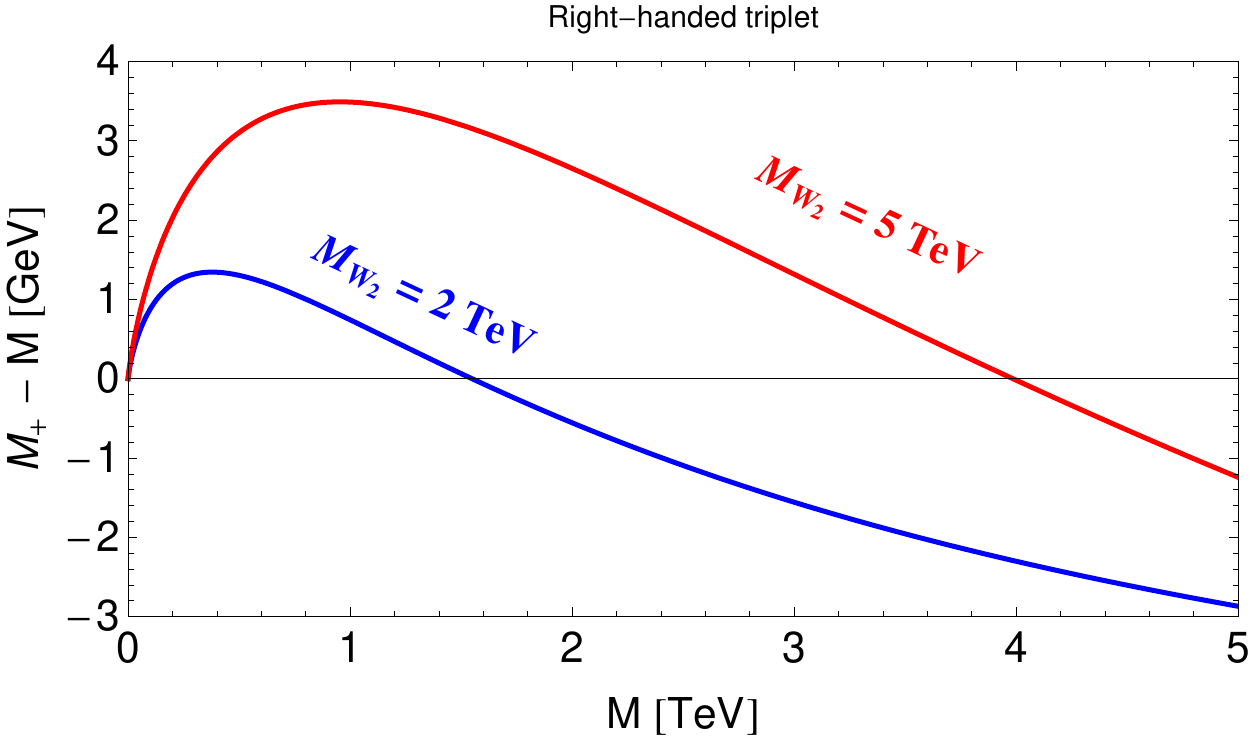}
\caption{Mass splitting $M_Q - M_0$ vs. $M$ for left-handed triplet (left) and right-handed triplet (right).}
\label{fig:mass-splitting}
\end{figure}

The mass splitting among the $\phi_L$ and $\phi_R$ components can be readily computed at one loop (see Appendix~\ref{app:mass_splitting}). Neglecting the gauge-boson mixing angles $\xi$ and $\phi$ for simplicity,\footnote{The gauge-boson mixing can be safely neglected because the mixing angles are of order $M_{Z_1}^2/M_{Z_2}^2$ and $f(r)$ only grows logarithmically for large $r$, $f(r)\to 6\log r$ (see Appendix~\ref{app:mass_splitting}).}
one obtains for the components of $\phi_L$~\cite{Cheng:1998hc,Feng:1999fu,Cirelli:2005uq}:
\begin{align}
\begin{split}
M_{\Psi_L^Q}-M_{\Psi_L^0} &\simeq \frac{\alpha_2}{4\pi} M Q^2 \left[ f(r_{W_1}) - c_W^2 f(r_{Z_1})-s_W^2 f(r_\gamma)\right]\\
&\simeq \alpha_2 Q^2 M_{W_1} \sin^2 (\theta_W/2) + \mathcal{O}(M_{W_1}^3/M^2) \,,
\end{split}
\end{align}
which is positive and evaluates to $Q^2\times \unit[167]{MeV} $, so the lightest (stable) component is indeed the neutral one.
Here and below, $r_V \equiv M_V/M$ for the vector bosons $V\in \{\gamma,Z_1,Z_2,W_1,W_2\}$.
Note that the mass splitting does not depend on $n$, but only on the electric charge $Q$ of the particles. A useful crosscheck in all of our mass-splitting formulae is invariance under a shift $f(r)\to f(r) +\text{const.}$, reflecting the result that the mass splitting is a finite loop effect.\footnote{An additional invariance for standard MDM without hypercharge is given by $f(r)\to f(r) +c_1 + c_2 r^2$ due to custodial symmetry, which is not present in the $SU(2)_R$ multiplets because of the $SU(2)_R$ breaking by scalar triplets.} 
For the right-handed multiplets we obtain~\cite{Heeck:2015qra}
\begin{align}
M_{\Psi_R^Q}-M_{\Psi_R^0} &\simeq \frac{\alpha_2}{4\pi} \frac{g_R^2}{g_L^2} M Q^2 \left[ f(r_{W_2}) - c_M^2 f(r_{Z_2})-s_W^2 s_M^2 f(r_{Z_1})- c_W^2 s_M^2 f(r_\gamma)\right] .
\label{eq:RH_mass_splitting}
\end{align}
In the limit $M\gg M_{W_2}$, this goes to $\alpha_2 Q^2 \left( M_{W_2} - c_M^2 M_{Z_2}\right)/2$, which is negative due to $M_{Z_2}/M_{W_2}\simeq \sqrt{2}/c_M\simeq 1.7$ and $c_M^2\simeq 0.7$. The lightest component of $\phi_R$ is neutral only for $M_{W_2}\geq 1.23\, M + \unit[0.1]{TeV}$ (see Fig.~\ref{fig:mass-splitting}), at least in our model with $g_L = g_R$ and $SU(2)_R$ broken by a right-handed triplet scalar $\Delta_R$. (LR models broken by doublets are safe from this constraint, but do not have a stabilizing symmetry~\cite{Ko:2015uma}.)
We will see in Sec.~\ref{sec:diboson} that this mass-splitting constraint weakens significantly or even disappears for $g_R<g_L$, as already noted in Ref.~\cite{Brehmer:2015cia}.

\subsubsection{Relic density}

Both $\Psi_L^0$ and $\Psi_R^0$ are separately stable and contribute to the DM density. In this work, we will assume that  scattering processes in the early Universe between the left- and right-handed sectors of the sort $\Psi_R \Psi_R \leftrightarrow \Psi_L \Psi_L$ are negligible because of the smallness of the relevant mass splittings. We stress, however, that this might not be always the case due to the thermal energy distributions and  potential non-perturbative effects similar to Sommerfeld enhancement. Under the assumption that such effects can be neglected, the two densities evolve independently of each other and the final abundance is hence simply the sum
\begin{align}
\Omega h^2 = \Omega_L h^2 + \Omega_R h^2\,.
\end{align}
For the experimental value we use the most recent result from Planck, $\Omega_\text{obs} h^2 = 0.1197\pm 0.0022$~\cite{Ade:2015xua}.
The abundance of $\Psi_L^0$, including the non-perturbative Sommerfeld effect, has been discussed in the literature both for the triplet (the wino case)~\cite{Hisano:2006nn} and quintuplet~\cite{Cirelli:2007xd,Cirelli:2009uv,Cirelli:2015bda}.  Instead of adapting these known results for $\Omega_L h^2$, we performed our own calculation in order to be able to compare our various candidates. For the right-handed contribution (and for the bi-multiplets discussed below) the corresponding calculation  has not been discussed in the literature and requires a dedicated analysis anyways. 

\begin{figure}[t]
\includegraphics[width=0.48\textwidth]{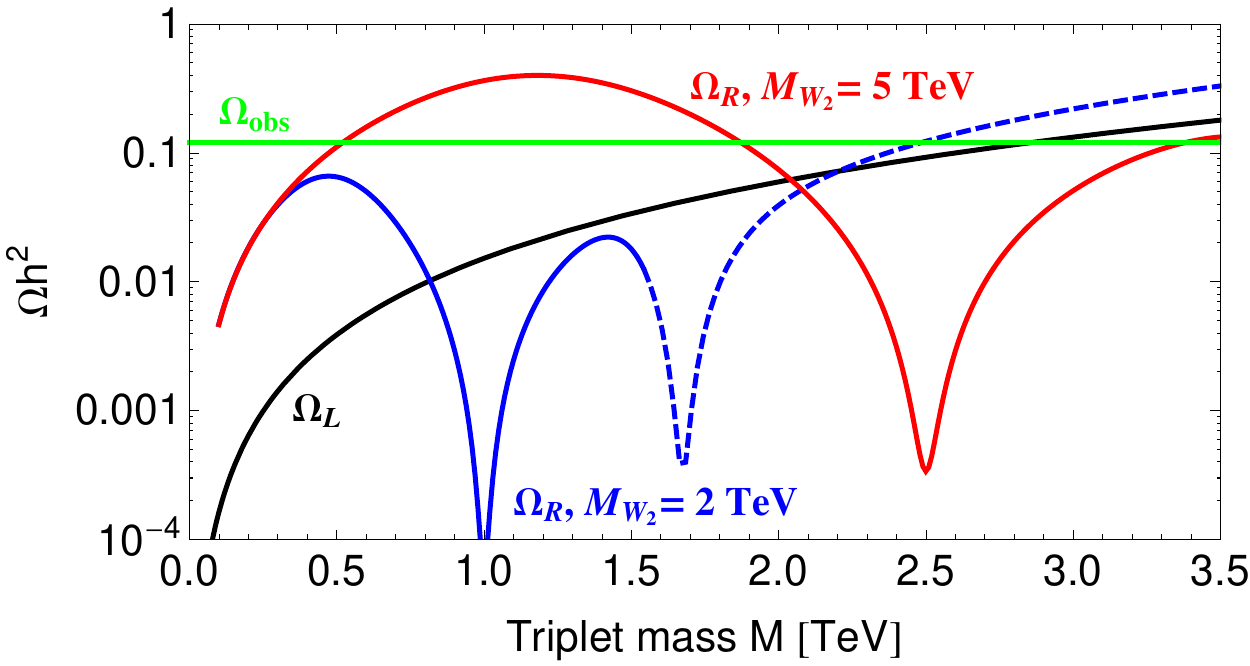}\hspace{2ex}
\includegraphics[width=0.48\textwidth]{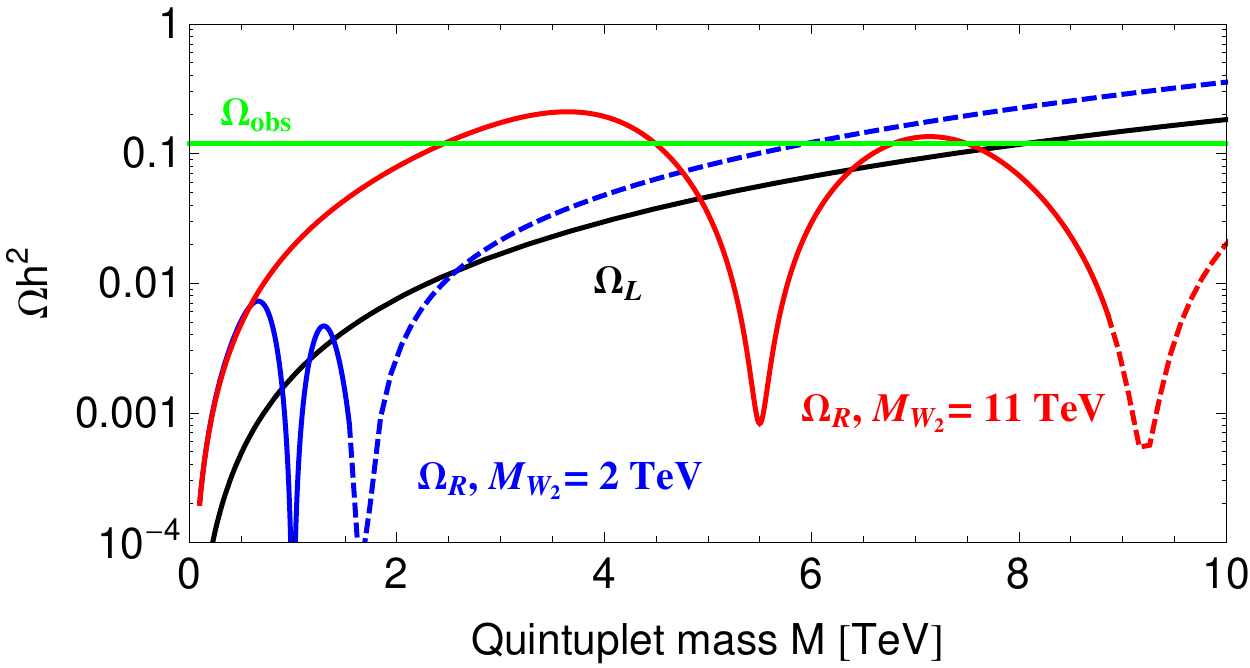}
\caption{Relic densities $\Omega_{L,R}$ for the LR fermion triplet (left) and quintuplet (right), obtained in the $SU(2)_L$-symmetric limit including Sommerfeld enhancement. $\Omega_R$ is shown for various $M_{W_2}$ masses in blue and red.
The dashed parts would have electrically charged DM and are hence excluded.
}
\label{fig:relic_densities}
\end{figure}

In this work, we consider the instantaneous freeze-out approximation for solving the corresponding Boltzmann equation in order to calculate these abundances. Furthermore, since our DM candidates are typically at the TeV scale, we must account for the Sommerfeld effect in the early Universe~\cite{Hisano:2006nn}. 
For simplicity, we will work in the $SU(2)_L$-symmetric limit~\cite{Cirelli:2009uv}, in which $W_1$ and $Z_1$ are massless and the mass splittings among the co-annihilating pairs are ignored. The calculation is detailed in Appendix~\ref{sec:SU2LL}. The key formula there is Eq.~\eqref{master}, which allows us to calculate the effective annihilation cross section at DM freeze-out and consequently the DM abundance by means of Eq.~\eqref{Omega}.

An additional approximation employed in this article is the omission of $W_2$ and $Z_2$ as final states of DM annihilation. This is obviously legitimate for $M < M_{W_2}$ (or $M<M_{W_2}/2$ if $W_L$--$W_R$ mixing is taken into account), which holds for most of our relevant parameter space. Let us note though that the opposite limit, $M\gg M_{W_2}$, allows us to perform calculations in the $SU(2)_L\times SU(2)_R$-symmetric limit, including \emph{right-handed} Sommerfeld enhancement. In this limit, and for $g_R=g_L$, the abundance $\Omega_R$ approaches $\Omega_L$ due to the enhanced symmetry. Since both the lower limit on $M_{W_2}$ and the upper limit on $M$ (from $\Omega_L$) are  in the TeV range, the limit $M\gg M_{W_2}$ is however difficult to achieve.

\begin{figure}[t]
\includegraphics[width=0.48\textwidth]{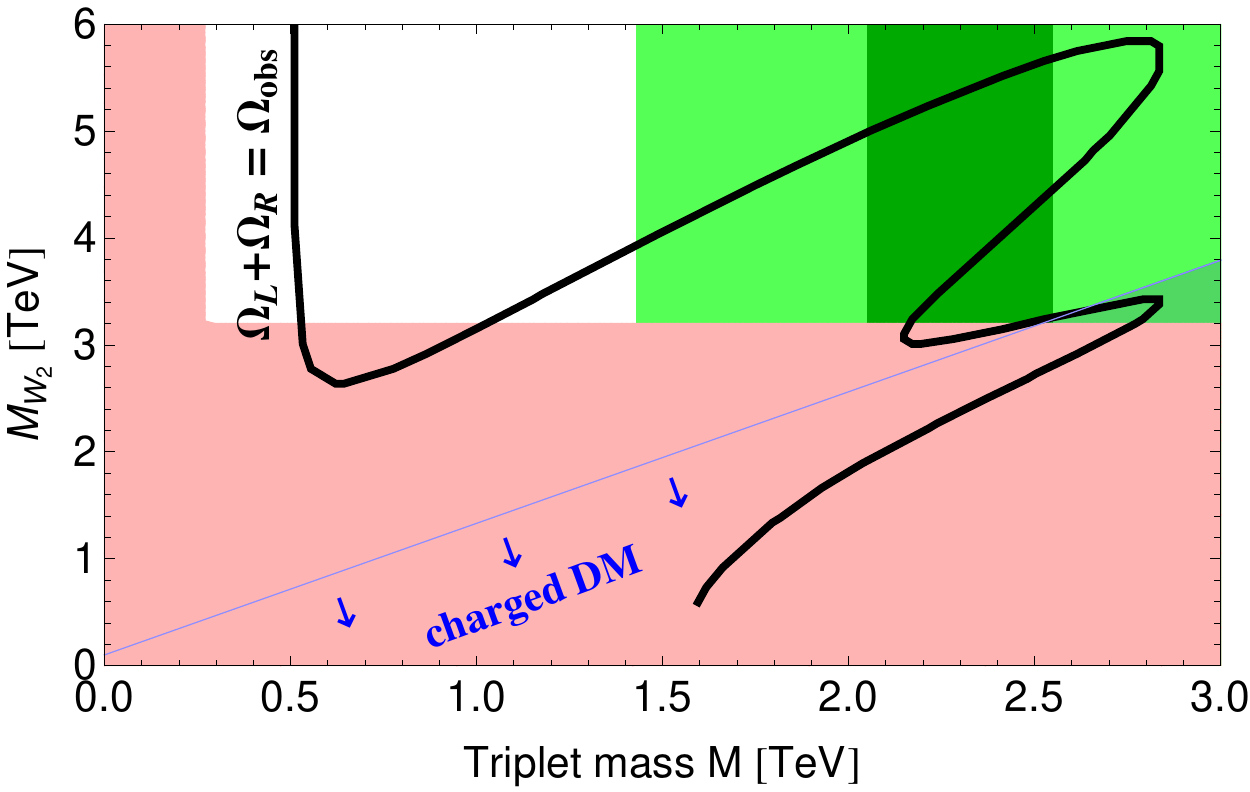}\hspace{2ex}
\includegraphics[width=0.48\textwidth]{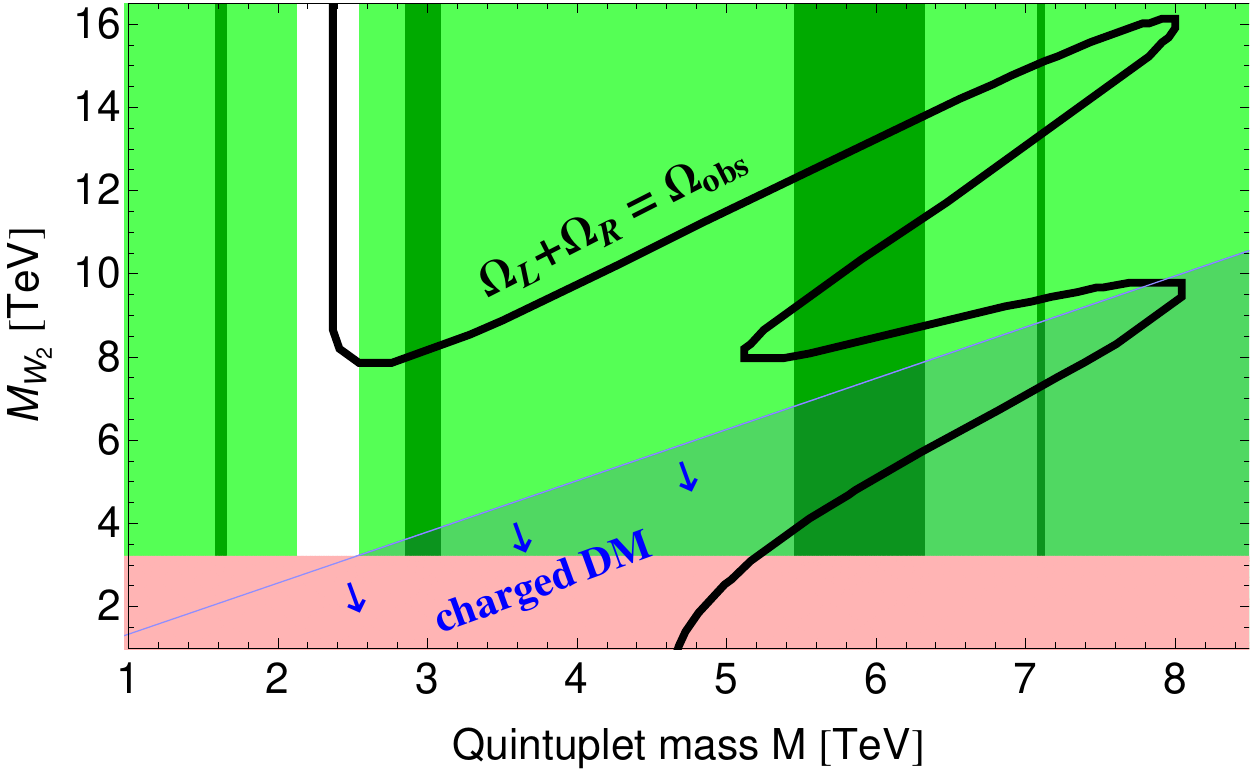}
\caption{Valid relic density $\Omega = \Omega_L + \Omega_R$ (black) for LR triplet (left) and quintuplet (right). The area to the upper right of the black curve would yield too much DM and is hence excluded.
The blue-shaded area is excluded by charged DM, the red region by LHC/meson data.
The (dark) green regions are excluded by gamma-ray line searches for an (isothermal) Einasto DM profile. 
}
\label{fig:MvsMWR}
\end{figure}

The resulting relic density for all our candidates in the $SU(2)_L$-symmetric limit depends only on the DM mass $M$, $M_{W_2}$, and very mildly on $\sin\beta$ (and on the ratio between $g_R$ and $g_L$ when they are different).
We show the relic density as function of DM mass for two choices of $M_{W_2}$ in Fig.~\ref{fig:relic_densities}, for the triplet and quintuplet representations, separating the left- and right-handed components for the sake of illustration. 
Notice that the dependence on $M_{W_2}$ only appears due to co-annihilation channels involving $W_2$ or $Z_2$ resonances (as follows from Tables~\ref{table:GVmultiplet},~\ref{table:bidoublet} and ~\ref{table:bitriplet} in Appendix~\ref{sec:SU2LL}). Consequently, points in the plane $M_{W_2}$\,vs.\,$M$ in agreement with the observed DM relic density can be divided into three regions: one that is away from the resonance lines in which the relic density is satisfied regardless of the $M_{W_2}$ value, and two of them associated to the resonances around $2 M = M_{W_2}$ and $2 M = M_{Z_2}$.  In Fig.~\ref{fig:MvsMWR}, we show these planes for the triplet and quintuplet representations. In addition, we show the different constraints that apply: LHC searches and meson observations as well as limits from indirect DM searches, to be discussed in the next subsections. 

One more comment is in order:
The charged components $\Psi^Q_X$ will decay via $W_X^-$ into $\Psi^{Q-1}_X$ plus SM particles until the neutral component $\Psi_X^0$ is reached. Assuming a sufficiently high reheating temperature $T_\mathrm{reh}\gg M$, the charged components will unavoidably be produced in the thermal bath via their coupling to photons, independently of the $Z_2$ and $W_2$ masses, thus making a freeze-in mechanism impossible. Since the lifetime of the right-handed state $\Psi_R^+$ will be of the form $M_{W_2}^4/\alpha_R^2 (M_{\Psi_R^+}-M_{\Psi_R^0})^5$, we can find an \emph{upper bound} on $M_{W_2}$ by demanding the charged states to have decayed e.g.~at the time of Big Bang nucleosynthesis~\cite{Mambrini:2015vna}. For $M=\unit[1]{TeV}$ ($\unit[10]{TeV}$), a lifetime below one second gives the rough bound $M_{W_2}< \unit[10^4]{TeV}$ ($\unit[10^5]{TeV}$). The decay is always fast enough for the freeze-out scenarios with low-scale LR symmetry discussed in this article, but would become problematic for higher LR scales, e.g.~at the scale of grand unification~\cite{Mambrini:2015vna}.
A more precise discussion of the resulting constraints goes beyond the scope of this paper.

\subsubsection{Indirect detection}

In this work, we consider the indirect detection limits from gamma-ray searches since they provide the most robust constraints for TeV-scale DM. In order to do so, we follow closely the procedure described in Ref.~\cite{Garcia-Cely:2015dda} and use the gamma-ray flux measured with the H.E.S.S.~telescope in a target region of a circle of $1^\circ$ radius centered in the Milky Way Center, excluding the Galactic Plane by requiring $|b|\geq 0.3^\circ$~\cite{Abramowski:2011hc, Abramowski:2013ax}. The expected gamma-ray flux from DM annihilations in that region of the sky is given by
\begin{align}
 \frac{\dd\phi_\gamma}{\dd E_\gamma}=\frac{J}{8 \pi M^2}\frac{\dd (\sigma v)  }{\dd E_\gamma}  \,,
\label{flux}
\end{align}
where  $J$ is an astrophysical factor, calculated by integrating the square of the DM density profile of the Milky Way over the line of sight and the region of interest.  In order to asses the astrophysical uncertainties associated with the DM distribution, we consider two DM halo profiles in this work: the isothermal profile, which describes a cored DM distribution and hence provides more conservative bounds, and the Einasto profile~\cite{Navarro:2003ew, Graham:2005xx}, which is more cuspy and thus leads to more constraining limits. In both cases we take the astrophysical parameters, in particular the $J$-factors, from Ref.~\cite{Garcia-Cely:2015dda}. Also, in our analysis we focus on gamma-rays with energies in-between $\unit[0.5]{TeV}$ and $\unit[20]{TeV}$. 

\begin{figure}[t]
\includegraphics[width=0.48\textwidth]{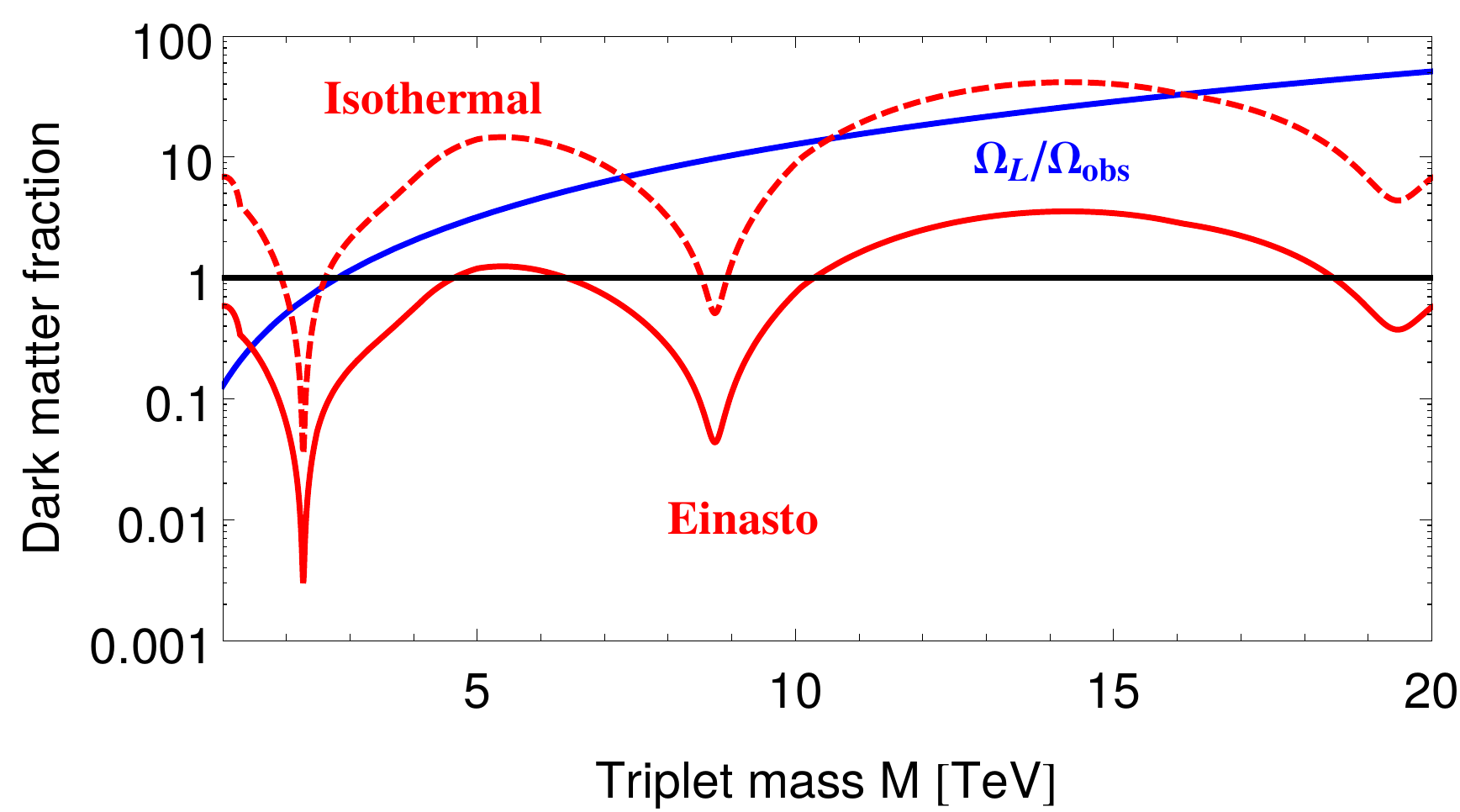}\hspace{2ex}
\includegraphics[width=0.48\textwidth]{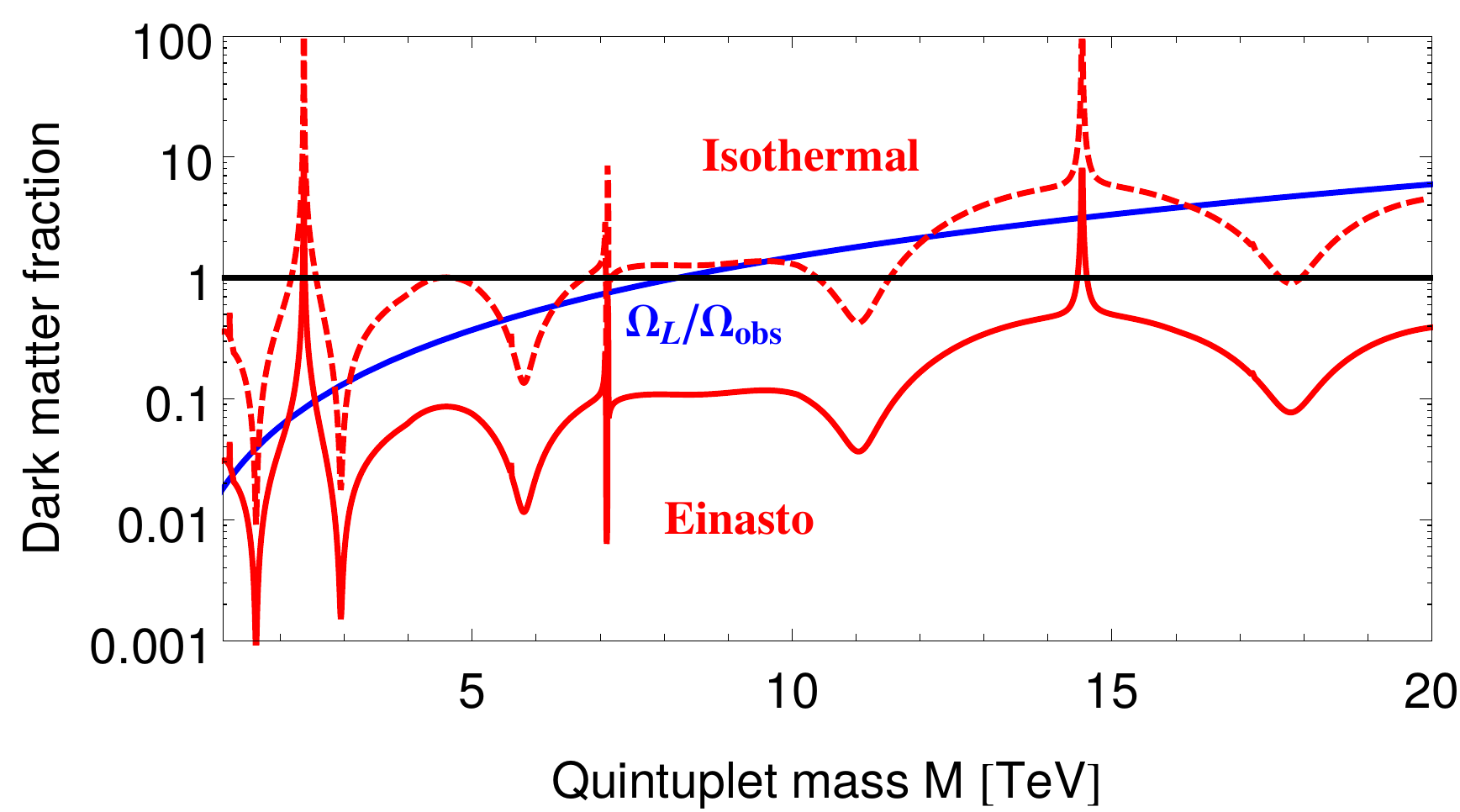}
\caption{
95\%~C.L.~limits on the DM fraction for fermionic triplet (left panel) and quintuplet (right panel) from the non-observation of monochromatic photons by H.E.S.S., assuming the Einasto profile (solid red line) and the isothermal profile (dashed red line). In blue we show our calculated thermal abundance $\Omega_L$ of the left-handed triplet and quintuplet.
}
\label{HESS:limits}
\end{figure}

Since we are dealing with TeV DM candidates, the cross sections entering in Eq.~\eqref{flux} must account for the Sommerfeld effect. This phenomenon arises because the DM particles are subject to long-range forces mediated by $W, Z$ and $\gamma$ exchange during the annihilation process, which modify their wave functions and therefore the corresponding cross sections. In Appendix \ref{sec:SEinGC}, we give all the details concerning the calculation of the Sommerfeld effect in the center of the galaxy, and describe the procedure we follow to calculate the cross section in Eq.~\eqref{flux}. This receives contributions from at least two parts: the featureless \emph{continuum} of gamma-rays arising in the decay and fragmentation of the $W$ and $Z$ bosons, and the \emph{monochromatic} photons associated to the annihilation into $\gamma\gamma$ and $\gamma\,Z$ final states.
The non-observation by H.E.S.S.~of a monochromatic spectral feature or an exotic featureless contribution to the gamma-ray flux allows us to set constraints on those cross sections.  We present these constraints in terms of the DM fraction, which is given  by the square root of the signal normalization factor that would exclude the signal at  95\%~C.L.  We find that  constraints coming from the lines are always more important than those from the continuum, and this is what we report in Fig.~\ref{HESS:limits} for the triplet and quintuplet representations. There, we also show the DM fraction associated to the corresponding left-handed component from our thermal freeze-out calculation. We stress that only  $\Psi_L$ gives rise to these gamma-ray limits because the annihilations of $\Psi_R$ are mediated by non-resonant processes involving $W_2$ and $Z_2$ bosons and are therefore highly suppressed. Modified LR models with only $\Psi_R$ would thus easily evade indirect detection constraints.

\begin{figure}[t]
\includegraphics[width=0.48\textwidth]{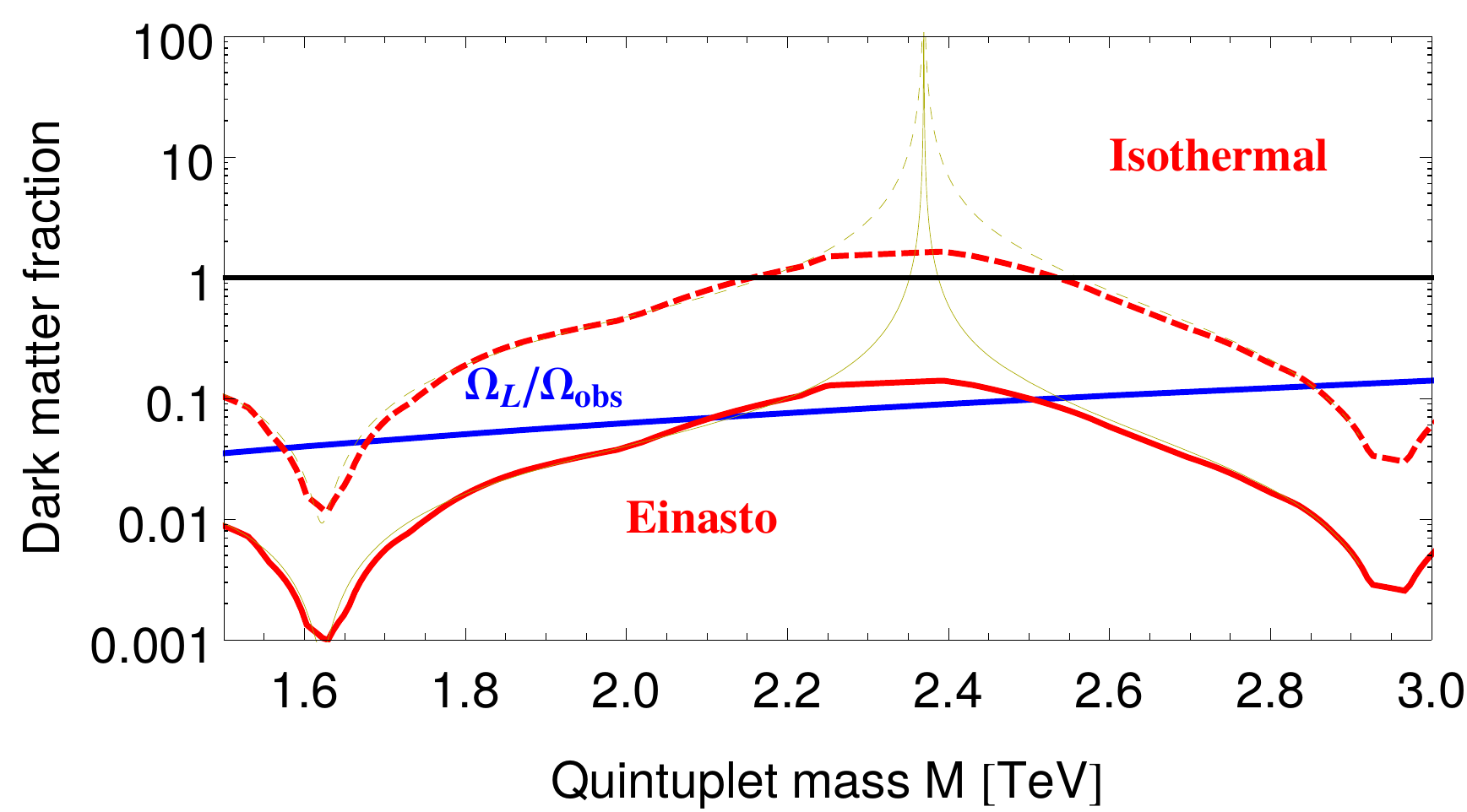}
\caption{
Zoom in of the right panel of Fig.~\ref{HESS:limits} around the region $2.1$--$\unit[2.5]{TeV}$. The red limit now shows H.E.S.S. limits after accounting for the contribution of virtual internal bremsstrahlung, following Ref.~\cite{Garcia-Cely:2015dda}, which softens the Ramsauer--Townsend dip (thin yellow line).
}
\label{HESS:limitsIB}
\end{figure}

For the triplet and the quintuplet we summarize these limits in Fig.~\ref{fig:MvsMWR}. As it is clear from the plot, the regions around the $W_2$ and $Z_2$ resonances  are excluded by indirect searches with the Einasto profile and only the mass $M\simeq\unit[0.5]{TeV}$ ($M\simeq\unit[2.4]{TeV}$) remains viable for the triplet (quintuplet). The remaining allowed region between $2.1$--$\unit[2.5]{TeV}$ for the quintuplet is of particular interest because, as illustrated in~Fig.~\ref{fig:MvsMWR}, the corresponding  limits from monochromatic photons are not constraining at all there. This is the  Ramsauer--Townsend effect~\cite{Chun:2015mka}, and appears because of a non-perturbative destructive interference between the Sommerfeld enhancement factors. It was shown in Ref.~\cite{Garcia-Cely:2015dda} that this effect can be circumvented by including the contribution from virtual internal bremsstrahlung  in the gamma-ray spectrum, which, for our models, corresponds to final states $WW\gamma$. These processes produce a line-like spectral feature which can mimic monochromatic photons for the energy resolutions of current telescopes~\cite{Bergstrom:1989jr,Flores:1989ru,Beacom:2004pe,Bergstrom:2004cy,Bergstrom:2005ss,Bringmann:2007nk,Garcia-Cely:2013zga}.  By adapting the procedure described in  Ref.~\cite{Garcia-Cely:2015dda}, we carefully re-derived the limits around $M\simeq\unit[2.4]{TeV}$ for the quintuplet, but this time taking into account the virtual internal bremsstrahlung. We find that region is still allowed, assuming the Einasto profile for the DM distribution, as shown in Fig.~\ref{HESS:limitsIB}. 
Furthermore, by employing the \unit[112]{h} prospect limits on line-like features for the upcoming Cherenkov Telescope Array (CTA)~\cite{Garcia-Cely:2015dda}, we find that this viable region can be reduced -- but not fully excluded -- assuming again the Einasto profile.

\subsubsection{Direct detection}

The charged-current interactions of the Majorana DM candidates lead to DM--nucleon scattering at loop level. (The mass splitting between the neutral and charged multiplet component being too large for inelastic scattering at tree level.) For the left-handed component, a careful analysis at next-to-leading order in the strong coupling constant gives the spin-independent cross section (off a proton)
\begin{align}
\sigma_\text{SI}^p \simeq \unit[2\times 10^{-47}]{cm^2}\left(\frac{n(n+1)}{2}\right)^2 ,
\label{eq:direct_detection}
\end{align}
for DM masses above $ \unit[300]{GeV}$~\cite{Hisano:2015rsa}.
For the triplet ($n=1$) and quintuplet ($n=2$) cases of interest here, this is below the current LUX limit~\cite{Akerib:2015rjg} but in principle testable since it is above the coherent neutrino scattering cross section~\cite{Hisano:2015rsa,Cushman:2013zza}. The direct-detection rate is maximal if the left-handed component dominates the DM abundance, i.e.~for masses around $\unit[3]{TeV}$ and $\unit[8]{TeV}$, respectively, 
which is however still too small for XENON1T~\cite{Aprile:2015uzo} and maybe even LZ~\cite{Malling:2011va}.
For lower masses, the cross section decreases by $\Omega_L/\Omega_\text{obs} \simeq (M/\unit[2.8]{TeV})^2$ and thus becomes even more difficult to detect. In particular, for the lowest triplet mass that still provides $100\%$ DM, $M\simeq \unit[500]{GeV}$, the fraction $\Omega_L/\Omega_\text{obs}$ is about $3\%$, which leads to an unobservably tiny cross section.
The scattering of $\Psi_R^0$ via \emph{right-handed} gauge bosons is more suppressed due to larger gauge boson masses and small gauge-mixing angles, so we expect no signal if the right-handed DM component dominates.

\subsubsection{Collider signatures}

For the left-handed triplet, ATLAS gives a lower limit of $M>\unit[270]{GeV}$ at $95\%$~C.L.~from the $\sqrt{s}=\unit[8]{TeV}$ run~\cite{Aad:2013yna}, and we expect a sensitivity to about $\unit[500]{GeV}$ at the high-luminosity (HL) LHC, and up to $\unit[3]{TeV}$ at a future $\unit[100]{TeV}$ collider~\cite{Cirelli:2014dsa}. Roughly the same limits hold for the left-handed quintuplet~\cite{Ostdiek:2015aga}.
As can be seen from Fig.~\ref{fig:MvsMWR}, the limit $M_{W_2} \gg M$ requires a triplet mass of $\unit[500]{GeV}$ and is thus potentially in reach of the HL LHC. Together with the correct relic density, this can put an \emph{upper} bound on $M_{W_2}$, as only the $M_{W_2}\sim 2 M$ resonance region would survive these collider constraints.
The left-handed quintuplet is unfortunately out of the LHC's reach if it provides all of the universes DM.

Additional constraints arise from the right-handed sector, most importantly from searches for $W_R$. We will go into more detail in Sec.~\ref{sec:diboson}, for now we only mention that ATLAS and CMS are expected to be sensitive to $M_{W_2}$ up to $\unit[6]{TeV}$~\cite{Ferrari:2000sp,Gninenko:2006br}, while future LHCb and Belle II data can reach $7$--$\unit[8]{TeV}$~\cite{Bertolini:2014sua} in indirect searches by studying mesons. The simplest model for left--right-symmetric dark matter -- the triplet -- is hence completely testable using accelerator experiments.

\subsection{Bi-doublet \texorpdfstring{$(\vec{2},\overline{\vec{2}},0)$}{(2,2,0)}}
\label{sec:bidoublet}

Chiral bi-multiplets $(\vec{n},\vec{n},0)$, $n\in \mathbb{N}$ also allow for a ``Majorana'' mass term and contain neutral components. We will only consider the two simplest examples, the fermion bi-doublet $(\vec{2},\overline{\vec{2}},0)$ and the bi-triplet $(\vec{3},\vec{3},0)$.\footnote{While finalizing this work we became aware of the preprint~\cite{Boucenna:2015sdg}, which discusses \emph{two} fermion bi-doublets in low-scale left--right models within $SO(10)$ theories.}

A chiral bi-doublet fermion $\sim (\vec{2},\overline{\vec{2}},0)$ gives -- at tree-level -- rise to one charged ($\Psi^+$) and one neutral ($\Psi^0$)  Dirac fermion, with degenerate mass $M$ and gauge interactions
\begin{align}
\begin{split}
\L &= \tfrac{1}{2} \overline{\Psi}^0 \left( g_L \slashed{W}^3_L- g_R \slashed{W}^3_R\right) \Psi^0 
+ \tfrac{1}{2} \overline{\Psi}^+ \left( g_L \slashed{W}^3_L+ g_R \slashed{W}^3_R\right) \Psi^+\\
&\quad+ \tfrac{1}{\sqrt{2}}\left[ g_R \overline{\Psi}^0 \slashed{W}^-_R \Psi^+ -  g_L\overline{\Psi}^- \slashed{W}^-_L \Psi^0 + \hc\right] .
\end{split}
\label{eq:bidoublet_lagrangian}
\end{align}
One can identify the parity symmetry $(\Psi^+, \Psi^0)\stackrel{\P}{\longleftrightarrow} (\Psi^+, (\Psi^0)^c)$.
The bi-doublet can be written as a self-conjugate ($\tilde \Psi \equiv \epsilon \Psi^c \epsilon = \Psi$) field in the following $2\times 2$ matrix form
\begin{align}
\Psi = \matrixx{\Psi^0 & \Psi^+ \\ \Psi^- & - (\Psi^0)^c}.
\end{align}
(Note that considering the representation $(\vec{2},{\vec{2}},0)$ instead of $(\vec{2},\overline{\vec{2}},0)$ is only a change in notation and gives the same physics.) 
The radiative mass splitting in the limit $M\gg M_{Z}$ is simply $M_{\Psi^+} - M_{\Psi^0} \simeq \alpha M_{Z}/2 \simeq \unit[356]{MeV}$, neglecting gauge-boson mixing.
This result can be understood as follows. Even after $SU(2)_R$ breaking, $SU(2)_L$ invariance forces the bi-doublet components to be degenerate, because the two $SU(2)_L$ doublets within $\Psi$ are conjugates of each other. The masses hence split only after $SU(2)_L$ is broken, with a value well known from other models with radiative mass splitting of doublets~\cite{Thomas:1998wy,Cirelli:2005uq}.
Since the fields carry hypercharge $\pm 1/2$, one finds a coupling $\sim g_2/2 c_W$ of $\Psi^0$ to the light $Z$ boson, which makes it -- at first sight -- difficult to consider it as a dominant DM component due to direct-detection constraints.

\begin{figure}[t]
\raisebox{+0.2\height}{\includegraphics[width=0.45\textwidth]{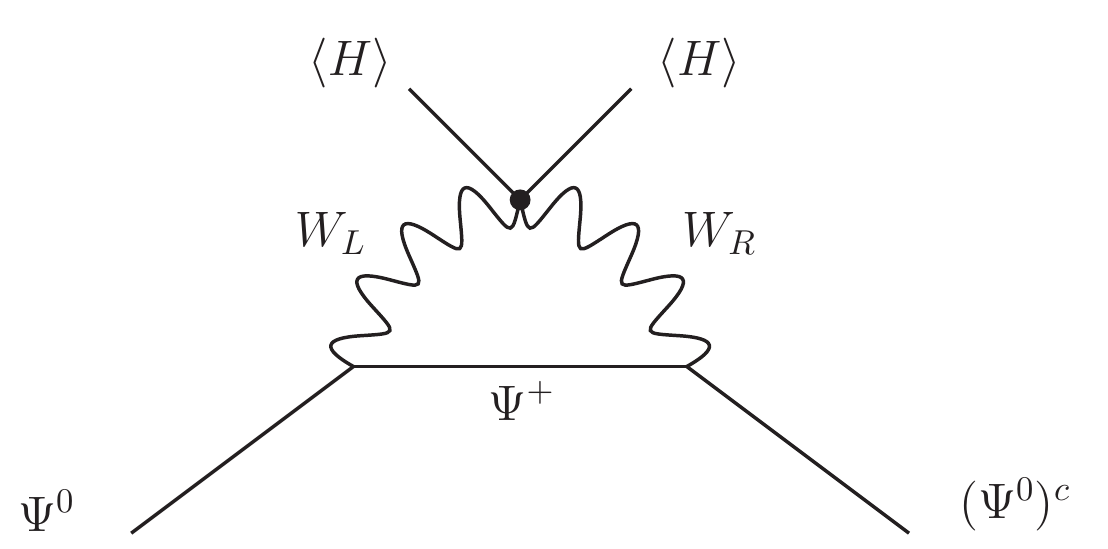} }
\includegraphics[width=0.5\textwidth]{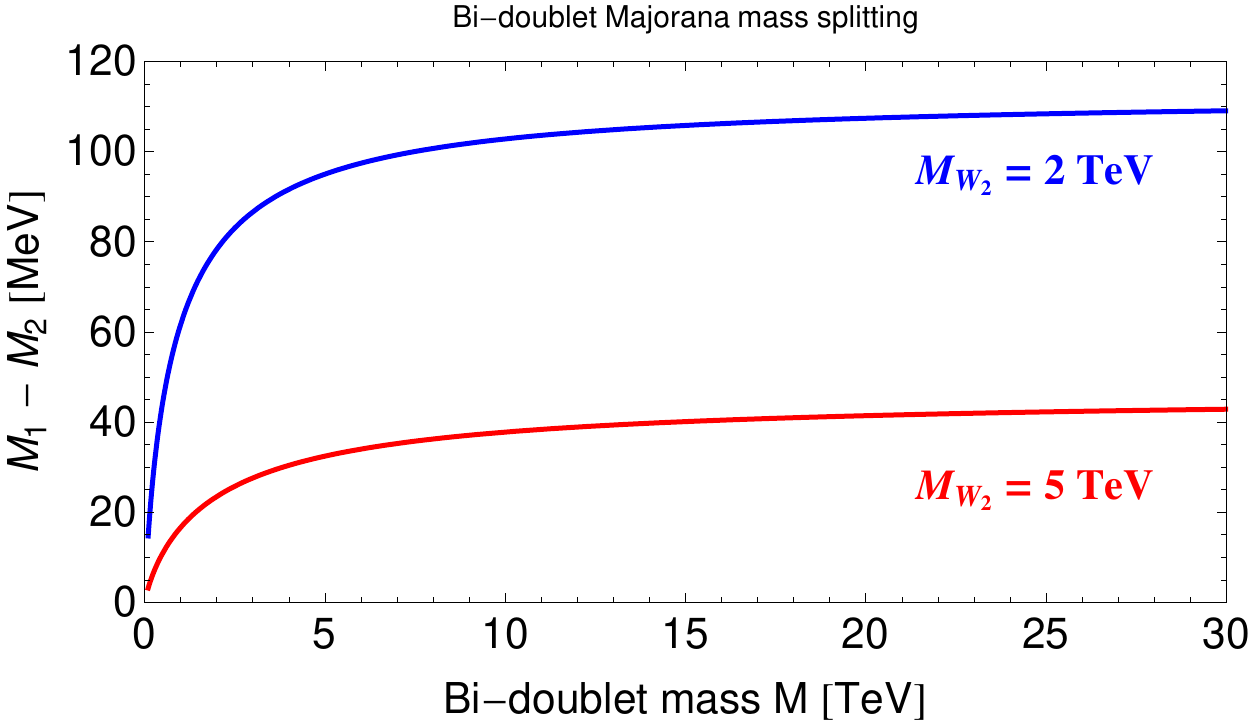} 
\caption{Left: Transition amplitude $\Psi^0 \to (\Psi^0)^c$ that splits the Dirac fermion $\Psi^0$ into two Majorana fermions $\Psi^0 = (\chi_1 + \chi_2)/\sqrt{2}$. The dot denotes $W_L$--$W_R$ mixing induced by the bi-doublet scalar $H$.
Right: Corresponding mass splitting of the two Majorana fermions $\chi_{1,2}$, taking $\kappa_1 = \kappa_2$.}
\label{fig:pseudo_dirac_feynman}
\end{figure}

Closer inspection of the Lagrangian in Eq.~\eqref{eq:bidoublet_lagrangian} reveals, however, that the Dirac nature of $\Psi^0$ is not protected by any symmetry and hence it actually splits into two quasi-degenerate Majorana fermions. The crucial observation here is that the Lagrangian still only has a $\mathbb{Z}_2$ symmetry among the new fermions, not a larger global $U(1)$ one would expect for a Dirac fermion. Writing $g_L\overline{\Psi}^- \slashed{W}^-_L \Psi^0 = - g_L\overline{(\Psi^0)^c} \slashed{W}^-_L \Psi^+$ shows that there is definitely no $U(1)$ associated with $\Psi$ number;
one can, however, still identify a global $U(1)$ symmetry that potentially protects the Dirac nature of $\Psi^0$ by exploiting the complex nature of the $W$~bosons,
\begin{align}
\Psi^0 \to e^{i \alpha} \Psi^0\,,\quad
\Psi^+ \to \Psi^+\,, \quad
W_R^- \to e^{i \alpha} W_R^-\,,\quad
W_L^- \to e^{-i \alpha} W_L^-\,.
\end{align}
This symmetry is obviously broken due to $W_L^-$--$W_R^-$ mixing (same as $\P$), and indeed one can draw a Feynman diagram that leads to a $\Psi^0\to (\Psi^0)^c$ transition, i.e.~splits the Dirac fermion into two Majorana fermions (see Fig.~\ref{fig:pseudo_dirac_feynman}). 
More accurately, we can decompose $\Psi^0 = (\chi_1 + \chi_2)/\sqrt{2}$, where $\chi_1 = -\chi_1^c$ and $\chi_2 = +\chi_2^c$ are degenerate Majorana fermions with opposite intrinsic CP charge. In terms of these fields the Lagrangian of Eq.~\eqref{eq:bidoublet_lagrangian} takes the form
\begin{align}
\begin{split}
\L &= \tfrac{1}{2} \overline{\chi}_1 \left( g_L \slashed{W}^3_L- g_R \slashed{W}^3_R\right) \chi_2 
+ \tfrac{1}{2} \overline{\Psi}^+ \left( g_L \slashed{W}^3_L+ g_R \slashed{W}^3_R\right) \Psi^+\\
&\quad+ \tfrac{1}{2}\left[ \overline{\chi}_1 \left(g_R \slashed{W}^-_R -g_L \slashed{W}^-_L \right) \Psi^+ + \overline{\chi}_2 \left(g_R \slashed{W}^-_R + g_L \slashed{W}^-_L \right) \Psi^+ + \hc\right] .
\end{split}
\end{align}
The mass splitting between $\chi_1$ and $\chi_2$ can then be calculated using the formulae from Appendix~\ref{app:mass_splitting},
\begin{align}
\Delta M_{12} \equiv M_1 - M_2 = \frac{\alpha_2}{4\pi} \sin (2\xi) M \left[ f (r_{W_1}) - f(r_{W_2})\right] ,
\end{align}
which vanishes in absence of $W_L^-$--$W_R^-$ mixing ($\xi\to 0$) in accordance to the above discussion.
Even though the splitting is suppressed by the small $\xi \simeq -\sin 2\beta M_{W_1}^2/M_{W_2}^2$, it can easily be of order MeV; in Fig.~\ref{fig:pseudo_dirac_feynman} (right) we show the mass splitting for two values of $M_{W_2}$, taking $\kappa_1 = \kappa_2$ ($\beta=\pi/4$) in order to maximize $\xi$ and hence $|\Delta M_{12}|$.
(The splitting between charged and neutral states changes to some degree now, we have $M_+ - M_2 \simeq (\alpha M_Z - \Delta M_{12})/2$.)

Direct-detection phenomenology is radically changed by this
 mass splitting $\Delta M_{12}$ -- together with the fact that the neutral-current interactions can only lead to transitions between  $\chi_1$ and $\chi_2$  (the interaction vertex is $\overline{\chi}_1 \slashed{Z}_{1,2} \chi_2 $). The scattering becomes inelastic~\cite{TuckerSmith:2001hy}, and for a mass splitting larger than about $\unit[200]{keV}$, none of the constraints apply anymore~\cite{Nagata:2014aoa}.\footnote{This limit applies to $M_\text{DM}\sim \unit[600]{GeV}$ and goes down to $\unit[150]{keV}$ for $M_\text{DM}\sim \unit[30]{TeV}$~\cite{Nagata:2014aoa}.} (The constraint $|\Delta M_{12}|> \unit[0.2]{MeV}$ by itself leads to an upper bound of $M_{W_2} <\unit[10^3]{TeV}$, but much better upper bounds arise in combination with the relic density discussed below.)
The self-conjugate bi-doublet is hence a viable (and quite minimal) candidate for dark matter within low-scale left--right models.

\subsubsection{Relic density and indirect detection}

\begin{figure}[t]
\includegraphics[width=0.48\textwidth]{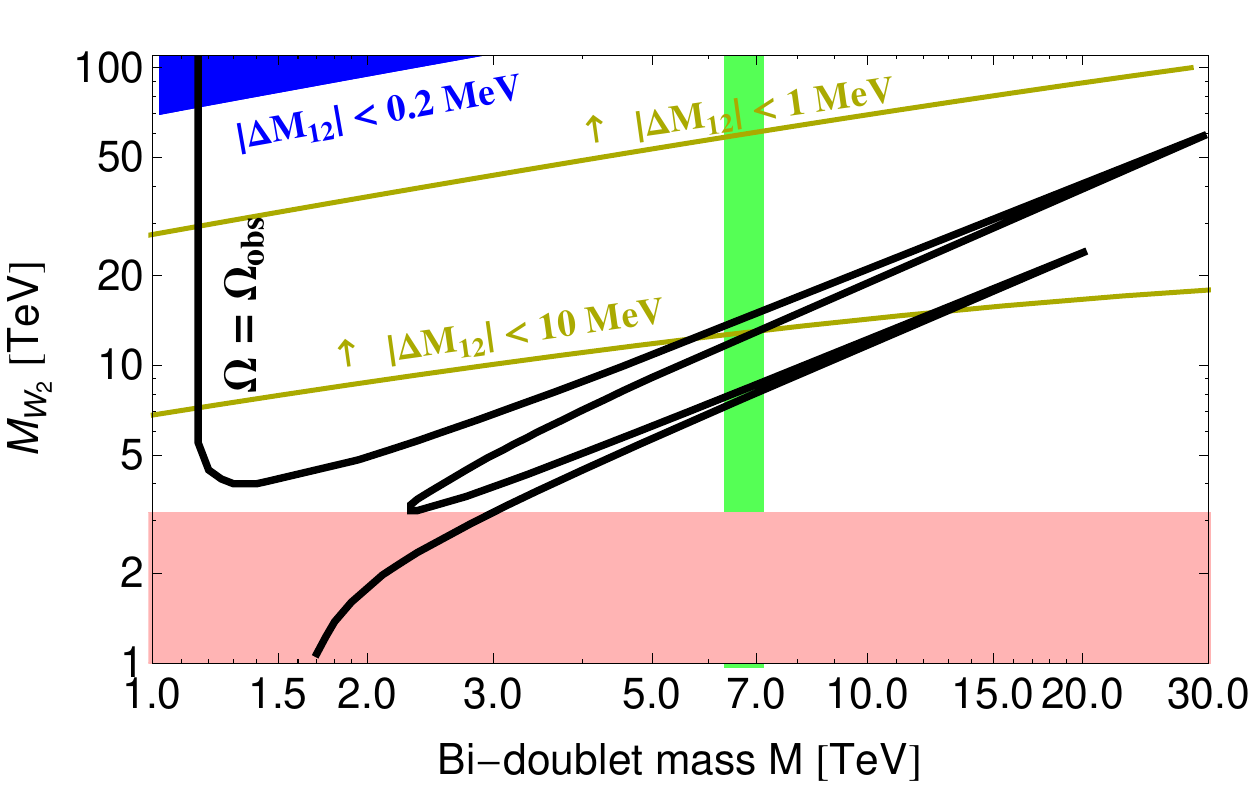}\hspace{2ex}
\includegraphics[width=0.48\textwidth]{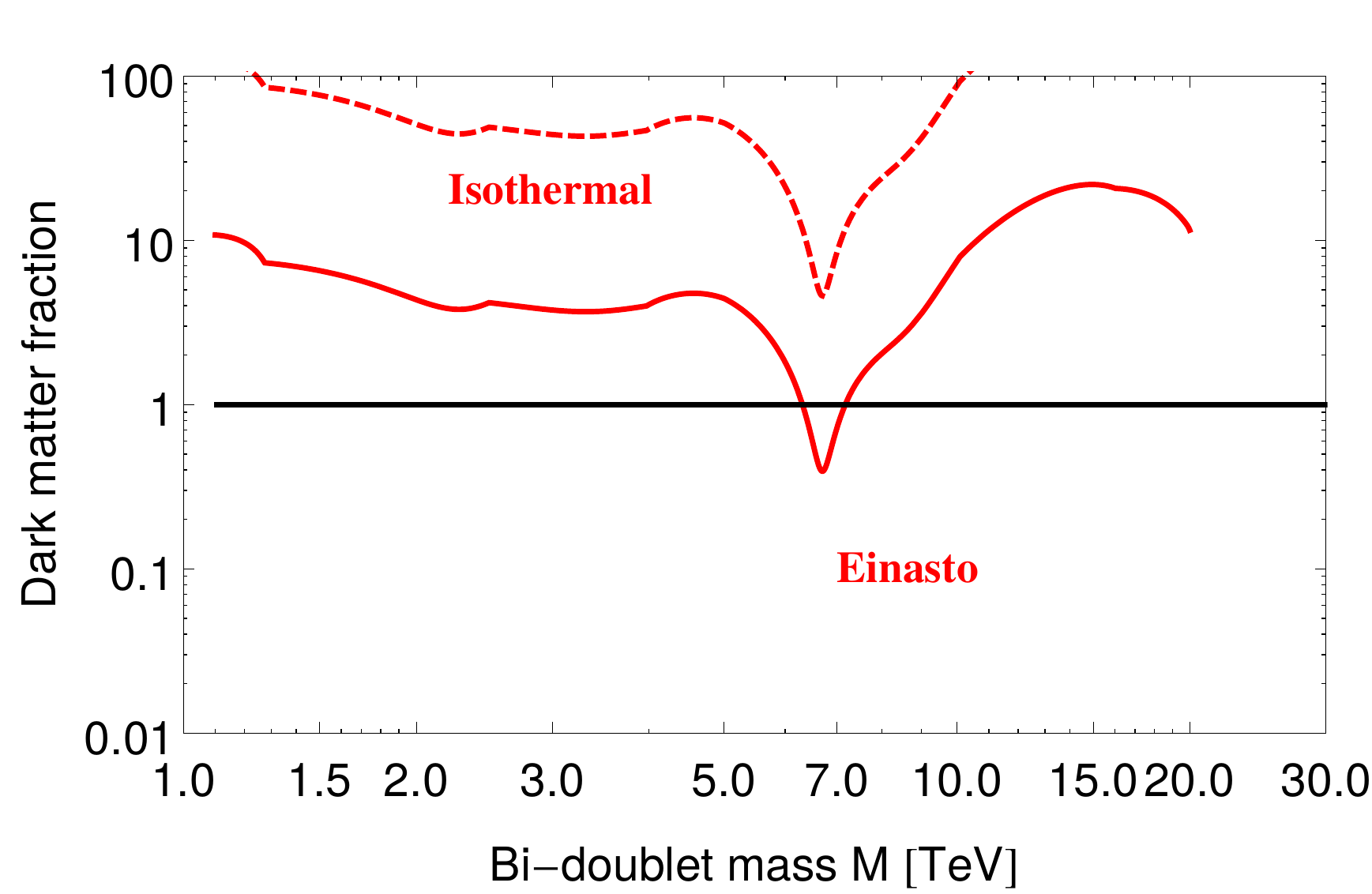}
\caption{
Left: Valid relic density for the fermion bi-doublet (black). The area to the upper right of the black line would yield too much DM and is hence excluded. The blue shaded region is excluded by direct-detection experiments because the mass-splitting $\Delta M_{12}$ between the two neutral Majorana fermions is smaller than $\unit[0.2]{MeV}$. In yellow we also show the regions with mass splitting $\Delta M_{12}<\unit[1]{MeV}$ and $\unit[10]{MeV}$ (depending on $\tan\beta$).
The green area is excluded by indirect detection for an Einasto DM profile.
Right: Same as Fig.~\ref{HESS:limits}, but for the bi-doublet. We assume $\Delta M_{12}=\unit[10]{MeV}$. The limits are however fairly insensitive to this value.} 
\label{fig:MvsMWR_bidoublet}
\end{figure}

As in the cases of the fermionic triplet and quintuplet, we calculate the relic density in the $SU(2)_L$-symmetric limit using formalism of Appendix~\ref{sec:SU2LL}. The corresponding results are shown in Fig.~\ref{fig:MvsMWR_bidoublet}. The DM mass can be as high as $M\sim\unit[30]{TeV}$ close to the $W_2$ or $Z_2$ co-annihilation resonances, out of reach of any terrestrial probe and even hard to search with indirect astrophysical methods.
The lower limit on the DM mass is $M\simeq \unit[1.2]{TeV}$ if we want to match the observed DM abundance -- a mass out of reach of the LHC. We can see from Fig.~\ref{fig:MvsMWR_bidoublet} (left) that the mass splitting $|\Delta M_{12}|$ between the two neutral Majorana fermions becomes smaller than $\unit[0.2]{MeV}$ for $M_{W_2}\gtrsim \unit[75]{TeV}$, which is then in conflict with direct detection experiments~\cite{Nagata:2014aoa}. 
So, even though the involved $W_2$ and $\Psi$ masses are very large, only a finite region of parameter space is viable.

For illustration purposes we also show other $\Delta M_{12}$ contours in Fig.~\ref{fig:MvsMWR_bidoublet} (left), which are obtained for $\beta = \pi/4$ in order to maximize the splitting. Nevertheless, the DM annihilation cross section, relevant for indirect detection, is barely sensitive to the mass splitting between the neutral components, and thus to the mixing angle $\beta$, because the most important splitting is the one between the charged and the neutral species, approximately equal to  $\unit[360]{MeV}$. Only the narrow region around $M\simeq 6$--$\unit[7]{TeV}$ is excluded at $95\%$~C.L.~by H.E.S.S.~line searches, assuming an Einasto profile. For an isothermal profile, no indirect-detection limits apply, as can be seen from Fig.~\ref{fig:MvsMWR_bidoublet} (right). Although CTA is expected to improve these limits by a factor of a few, it will also be able to probe only the region around the indirect detection resonance~\cite{Garcia-Cely:2015dda, Ibarra:2015tya}.

We should mention that the bi-doublet phenomenology is remarkably similar to the (split) supersymmetric Higgsino, i.e.~the fermionic superpartners of the two scalar doublets $H_{u,d}$ required for electroweak symmetry breaking~\cite{Giudice:2004tc}. These fermions obtain a common mass from the superpotential term $\mu H_u H_d$ and are split by radiative corrections exactly as in our case, assuming all additional non-SM fields are much heavier. In this Higgsino-like neutralino limit, the correct relic density is indeed obtained for $M\simeq \unit[1]{TeV}$~\cite{Hisano:2004ds,Cheung:2005pv,Cirelli:2007xd,Essig:2007az}, just as for our bi-doublet in the limit $M_{W_2} \gg M$. 
Furthermore, the bi-doublet indirect detection signatures are also similar to those of Higgsino DM. In particular, the Sommerfeld-effect matrices are the same and the Sommerfeld peak around $M \sim \unit[7]{TeV}$ (Fig.~\ref{fig:MvsMWR_bidoublet} (right)) is the one found in Ref.~\cite{Hisano:2004ds}.
Direct detection cross sections also follow from a recent Higgsino analysis and are found to be deep in the neutrino-background region, arguable impossible to probe~\cite{Hisano:2015rsa}.
The main phenomenological difference between bi-doublet and Higgsino is the mass splitting between the neutral fermions, $\Delta M_{12}$, which is purely radiative in our case.

\subsubsection{Bi-doublet decays}
\label{sec:bidoublet_decays}

To complete the discussion of the bi-doublet we collect the possible decay modes relevant for collider searches.
The charged component $\Psi^+$ will decay to the stable neutral component in complete analogy to MDM or Higgsinos, with a dominant rate into (soft) charged pions~\cite{Thomas:1998wy}
\begin{align}
\Gamma (\Psi^+\to \chi_j \pi^+) \simeq \frac{1}{2\pi} G_F^2 V_{ud}^2 (M_+ - M_j)^3 f_\pi^2 \sqrt{1-\frac{M_\pi^2}{(M_+ - M_j)^2}}\,.
\end{align}
Even though the coupling to $W_1^+$ is a factor $\sqrt{2}$ smaller, the decay rate of the bi-doublet is typically larger than the corresponding triplet decay rate~\cite{Cirelli:2005uq} because of the larger mass splitting. Neglecting $\Delta M_{12}$, this gives a $\Psi^+$ decay rate into pions roughly seven times larger than for the wino, i.e.~a decay length of about $\unit[0.7]{cm}$. The factor-two smaller production cross section further reduces the disappearing-track search sensitivity and puts it far below our region of interest. We refer the interested reader to the literature on Higgsino searches for further details.

\begin{figure}[t]
\includegraphics[width=0.8\textwidth]{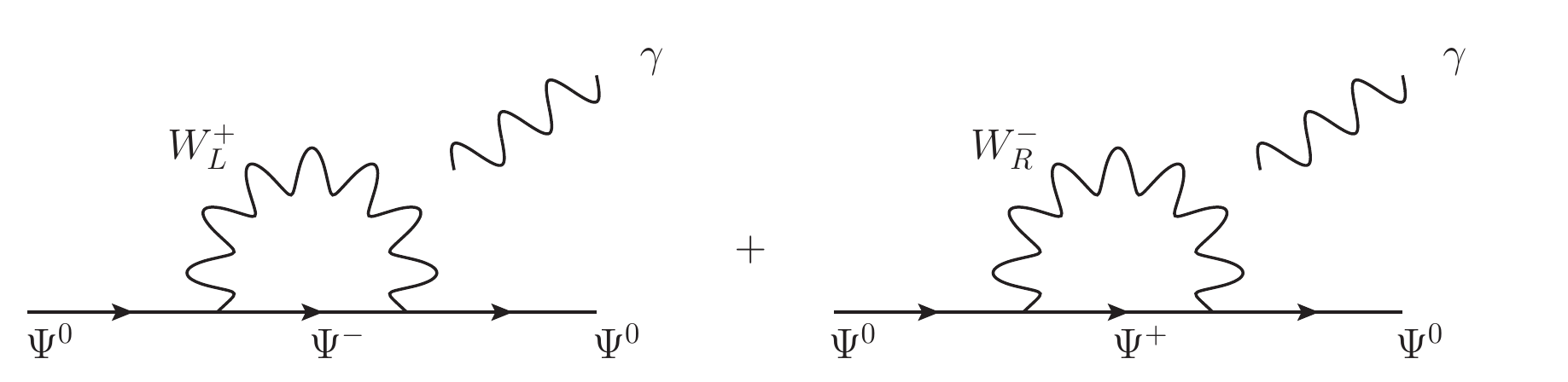} 
\caption{Feynman diagrams relevant for the magnetic moment of the Dirac fermion $\Psi^0$ of the bi-doublet, neglecting $W_L$--$W_R$ mixing. These operators also give rise to $\chi_2\to\chi_1\gamma$. The photon with four-momentum $q$ couples to the charged particles in the loop.}
\label{fig:magnetic_moment_Psi0}
\end{figure}

The heavier neutral state, say $\chi_2$, will decay at tree level into the DM state $\chi_1$ via neutral currents. For a mass splitting below MeV (twice the electron mass), only the three light active neutrinos are accessible, leading to a rate
\begin{align}
\Gamma (\chi_2 \to \chi_1 \nu\nu) \simeq \frac{7}{320\pi} \frac{\alpha_2^2}{c_W^4} \frac{|\Delta M_{12}|^5}{M_{Z_1}^4}\simeq \frac{1}{\unit[3\times 10^3]{s}} \left(\frac{\Delta M_{12}}{\unit[1]{MeV}}\right)^5 .
\end{align}
For $\Delta M_{12}\sim \unit[100]{keV}$, the lifetime is about eleven years, so the heavier neutral component will not be long-lived on cosmological scales if we want to satisfy constraints from direct-detection experiments. At loop-level the decay $\chi_2\to\chi_1\gamma$ opens up, with photon energy $E_\gamma \simeq |\Delta M_{12}|$.
The amplitude for this process can be conveniently derived by neglecting mass splittings and calculating the magnetic-moment form factor $F_2(q^2)$ of the neutral Dirac fermion $\Psi^0$ (see Fig.~\ref{fig:magnetic_moment_Psi0}). Without $W_L$--$W_R$ mixing, only four diagrams contribute in unitary gauge, courtesy of the global $U(1)$ symmetry described above. This yields $F_2(0) = e \alpha_2 [h(r_{W_1}) -h(r_{W_2})]/(8\pi)$, and ultimately the decay width
\begin{align}
\begin{split}
\Gamma (\chi_2 \to \chi_1 \gamma) &\simeq \frac{\alpha \alpha_2^2}{64\pi^2}  \frac{|\Delta M_{12}|^3}{M^2}\ \left|h\left(r_{W_1}\right)-h\left(r_{W_2}\right)\right|^2 ,
\end{split}
\label{eq:radiative_bidoublet_decay}
\end{align}
which coincides with a straightforward calculation of $\chi_2\to\chi_1\gamma$ following Ref.~\cite{Lavoura:2003xp} in the limit of small mass splitting. Here we have employed the loop function $h(r)$ 
\begin{align}
\begin{split}
h(r) &\equiv \frac{4}{r^2-4} \left[4-r^2+\left(4-5 r^2+r^4\right) \log r -r \sqrt{r^2-4} \left(r^2-3\right) \log \left(\frac{r+\sqrt{r^2-4}}{2}\right)\right]\\
&= -4 -4\log r+3 \pi  r+(-3+4 \log r) r^2 +\mathcal{O}(r^3)\,,
\end{split}
\end{align}
which is monotonically decreasing and approaches $h(r)\to 2/r^2$ for large $r$.
For $M=\unit[1]{TeV}$, Eq.~\eqref{eq:radiative_bidoublet_decay} gives a very short lifetime of $\tau (\chi_2\to\chi_1\gamma) \simeq \unit[10^{-3}]{s}\times (\unit[1]{MeV}/|\Delta M_{12}|)^3$, only weakly dependent on the $M_{W_2}$ values of interest here.
The radiative decay channel can hence easily dominate over the tree-level decay for a TeV DM mass and small mass splitting.

\subsection{Bi-triplet \texorpdfstring{$(\vec{3},\vec{3},0)$}{(3,3,0)}}

\begin{figure}[t]
\includegraphics[width=0.47\textwidth]{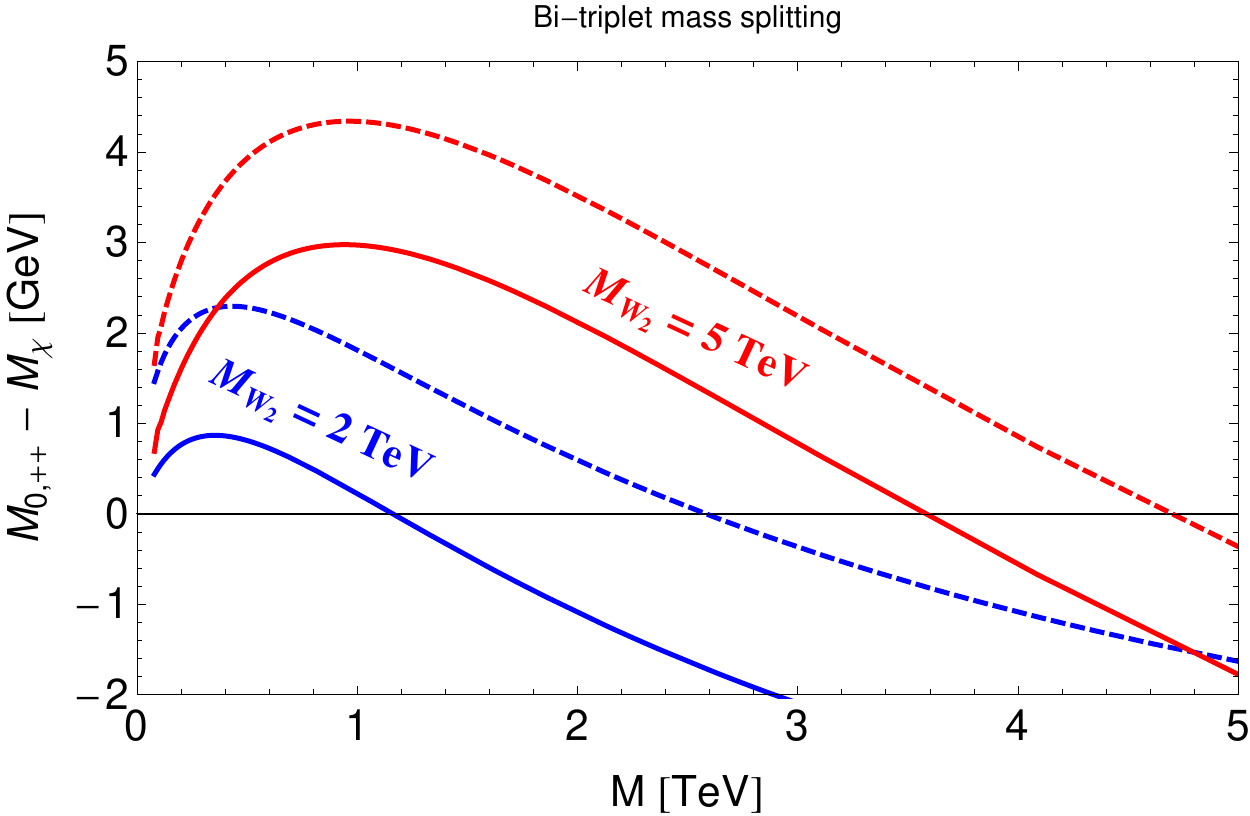} \hspace{2ex}
\includegraphics[width=0.47\textwidth]{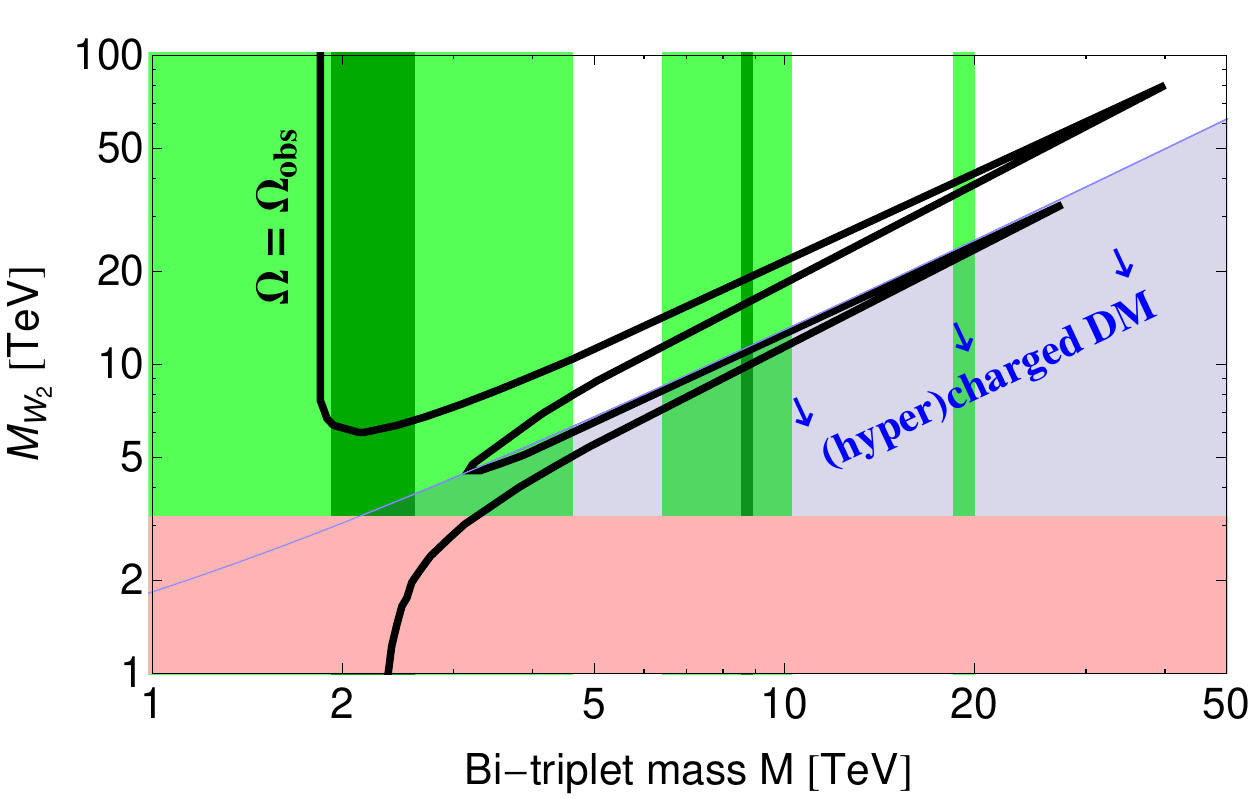}
\caption{Left: Mass splitting $M_{\Psi^{0,++}}-M_{\chi}$ vs. $M$ for the bi-triplet components $\Psi^0$ (solid lines) and $\Psi^{++}$ (dashed). Blue lines are for $M_{W_2} =\unit[2]{TeV}$, red lines for $\unit[5]{TeV}$.
Right: Valid relic density for the bi-triplet (black), using the Sommerfeld calculation. The blue-shaded area is excluded by direct detection experiments or charged DM, while the red regions are excluded by LHC searches and low-energy meson data. The green area is excluded by $\gamma$-ray line searches, (dark) green for (isothermal) Einasto profiles. Limits from dwarf galaxies further exclude the region $M \leq \unit[2.84]{TeV}$.}
\label{fig:bitriplet}
\end{figure}

Bi-multiplets of the form $(\vec{2n+1},\vec{2n+1},0)$ contain a neutral fermion without hypercharge and are thus naturally safe from direct detection constraints, without having to rely on inelastic scattering as in the bi-doublet case.
Here we will only discuss the simplest possibility, namely the bi-triplet $(\vec{3},\vec{3},0)$, but the phenomenology of the more general bi-multiplet $(\vec{2n+1},\vec{2n+1},0)$ will be similar.
Counting degrees of freedom already shows that one neutral component of $(\vec{3},\vec{3},0)$ is a Majorana fermion $\chi$, while the rest comes as Dirac fermions $\Psi^{++}$, $\Psi^{+}_{1,2}$, $\Psi^0$.
One can describe the bi-triplet $(\vec{3},\vec{3},0)$ as a matrix
\begin{align}
\Psi= \matrixx{\Psi^{++} & \Psi^+_1 & \Psi^0 \\ \Psi^+_2 & \chi & -\Psi^-_2 \\ (\Psi^0)^c & -\Psi^-_1 & \Psi^{--}} ,
\end{align}
where $SU(2)_L$ ($SU(2)_R$) acts in the vertical (horizontal) direction. It is self-conjugated because it fulfills the relation $\epsilon \Psi^c \epsilon = \Psi$ (see Appendix~\ref{app:real_reps}).
The gauge interactions are
\begin{align}
\begin{split}
\L &= \overline{\Psi}^{++}\left( g_L \slashed{W}^3_L+ g_R \slashed{W}^3_R\right) \Psi^{++} 
+g_L \overline{\Psi}^+_1 \slashed{W}^3_L \Psi^+_1 \\
&\quad+ \overline{\Psi}^0 \left( g_L \slashed{W}^3_L- g_R \slashed{W}^3_R\right) \Psi^0 
-g_R \overline{\Psi}^-_2 \slashed{W}^3_R \Psi^-_2 \\
&\quad+g_L \left( \overline{\chi} \slashed{W}^-_L \Psi^+_1 + \overline{\Psi}^-_2 \slashed{W}^-_L \Psi^0 + \overline{\Psi}^{--} \slashed{W}^-_L \Psi^{-}_2 + \hc\right) \\
&\quad+g_R \left( \overline{\Psi}^-_2 \slashed{W}^-_R \chi + \overline{\Psi}^0 \slashed{W}^-_R \Psi^+_1 + \overline{\Psi}^+_1 \slashed{W}^-_R \Psi^{++} + \hc\right),
\end{split}
\label{eq:bitriplet_lagrangian}
\end{align}
and parity can be realized as 
\begin{align}
(\Psi^{++},\Psi^{+}_1,\Psi^{+}_2,\Psi^{0},\chi) \stackrel{\P}{\longleftrightarrow} (\Psi^{++},-\Psi^{+}_2,-\Psi^{+}_1,(\Psi^{0})^c,\chi) \,.
\end{align}
The mass splittings can be easily obtained by noticing that $(\Psi_1^+,\chi,-\Psi_1^-)$ and $(\Psi^0,-\Psi_2^-,\Psi^{--})$ form $SU(2)_L$ triplets with hypercharge $0$ and $-1$, respectively, so we can use the MDM formula from Ref.~\cite{Cirelli:2005uq}. Since $(\Psi_2^+,\chi,-\Psi_2^-)$ form an $SU(2)_R$ triplet, we can use the formula from Ref.~\cite{Heeck:2015qra} to obtain the fourth mass splitting we need to fully describe the system. We are of course more interested in the splittings relative to the DM candidate $\chi$, which take the form
\begin{align}
\begin{split}
M_{\Psi^0}-M_{\chi} &\simeq \frac{\alpha_2}{4\pi} M \left[ f(r_{W_1})+f(r_{W_2}) - f(r_{Z_1})/c_W^2-c_M^2 f(r_{Z_2})\right] ,\\
M_{\Psi^+_1}-M_{\chi} &\simeq \frac{\alpha_2}{4\pi} M \left[ f(r_{W_1}) -c_W^2 f(r_{Z_1})-s_W^2 f(r_\gamma)\right] ,\\
M_{\Psi^+_2}-M_{\chi} &\simeq \frac{\alpha_2}{4\pi} M \left[ f(r_{W_2}) -s_W^2 s_M^2 f(r_{Z_1})-c_M^2 f(r_{Z_2}) - s_W^2 f(r_\gamma)\right] ,\\
M_{\Psi^{++}}-M_{\chi} &\simeq \frac{\alpha_2}{4\pi} M \left[ f(r_{W_1})+f(r_{W_2}) -\cos^2(2\theta_W)f(r_{Z_1})/c_W^2-c_M^2 f(r_{Z_2}) -4 s_W^2 f(r_\gamma)\right] ,
\end{split}
\end{align}
again in the limit $\xi = \phi = 0$ (see Appendix~\ref{app:mass_splitting} for definitions). 
We recognize the mass splittings of $\Psi^+_1$ and $\Psi^+_2$ as those of the purely left-handed and right-handed triplet, respectively, already plotted in Fig.~\ref{fig:mass-splitting}.
The splittings $M_{\Psi^{0,++}}-M_{\chi}$ are shown in Fig.~\ref{fig:bitriplet} (left).
Since $\Psi^0$ carries hypercharge and thus couples to the light $Z_1$ boson, we have to demand $M_{\Psi^{0}}>M_{\chi}$.
to evade direct-detection limits, even though both fields are electrically neutral. This gives a slightly more restrictive mass splitting constraint than for the triplet and quintuplet discussed in Sec.~\ref{sec:MLRDM}, approximately $M_{W_2}\gtrsim 1.23\, M + \unit[0.6]{TeV}$ for large $M$ (see Fig.~\ref{fig:bitriplet} (right)).
This in particular excludes the entire $Z_2$-resonance region.
For non-vanishing $W_L^-$--$W_R^-$ mixing the states $\Psi_1^+$ and $\Psi_2^+$ will mix, and at two-loop order $\Psi^0$ will split into two quasi-degenerate Majorana fermions similar to the bi-doublet case.

In Fig.~\ref{fig:bitriplet} (right) we show the valid points in the $M$--$M_{W_2}$ plane using our calculation of the relic density including Sommerfeld enhancement. The allowed region spans $\unit[1.8]{TeV} \lesssim M \lesssim \unit[40]{TeV}$ and $\unit[4.5]{TeV} \lesssim M_{W_2}$ in order to make the neutral Majorana fermion $\chi$ the lightest multiplet component. The valid points are either at $M\simeq \unit[1.8]{TeV}$ or around the $W_2$ resonance $2 M\simeq M_{W_2}$.
The DM abundance today consists only of the Majorana fermion $\chi$, so resonant  processes involving $W_2$  do not take place, for instance, in the Galactic Center, even if they are important at freeze-out. Also, since $W_2$ is so much heavier than $W_1$, the states $\Psi^\pm_2$ are effectively inaccessible, even though the mass splitting between the states is small. By extension, the states $\Psi_0$ and $\Psi^{++}$ are even more difficult to produce from $\chi$. Hence, to a good approximation, the bi-triplet behaves today just like an $SU(2)_L$ triplet $(\Psi_1^+,\chi,-\Psi_1^-)$, i.e.~a wino, except for the different $\Omega h^2$--$M$ dependence. In particular, indirect detection signatures of the wino are hence directly applicable to our bi-triplet, see Fig.~\ref{HESS:limits} (left), excluding already much of the parameter space. We show these limits in Fig.~\ref{fig:bitriplet} for the Einasto and the isothermal profile.  As for the other candidates, CTA will improve these limits leading to smaller viable regions, specially around the resonances. Moreover, the mass range $M\gtrsim \unit[20]{TeV}$ is barely constrained by H.E.S.S.~data, but is in reach of CTA.

The bi-triplet also behaves like a wino with regards to direct detection, so we expect a spin-independent cross section off a proton of $\sigma_\text{SI}^p \simeq \unit[2\times 10^{-47}]{cm^2}$~\cite{Hisano:2015rsa} over the entire mass range $\unit[1.8]{TeV} \lesssim M \lesssim \unit[40]{TeV}$. This is small but more promising than the prospects for the triplet of Sec.~\ref{sec:MLRDM}, where this cross section was reduced by $\Omega_L/\Omega_\text{obs}$.
An important difference to the wino case is the collider signature. While \emph{indirect} detection implicitly probes the bi-triplet part $(\Psi_1^+,\chi,-\Psi_1^-)$, colliders will probe the triplet that carries hypercharge, $(\Psi^0,-\Psi_2^-,\Psi^{--})$. The coupling to hypercharge significantly increases the production cross section and makes it possible to probe the $M\simeq \unit[1.8]{TeV}$ threshold at the (HL)LHC~\cite{Cirelli:2005uq}.

\section{Scalar dark matter}
\label{sec:real}

In this section we discuss scalar DM candidates that are either stable in the MDM spirit, namely the 7-plet $(\vec{7},\vec{1},0)\oplus (\vec{1},\vec{7},0)$, or absolutely stable due to matter parity $\mathbb{Z}_2^{B-L}$, namely the doublet $(\vec{2},\vec{1},-1)\oplus (\vec{1},\vec{2},-1)$.

\subsection{7-plet \texorpdfstring{$(\vec{7},\vec{1},0 )\oplus (\vec{1},\vec{7},0)$}{(7,1,0)+(1,7,0)}}
\label{sec:7-plet}

The LR real-scalar 7-plet is stable unless we consider dimension-five operators, in complete analogy to the MDM case. We will have two-component DM due to the LR exchange symmetry $\phi_L \stackrel{\P}{\longleftrightarrow} \phi_R$, where $\phi_L$ acts just like standard MDM~\cite{Cirelli:2005uq,Cirelli:2007xd,Cirelli:2009uv,Garcia-Cely:2015dda}. 
Each 7-plet 
\begin{align}
\phi_L \sim (\vec{7},\vec{1},0) \,, &&
\phi_R \sim (\vec{1},\vec{7},0)
\end{align}
can be written as a self-conjugate multiplet\footnote{Since $\vec{7}$ is a real representation of $SU(2)$ one can also define a real 7-plet with Lagrangian $\tfrac12 (D_\mu\phi)^T D^\mu \phi -\tfrac12 M^2 \phi^T\phi$ and three hermitian generator $7\times 7$ matrices $(T^a)^* = - T^a$. This leads to the same mass eigenstates and gauge interactions we gave in the text (see Appendix~\ref{app:real_reps}).}
\begin{align}
\phi_X = (\phi_X^{+++},\,\phi_X^{++},\,\phi_X^{+},\,\phi_X^{0},\,-\phi_X^{-},\,\phi_X^{--},\,-\phi_X^{---})^T
\end{align}
 with gauge interactions (defining $A\overset{\longleftrightarrow}{\del^\mu}B \equiv A\del^\mu B - B\del^\mu A$)
\begin{align}
\L_{\phi_X} &= i g_X W^3_{X,\mu} (\phi^{-}_X \overset{\longleftrightarrow}{\del^\mu} \phi_X^{+} +2 \,\phi^{--}_X \overset{\longleftrightarrow}{\del^\mu} \phi_X^{++} + 3\, \phi^{---}_X \overset{\longleftrightarrow}{\del^\mu} \phi_X^{+++} )\\
&\quad+ \left[ i g_X W^-_{X,\mu} (\sqrt{6}\, \phi^{0}_X \overset{\longleftrightarrow}{\del^\mu} \phi_X^{+} +\sqrt{5} \,\phi^{-}_X \overset{\longleftrightarrow}{\del^\mu} \phi_X^{++} + \sqrt{3}\, \phi^{--}_X \overset{\longleftrightarrow}{\del^\mu} \phi_X^{+++} ) +\hc \right]\\
&\quad+ g_X^2 W^3_{X,\mu}W^{3,\mu}_{X}\left( |\phi_X^+|^2 + 4\, |\phi_X^{++}|^2 + 9\, |\phi_X^{+++}|^2\right)\\
&\quad+ g_X^2 W^+_{X,\mu}W^{-,\mu}_{X}\left(6\, (\phi_X^0)^2 +11\, |\phi_X^+|^2 + 8\, |\phi_X^{++}|^2 + 3\, |\phi_X^{+++}|^2\right)\\
&\quad+ g_X^2 \left[ W^-_{X,\mu}W^{-,\mu}_{X}\left(\sqrt{30}\, \phi_X^0 \phi_X^{++} + \sqrt{15}\, \phi_X^- \phi_X^{+++}-3\, \phi_X^+ \phi_X^{+}\right) +\hc \right]\\
&\quad+ g_X^2 \left[ W^3_{X,\mu}W^{-,\mu}_{X}\left(\sqrt{6}\, \phi_X^0 \phi_X^{+} + \sqrt{45}\, \phi_X^- \phi_X^{++}+\sqrt{75}\, \phi_X^{--} \phi_X^{+++}\right) +\hc \right] .
\end{align}
The full $\P$-symmetric Lagrangian includes scalar--scalar interactions and is given by
\begin{align}
\L &= \sum_{X = L,R} \left[\frac12 (D_\mu \phi_X)^\dagger D^\mu \phi_X -\frac12 M^2 |\phi_X|^2 -\sum_{k=1,2} \lambda_{k} [\phi_X]^4_k \right] -\lambda_{LR} 	|\phi_L|^2 |\phi_R|^2\\
&\quad- \lambda_{\Delta_1} (|\Delta_L|^2 |\phi_R|^2+|\Delta_R|^2 |\phi_L|^2) \\
&\quad- \sum_{X = L,R} \left[\lambda_H |H|^2 |\phi_X|^2 +\lambda_{\Delta_2} |\Delta_X|^2 |\phi_X|^2+\lambda_{\Delta_3} (\Delta_X^\dagger \Delta_X)_\vec{5} (\phi_X^\dagger \phi_X)_\vec{5} \right] ,
\label{eq:7-plet_lagrangian}
\end{align}
where $[\phi_X]^4_{1,2}$ denotes two linearly independent ways to couple $\vec{7}^4$ to a singlet~\cite{Aoki:2015nza}, for example $(|\phi_X|^2)^2$ and $(\phi_X^\dagger \phi_X)_\vec{5}(\phi_X^\dagger \phi_X)_\vec{5}$.
$(\phi_X^\dagger \phi_X)_\vec{5}$ denotes the $\vec{7}\otimes\vec{7} \to \vec{5}$ coupling to a quintuplet which, unlike the coupling to a triplet $(\phi_X^\dagger \phi_X)_\vec{3}$~\cite{Hambye:2009pw}, does not vanish and leads to a tree-level mass splitting
\begin{align}
\L \ \supset\ -\frac{\lambda_{\Delta_3}}{6\sqrt{14}} v_X^2 \left[ 2 (\phi_X^0)^2 + 3 |\phi_X^+|^2 -5 |\phi_X^{+++}|^2\right] .
\end{align}
For the left-handed components this hardly matters, seeing as $v_L$ is tiny (or even zero), but for the right-handed 7-plet this mass splitting can obviously be sizable. Assuming $\delta M^2 \equiv -\lambda_{\Delta_3} v_R^2/(12\sqrt{14})$ to be much smaller in magnitude than $M^2$, we obtain the splitting
\begin{align}
M_{\phi_R^Q}-M_{\phi_R^0}  \simeq Q^2 \frac{\delta M^2}{ M} \,,
\end{align}
which features the same $Q^2$ dependence as the radiative corrections (see Eq.~\eqref{eq:7Lsplitting} below). The one-loop contributions of the gauge bosons to the masses are hence no longer finite, as $\lambda_{\Delta_3}$ provides a counterterm (see Appendix~\ref{app:mass_splitting}).
Therefore, we can neglect the radiative corrections to $\phi_R$ masses and consider arbitrary values $|\delta M^2| \ll M^2$, the actual value being irrelevant for our relic-density calculation.

The radiative mass splitting for the components of $\phi_L$ is also divergent for $v_L\neq 0$ and merely renormalizes the tree-level term. Since we work in the limit of tiny or even vanishing $v_L$, we will assume the mass splitting to still be of the form~\cite{Cirelli:2005uq},\footnote{Notice that even though the result appears to be finite, i.e.~invariant under $g(r)\to g(r) + k r^2$, it is actually divergent for $v_L\neq 0$ because the custodial symmetry is broken, so $M_{W_1} \neq c_W M_{Z_1}$ (see Appendix~\ref{app:mass_splitting}).}
\begin{align}
\begin{split}
M_{\phi_L^Q}-M_{\phi_L^0} &\simeq \frac{\alpha_2}{4\pi} M Q^2 \left[ g(r_{W_1}) - c_W^2 g(r_{Z_1})-s_W^2 g(r_\gamma)\right]\\
&\simeq \alpha_2 Q^2 M_{W_1} \sin^2 (\theta_W/2) + \mathcal{O}(M_{W_1}^2/M) \,,
\end{split}
\label{eq:7Lsplitting}
\end{align}
which evaluates to $Q^2 \times \unit[167]{MeV}$, so the lightest stable component is indeed neutral.

As is common in MDM, we would actually like to neglect all the quartic couplings $\lambda_x$ in Eq.~\eqref{eq:7-plet_lagrangian} in order to simplify the discussion and keep the number of free parameters small. Some comments are in order though: first, small self-couplings are not technically natural and will thus receive large radiative corrections that can even lead to low-scale Landau poles~\cite{Hamada:2015bra}; second, the couplings of Eq.~\eqref{eq:7-plet_lagrangian} will lead to cubic couplings $v \varphi |\phi_X|^2$, where $v\in \{v_{L,R},\kappa_{1,2}\}$ is a VEV and $\varphi$ a neutral scalar field, which open up new annihilation channels and can drastically change the DM phenomenology~\cite{Hambye:2009pw}, e.g.~for $M$ close to a $\varphi$ resonance; third, since $\lambda_{\Delta_3}$ is the origin of the $\phi_R$ mass splitting, it is in principle inconsistent to neglect it. Only because we assume the coupling constant $\lambda_{\Delta_3}$ to be much smaller than all the gauge couplings can we justify to drop it in the phenomenology of $\phi_R$.

\begin{figure}[t]
\includegraphics[width=0.48\textwidth]{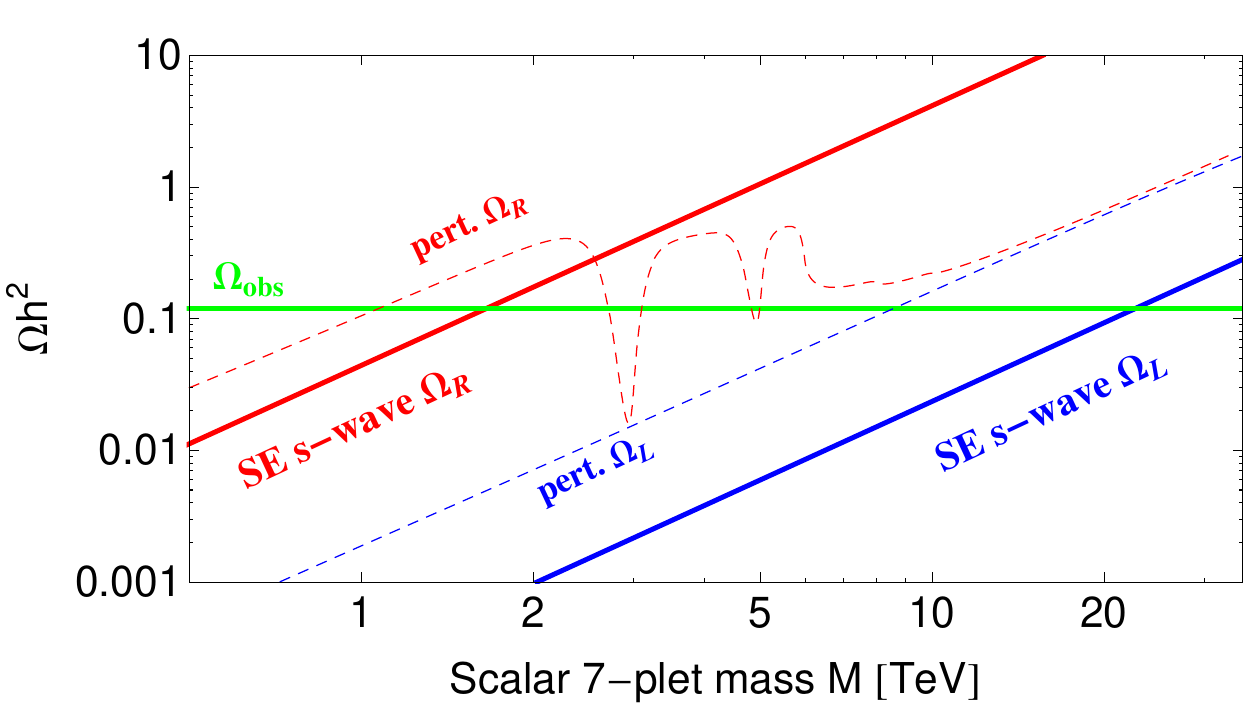}
\includegraphics[width=0.48\textwidth]{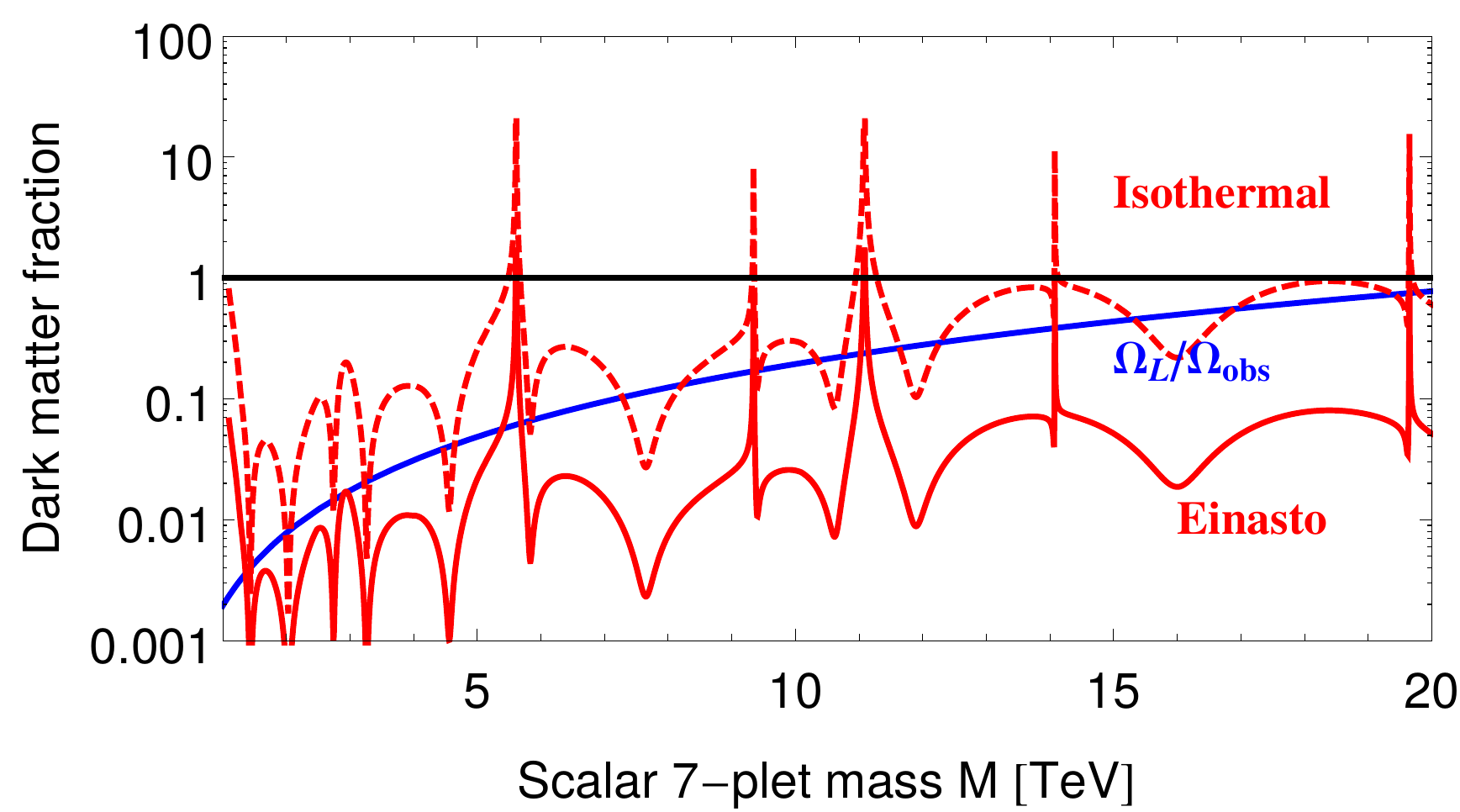}
\caption{Left: Relic densities $\Omega_{L}$ (blue) and $\Omega_{R}$ (red) for the LR scalar 7-plet, ignoring all scalar--scalar interactions and mass splittings. The solid lines are from our ($s$-wave) Sommerfeld calculation, the dashed lines from a perturbative \texttt{micrOMEGAs} calculation (setting $M_{W_2}=\unit[6]{TeV}$). Right: Same as Fig.~\ref{HESS:limits}, but for the scalar 7-plet.
}
\label{fig:relic_density_7plet}
\end{figure}

Only taking the gauge interactions of $\phi_{L,R}$ into account simplifies matters enormously. $\phi_L$ behaves as normal MDM~\cite{Cirelli:2005uq,Cirelli:2007xd} and is subject to significant Sommerfeld enhancement in the early Universe, which leads to an upper bound of $M\lesssim\unit[22.7]{TeV}$ (in our $SU(2)_L$-symmetric approximation) in order not to overclose the Universe (see the left panel of Fig.~\ref{fig:relic_density_7plet}).
$\phi_R$ evolves independently -- assuming again negligible quartic $\lambda_x$ couplings -- and is also subject to Sommerfeld-enhancement because $\phi_R^{+,++,+++}$ couple to photon and $Z_1$.
Notice that our relic-density calculation only takes into account the $s$-wave part of the annihilation cross section. Although this is a good approximation for the fermion DM candidates of Sec.~\ref{sec:majorana}, for scalars the $s$-wave processes poorly describe the DM production  around the $W_2$ and $Z_2$ resonances. This is because the latter carry a non-vanishing angular momentum, and if they are produced by annihilating scalars, such momentum can only come from the orbital part. Consequently, in spite of its velocity suppression, the $p$-wave part of the cross section is resonantly enhanced. We can also see this from the fact that  our ($s$-wave) Sommerfeld-enhanced relic density does not depend on $M_{W_2}$ at all, as shown in the left panel of Fig.~\ref{fig:relic_density_7plet}. Our calculation of Appendix~\ref{sec:SU2LL} is hence only valid far away from the resonances, i.e.~for $M\ll M_{W_2}$ (the region $M\gg M_{W_2}$ requires the addition of $W_2$ and $Z_2$ final states in $\phi_R \phi_R$ annihilation that we have omitted). Nevertheless, in order to check its consistency, we have checked that our $s$-wave calculation (without Sommerfeld effect) agrees to good accuracy with the perturbative calculation of \texttt{micrOMEGAs}~\cite{Belanger:2006is,Belanger:2013oya} in the limit $M\ll M_{W_2}$. In conclusion, an accurate calculation of $\Omega_R$ would require to account for the Sommerfeld effect also on the $p$ waves, which is unfortunately beyond the scope of this article.

Indirect detection constraints arise again from the left-handed component, similar to the triplet and quintuplet cases of Sec.~\ref{sec:MLRDM}. They are shown in the right panel of Fig.~\ref{fig:relic_density_7plet}. For the Einasto profile, this will exclude almost the entire mass region above $\unit[1]{TeV}$~\cite{Garcia-Cely:2015dda}, with the possible exception of the dips associated to Ramsauer--Townsend effect, which are expected to disappear once the internal bremsstrahlung contribution is accounted for.

Direct detection and collider constraints are again inherited from the left-handed MDM component ~\cite{Cirelli:2005uq}. In particular, the large spin-independent cross section $\sigma_\text{SI}^p \simeq \unit[3\times 10^{-44}]{cm^2}$ that arises at one loop -- ignoring again scalar--scalar interactions -- has be scaled down by $\Omega_L/\Omega_\text{obs} = (M/\unit[22.7]{TeV})^2$ for smaller masses and thus survives current bounds.

\subsection{Inert doublet \texorpdfstring{$(\vec{2},\vec{1},-1 )\oplus (\vec{1},\vec{2},-1)$}{(2,1,-1)+(1,2,-1)}}
\label{sec:inert_doublet}

As a last candidate for LR DM, we consider a scalar that is stabilized by matter parity. Since it shares many features of well-known models and brings with it a comparatively large number of free parameters, we will not go into details here but mainly outline the qualitative phenomenology.
We take complex scalars in the lepton-like representation~\cite{Heeck:2015qra}
\begin{align}
\phi_L = \matrixx{\phi^0_L\\ \phi^-_L} \sim (\vec{2},\vec{1},-1 ) \,, &&
\phi_R = \matrixx{\phi^0_R\\ \phi^-_R} \sim (\vec{1},\vec{2},-1) \,,
\end{align}
which will actually lead to a \emph{real} scalar DM candidate in the end, so it can be safe from direct-detection bounds. These representation are not only reminiscent of leptons (see Eq.~\eqref{eq:LRleptons}), but are actually used in LR models without scalar triplets $\Delta_{L,R}$ to break the LR gauge symmetry~\cite{Pati:1974yy,Mohapatra:1974gc,Senjanovic:1975rk,Senjanovic:1978ev}. While they carry the same quantum numbers as in our case, we stress that our $\phi_{L,R}$ do not acquire VEVs -- making them \emph{inert} doublets -- which is a stable solution of the minimization conditions of the potential.
The $\P$-invariant Lagrangian contains many terms and takes the form
\begin{align}
\L &= \sum_{X = L,R} \left[ (D_\mu \phi_X)^\dagger D^\mu \phi_X - M^2 |\phi_X|^2 - \lambda_{\phi} |\phi_X|^4\right] -\lambda_{LR} 	|\phi_L|^2 |\phi_R|^2
\label{eq:slepton_lagrangian1}\\
&\quad- \lambda_{\Delta_1} (|\Delta_L|^2 |\phi_R|^2+|\Delta_R|^2 |\phi_L|^2) 
- \sum_{X = L,R} \left[\lambda_H |H|^2 |\phi_X|^2 +\lambda_{\Delta_2} |\Delta_X|^2 |\phi_X|^2\right] \\
&\quad -\left[ \lambda_{\phi\Delta H \phi} \left(\phi_L^T  i\sigma_2\Delta_L H \phi_R +\phi_R^T  i\sigma_2\Delta_R H^\dagger \phi_L\right) + \lambda_{\phi\Delta \tilde{H} \phi} (H\to\tilde H)+\hc\right] \\
&\quad -\left[ \lambda_{\phi H\Delta \phi} \left(\phi_L^T  i\sigma_2 H \Delta_R \phi_R + \phi_R^T  i\sigma_2 H^\dagger \Delta_L \phi_L\right)+\lambda_{\phi \tilde{H}\Delta \phi} (H\to\tilde{H}) + \hc\right]\\
&\quad -\lambda_{\phi \Delta \Delta \phi}\left( \phi_L^\dagger \Delta_L^\dagger \Delta_L \phi_L +\phi_R^\dagger \Delta_R^\dagger \Delta_R \phi_R\right)\\
&\quad -\lambda_{\phi H H \phi}\left( \phi_L^\dagger H H^\dagger \phi_L +\phi_R^\dagger H^\dagger H \phi_R\right)
-\lambda_{\phi \tilde{H} \tilde{H} \phi}\left( \phi_L^\dagger \tilde{H} \tilde{H}^\dagger \phi_L +\phi_R^\dagger \tilde{H}^\dagger \tilde{H} \phi_R\right)\\
&\quad- \sqrt{2} \left[\phi_L^\dagger (\mu_H H + \mu_{\tilde{H}} \tilde{H}) \phi_R - \sum_{X = L,R}\mu_\Delta \phi_X^T i\sigma_2 \Delta_X \phi_X +\hc\right] ,
\label{eq:slepton_lagrangian}
\end{align}
with gauge interactions contained in the covariant derivative
\begin{align}
\hspace{-2ex}(D_\mu \phi_X)^\dagger D^\mu \phi_X  &= (\del_\mu \phi_X)^\dagger \del^\mu \phi_X \\
&\quad+ \frac{i}{2} (g_{BL} B_\mu - g_X W^3_{X,\mu}) \phi^0_X \overset{\longleftrightarrow}{\del^\mu} \overline{\phi}_X^0 + \frac{i}{2} (g_{BL} B_\mu + g_X W^3_{X,\mu}) \phi^-_X \overset{\longleftrightarrow}{\del^\mu} \overline{\phi}_X^- \\
&\quad+ \left(i\frac{g_X}{\sqrt{2}} W^-_{X,\mu} \overline{\phi}_X^-\overset{\longleftrightarrow}{\del^\mu}  \phi^0_X  + \hc\right)
-\left(\frac{g_X g_{BL}}{\sqrt{2}} B^\mu W^-_{X,\mu} \overline{\phi}_X^- \phi^0_X + \hc \right)\\
&\quad+\frac{1}{4} (g_{BL} B_\mu - g_X W^3_{X,\mu})(g_{BL} B^\mu - g_X W^{3,\mu}_{X})|\phi^0_X|^2 +\frac{g_X^2}{2} W^-_{X,\mu} W_{X}^{+,\mu} |\phi^0_X|^2\\
&\quad+\frac{1}{4} (g_{BL} B_\mu + g_X W^3_{X,\mu})(g_{BL} B^\mu + g_X W^{3,\mu}_{X})|\phi^-_X|^2 +\frac{g_X^2}{2} W^-_{X,\mu} W_{X}^{+,\mu} |\phi^-_X|^2 \,.
\end{align}

Parity $\P$ ensures that $\mu_{H,\tilde H}$ are real parameters (with dimension of mass). 
Some phenomenology can already be read off from the scalar potential: 1) terms that contain both $\phi_L$ and $\phi_R$ linearly, e.g.~the $\mu_{H,\tilde H}$ terms, induce a mixing between the left- and right-handed doublets, ensuring that only \emph{one} of the neutral scalars will be exactly stable; 2) terms that contain one scalar triplet $\Delta$ as well as two doublets $\phi$, e.g.~the $\mu_\Delta$ or the $\lambda_{\phi\Delta H \phi}$ terms, split the complex neutral scalars into \emph{real} scalars, because they break the global $U(1)$ symmetry that protects the complex nature, $\phi_{L,R}\to e^{i\alpha} \phi_{L,R}$, which is actually just lepton number.
Both effects will be helpful to avoid direct detection constraints on our DM, because they allow to either make DM dominantly $\phi_R^0$, i.e.~singlet-like without hypercharge, or to make the mass splitting of $\Re(\phi_L^0)$ and $\Im(\phi_L^0)$ large enough to evade $Z_1$-mediated detection via inelastic scattering~\cite{TuckerSmith:2001hy}.

To determine the mass eigenstates, we insert the VEVs of $H$ and $\Delta_{L,R}$ into the Lagrangian of Eqs.~\eqref{eq:slepton_lagrangian1}--\eqref{eq:slepton_lagrangian}.
Ignoring for simplicity all the quartic interactions, we arrive at a symmetric mass (squared) matrix for the charged scalars $(\phi^-_L,\phi^-_R)$,
\begin{align}
\mathcal{M}^2_- = \matrixx{M^2 & \mu_H\kappa_2+ \mu_{\tilde H} \kappa_1\\ \mu_H\kappa_2+ \mu_{\tilde H} \kappa_1 & M^2} .
\end{align}
The complex neutral scalars will be split into four real scalars by $v_{L,R}$, so we parametrize $\phi_{L,R}^0 = (\phi_{L,R}^{0,r}+i \phi_{L,R}^{0,i})/\sqrt{2}$. In the basis $(\phi_{L}^{0,r},\phi_{R}^{0,r},\phi_{L}^{0,i},\phi_{R}^{0,i})$, the mass matrix takes the form
\begin{align}
\mathcal{M}^2_0 = \matrixx{M^2 - 2\mu_\Delta v_L &  \delta m^2 & 0 & 0 \\ \delta m^2 & M^2 - 2\mu_\Delta v_R & 0 & 0 \\ 0 & 0 & M^2 +2 \mu_\Delta v_L & \delta m^2 \\ 0 & 0 & \delta m^2 &  M^2 +2 \mu_\Delta v_R} ,
\end{align}
with $\delta m^2 \equiv \mu_H \kappa_1 + \mu_{\tilde H} \kappa_2$. 
(Note that the scalars $\phi_X^i$ and $\phi_X^r$ will mix if we allow for CP-violating phases in the scalar potential.)
We have to demand $M^2 >2 |\mu_\Delta| v_{L,R}$ to not induce a VEV in our new scalars. For positive $\mu_\Delta$ and small $\delta m^2$, the lightest state will be dominantly $\phi_{R}^{0,r}$, which has no coupling to $Z_1$. The couplings to $Z_1$ are always of the form $Z_1^\mu (\phi^i \del_\mu \phi^r-\phi^r \del_\mu \phi^i)$, so a large mass splitting of $\phi_{L}^{0,r}$ and $\phi_{L}^{0,i}$ will kill the $Z_1$-mediated inelastic direct-detection process in case $\phi_L$ is the DM candidate~\cite{TuckerSmith:2001hy}. 

\begin{figure}[t]
\includegraphics[width=0.6\textwidth]{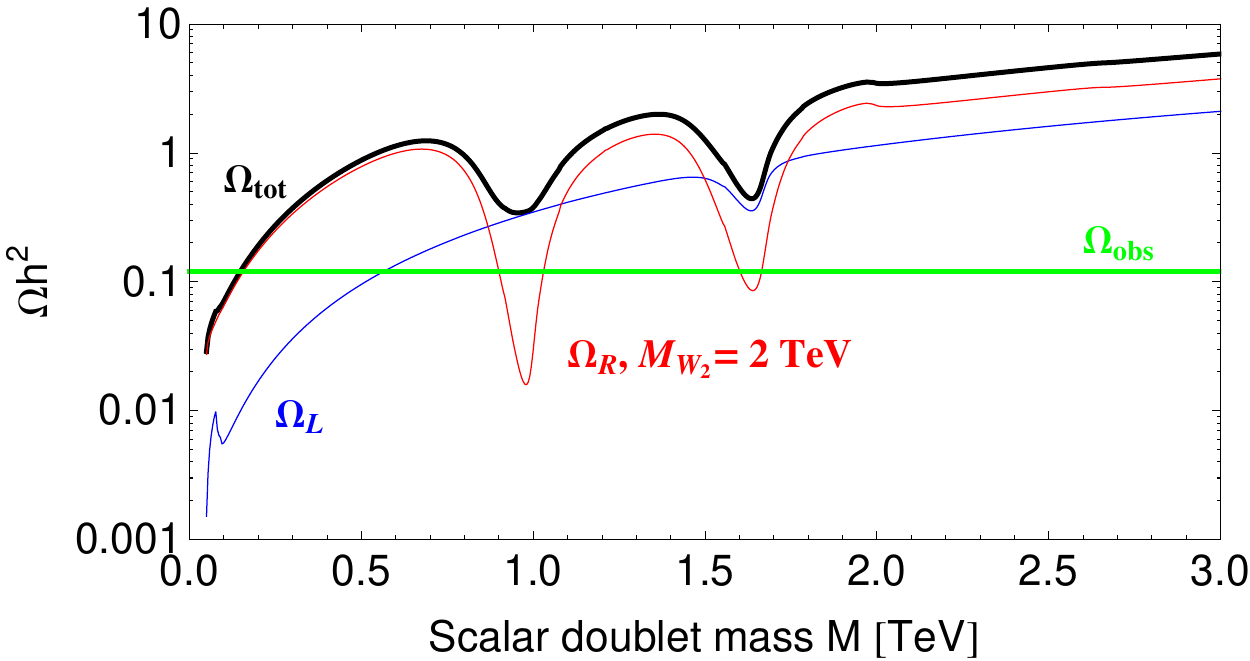}
\caption{Relic densities $\Omega_{L,R}$ for the LR scalar doublet, ignoring all scalar--scalar interactions and mass splittings. $\Omega_\text{tot}=\Omega_L+\Omega_R$ (black) matches the observed $\Omega_\text{obs}$ for $M\simeq \unit[150]{GeV}$.
}
\label{fig:relic_density_scalar_doublet_2TeV}
\end{figure}

Still ignoring the quartic interactions, one gains a global $U(1)_{\phi_L} \times U(1)_{\phi_R}$ for $\mu_{H,\tilde H,\Delta}\to 0$, under which $\phi_X \to e^{i \alpha_X} \phi_X$. This symmetry is broken down to $U(1)_{\phi}$ (lepton number) by $\mu_{H,\tilde H}$, and to $\mathbb{Z}_2^{\phi_L}\times \mathbb{Z}_2^{\phi_R}$ by $\mu_\Delta$. Both terms together only leave a $\mathbb{Z}_2$ symmetry, which is nothing but our stabilizing matter parity.
In view of this, it is technically natural~\cite{tHooft:1979bh} to take $\mu_{H,\tilde H,\Delta}$ to be small, and a similar argument can be made taking some of the quartic couplings into account. This simplifies the phenomenology of our doublets because it again leads to quasi-degenerate multiplets, similar to the other LR DM candidates discussed so far.
The relic density for this quasi-degenerate case is shown in Fig.~\ref{fig:relic_density_scalar_doublet_2TeV} -- calculated using \texttt{micrOMEGAs} because Sommerfeld enhancement is negligible here -- which matches the observed one for $M\simeq \unit[150]{GeV}$ (with $\Omega_L/\Omega_R \simeq 8\%$), valid for all $M_{W_2}>\unit[2]{TeV}$ and $g_R\leq g_L$.
Small mass splittings are, of course, necessary to avoid direct detection bounds along the lines outlined above, which will also determine collider and indirect-detection constraints on our scenario.
Larger DM masses are possible if $\phi_L$--$\phi_R$ mixing terms such as $\mu_{H,\tilde H}$ are turned on, for example around the $W_2$ co-annihilation resonance in case $\phi_R$ is the lightest DM particle.

Let us make the connection to other models with similar phenomenology.
Splitting of the neutral components of a stable scalar doublet is reminiscent of the Inert Doublet Model~\cite{Deshpande:1977rw,Barbieri:2006dq,LopezHonorez:2006gr,Ma:2006km}, and indeed we can reproduce this model in a certain parameter space if $\phi_L$ is lighter than $\phi_R$. The correct relic density then requires $M\simeq \unit[535]{GeV}$ in the pure gauge case~\cite{Hambye:2009pw}, with a one-loop direct-detection cross section of $\sigma_\mathrm{SI}^p\simeq\unit[2\times 10^{-46}]{cm^2}$ if scalar--scalar couplings are neglected~\cite{Cirelli:2005uq,Klasen:2013btp}. See Refs.~\cite{Queiroz:2015utg,Garcia-Cely:2015khw} for the indirect-detection prospects in this case.
If $\phi_R$ is the lighter multiplet, the model looks very different and is reminiscent of right-handed sneutrino DM in supersymmetric models~\cite{Hall:1997ah,ArkaniHamed:2000bq,TuckerSmith:2001hy}, which have been discussed both in the lepton-number conserving case (corresponding to negligible couplings to scalar triplets in our case) as well as the lepton-number violating case (featuring inelastic DM). There is a vast body of literature on this topic that can unfortunately not be listed here. We will leave a more detailed discussion of the rich phenomenology of this model for future work, in particular the collider phenomenology of these fairly light scalars.

\section{Diboson excess}
\label{sec:diboson}

The fermion DM candidates presented in Sec.~\ref{sec:majorana} are minimal in the sense that they only introduce \emph{one} additional parameter to LR models -- their mass $M$ -- which is fixed to obtain the observed relic density. Similar to MDM~\cite{Cirelli:2005uq}, the theory thus becomes fully predictive as soon as the gauge boson mass $M_{W_2}$ is fixed (as well as $g_R$ in more general LR realizations). Even then, $M$ is not necessarily uniquely determined; due to the co-annihilation resonances, there can actually be up to \emph{five} values for $M$ that yield the correct relic density, depending on $M_{W_2}$ (see Figs.~\ref{fig:MvsMWR}, \ref{fig:MvsMWR_bidoublet}, \ref{fig:bitriplet}). Nevertheless, this yields a predictive and testable realization of DM within LR models.

Recently, a number of excesses have appeared in analyses by both ATLAS and CMS that can potentially be interpreted as LR gauge bosons. The excesses hint at a $W_2$ mass of $\unit[2]{TeV}$, which provides the last parameter we need to make our DM candidates predictive. Even though these excesses are not yet statistically relevant, we are compelled to speculate about the implications for DM should they turn out to be real. We find below that our (fermionic) DM multiplets can indeed consistently give the observed relic density for $M_{W_2}\simeq \unit[2]{TeV}$, either using $g_R< g_L$ -- which coincidentally relaxes the mass-splitting constraint we found for the triplet, quintuplet, and bi-triplet -- or by opening up new decay channels $W_2\to \text{DM}$ that lower the relevant branching ratios $W_2\to\text{SM}$ even for $g_R=g_L$.

\subsection{Diboson excess with \texorpdfstring{$g_R<g_L$}{gR<gL}}

Recent excesses seen at ATLAS~\cite{Aad:2015owa} and CMS~\cite{Khachatryan:2014hpa,Khachatryan:2014gha} 
experiments point towards a $\unit[2]{TeV}$ mass of $W_2$, which can be consistently accommodated in left--right 
symmetric models with $g_R < g_L$ and a rather large $W_L^-$--$W_R^-$ mixing angle $\xi$. The gauge coupling $g_R$ is essentially fixed using the (small) dijet excess $pp\to W_2\to jj$~\cite{Aad:2014aqa,Khachatryan:2015sja}, while $\xi$ is set by the excess $pp\to W_2\to W_1 Z_1$~\cite{Khachatryan:2014hpa,Khachatryan:2014gha,Aad:2015owa}.
We will not attempt to review all possible models and analyses of these tantalizing hints.
A fit to all available relevant cross sections was recently performed in Ref.~\cite{Brehmer:2015cia}, quoting the preferred values for the standard LR model (broken by scalar triplets) as $M_{W_2}\simeq \unit[1.9]{TeV}$, $g_R/g_L\simeq 0.55$--$0.7$, and $1.1\times 10^{-3} \lesssim \sin \xi \lesssim 1.8 \times 10^{-3}$, assuming right-handed neutrino masses $M_N> M_{W_2}$.\footnote{Opening the decay mode $W_2\to \ell N$ allows, in principle, to also explain the $pp\to W_2 \to e e j j$ excess seen in CMS~\cite{Khachatryan:2014dka}, but requires non-minimal models~\cite{Deppisch:2014qpa,Deppisch:2015cua,Dobrescu:2015qna,Dev:2015pga} or severe finetuning~\cite{Gluza:2015goa}.} 
Taken together, these excesses deviate from the SM by about $2.9\sigma$~\cite{Brehmer:2015cia}.
Other analyses yield similar values~\cite{Cheung:2015nha,Dobrescu:2015qna,Deppisch:2015cua}.
It is important to note that $\xi$ can not be arbitrarily large, but rather satisfies $|\xi| \lesssim (g_R M_{W_1}^2)/(g_L M_{W_2}^2)$ with regards to the $\beta$ dependence (see Eq.~\eqref{eq:xi}). For the benchmark values adopted by us, $M_{W_2} = \unit[1.9]{TeV}$ and $g_R/g_L = 0.65$, this gives $|\xi|\lesssim 1.15\times 10^{-3}$, which is within the preferred region of the diboson excess. The $Z_2$ mass can be calculated to be $\unit[4.7]{TeV}$ for these parameters (Eq.~\eqref{eq:neutral_masses}), much larger than the $\unit[3.2]{TeV}$ one would obtain for $g_R=g_L$. In particular, the decay channel $Z_2\to W_2 W_2$ opens up, albeit of little importance here.

Let us study the implications of these experimental hints on our DM models.\footnote{DM in the context of the diboson anomaly has been previously mentioned or discussed in Refs.~\cite{Heeck:2015qra,Brehmer:2015cia,Dev:2015pga,Ko:2015uma}.}
LR models with $g_R\neq g_L$ obviously break generalized parity $\P$ (or $\C$), at least at some high scale. In our construction of LR DM (Sec.~\ref{sec:stability}) it is then, strictly speaking, no longer necessary to introduce degenerate multiplets in the $\P$-symmetric form $\Psi_L\oplus \Psi_R$. Omitting the left-handed component -- or changing its mass independently of the right-handed one -- will open up more parameter space, especially in the simple fermion cases of Sec.~\ref{sec:MLRDM}, where the strongest constraints come from the left-handed DM component. We will nevertheless stick to our framework with $\P$-symmetric DM multiplets in the following, not least because they \emph{are} experimentally testable.

\begin{figure}[t]
\includegraphics[width=0.45\textwidth]{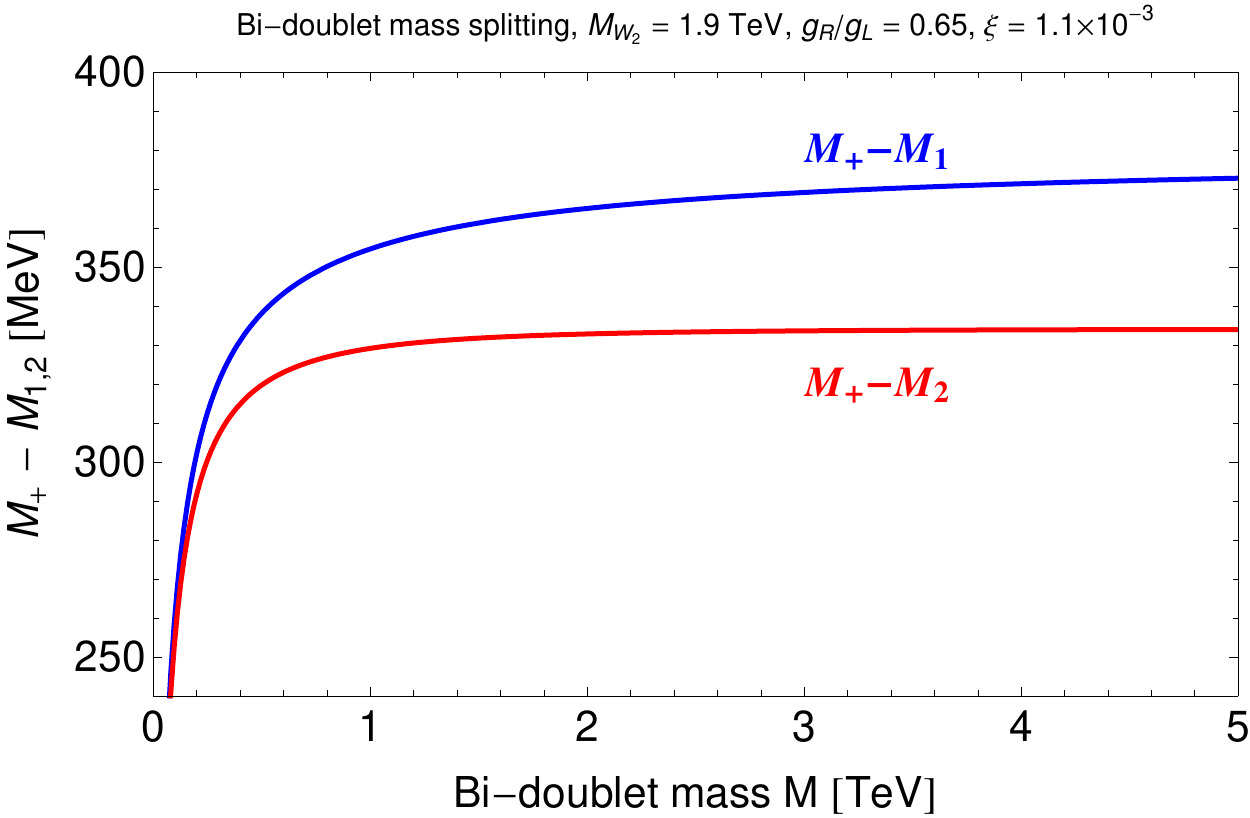}\hspace{2ex}
\includegraphics[width=0.45\textwidth]{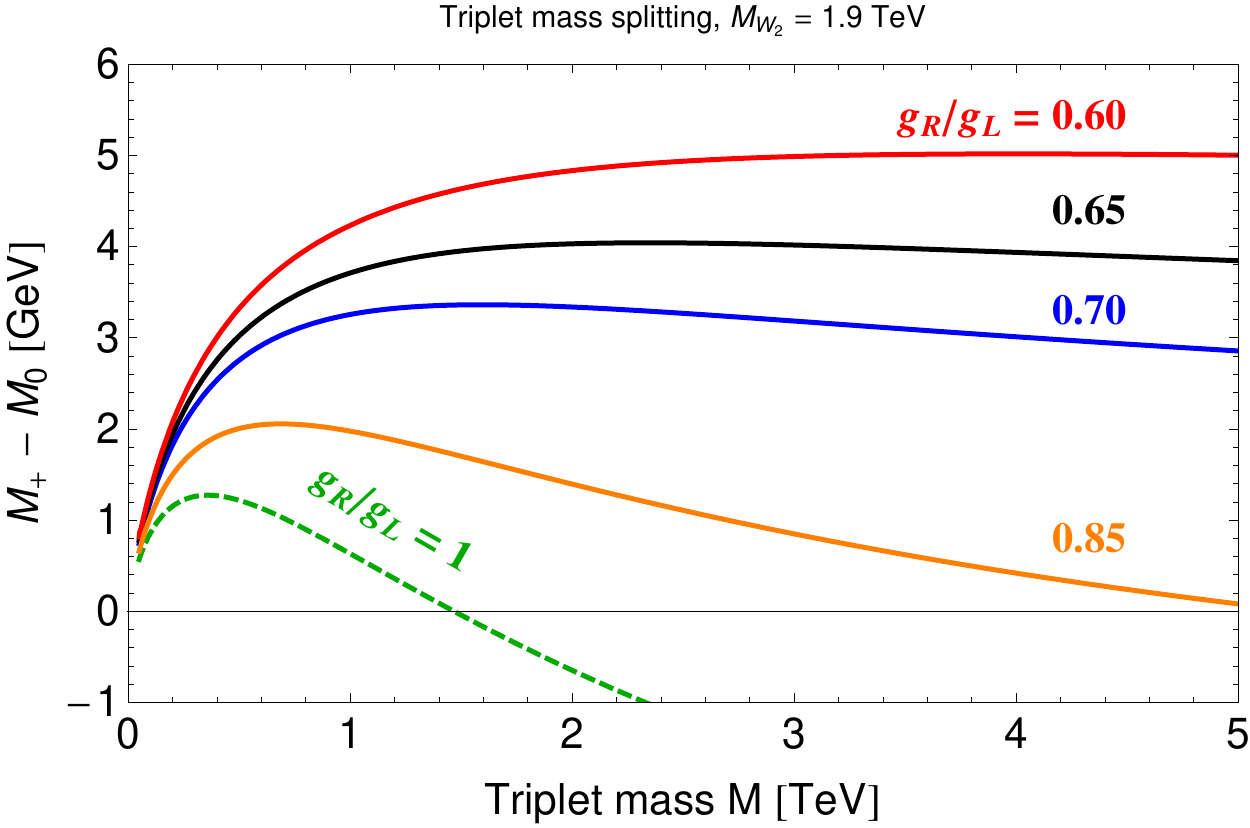}
\caption{Mass splitting $M_+ - M_0$ vs.~$M$ for the bi-doublet (left) and the right-handed triplet (right), setting $M_{W_2}=\unit[1.9]{TeV}$ and using several values for $g_R/g_L$ relevant for the diboson excess.}
\label{fig:mass-splitting_gRneqgL}
\end{figure}

The DM relic density is different for $g_R < g_L$, because a smaller value for $g_R$ leads to a suppression in the annihilation cross sections associated to the $W_2$ and $Z_2$ resonances (this  follows from Eqs.~\eqref{eq:AZ2} and \eqref{eq:AW2}). We will here only comment on the fermionic DM candidates, the fermion triplet, quintuplet, bi-doublet, and bi-triplet. These form predictive models because their only new parameter (the DM mass $M$) is fixed to obtain the observed relic density. As far as the mass splitting is concerned, the bi-doublet remains almost unaffected by $g_R < g_L$, and we find
\begin{align}
M_{\Psi^+} - M_{\chi_{1,2}} \simeq \frac{\alpha}{2} M_{Z_1} \pm \frac{\alpha_2}{8\pi} \frac{g_R}{g_L} M \sin (2\xi) \left[ f(r_{W_2})-f(r_{W_1})\right]
\end{align}
for the mass splitting of the charged component with respect to one of the two neutral Majorana components $\chi_{1,2}$ (Fig.~\ref{fig:mass-splitting_gRneqgL}). The mass splitting between the neutral components is hence suppressed by $g_R/g_L$, but still large enough to evade direct detection bounds via inelastic scattering. 

The situation is different for the triplet and quintuplet (similar for the bi-triplet), where the smaller $g_R$ significantly changes the mass splitting.
From Eq.~\eqref{eq:RH_mass_splitting} we see that the mass splitting between right-handed charged and neutral components is proportional to $M_{W_2}- c_M^2 M_{Z_2}$ for large $M$, which is \emph{positive} for $c_M <1/\sqrt{2}$, i.e.~$g_R/g_L < 0.76$. For the $g_R$ values of interest for the diboson excess, our stable DM candidates are hence automatically neutral, nullifying the mass splitting constraints we found in Sec.~\ref{sec:majorana}. $M_{\Psi^Q}-M_{\Psi^0}$ now remains positive and of order of GeV even for $M> M_{W_2}$ (Fig.~\ref{fig:mass-splitting_gRneqgL} (right)), as already noted in Refs.~\cite{Brehmer:2015cia}. This opens up previously inaccessible parameter space and in particular allows us to consider triplet and quintuplet as DM candidates for the $\unit[2]{TeV}$ $W_R$ gauge-boson explanation of the diboson excess.

\begin{figure}[t]
\includegraphics[width=0.48\textwidth]{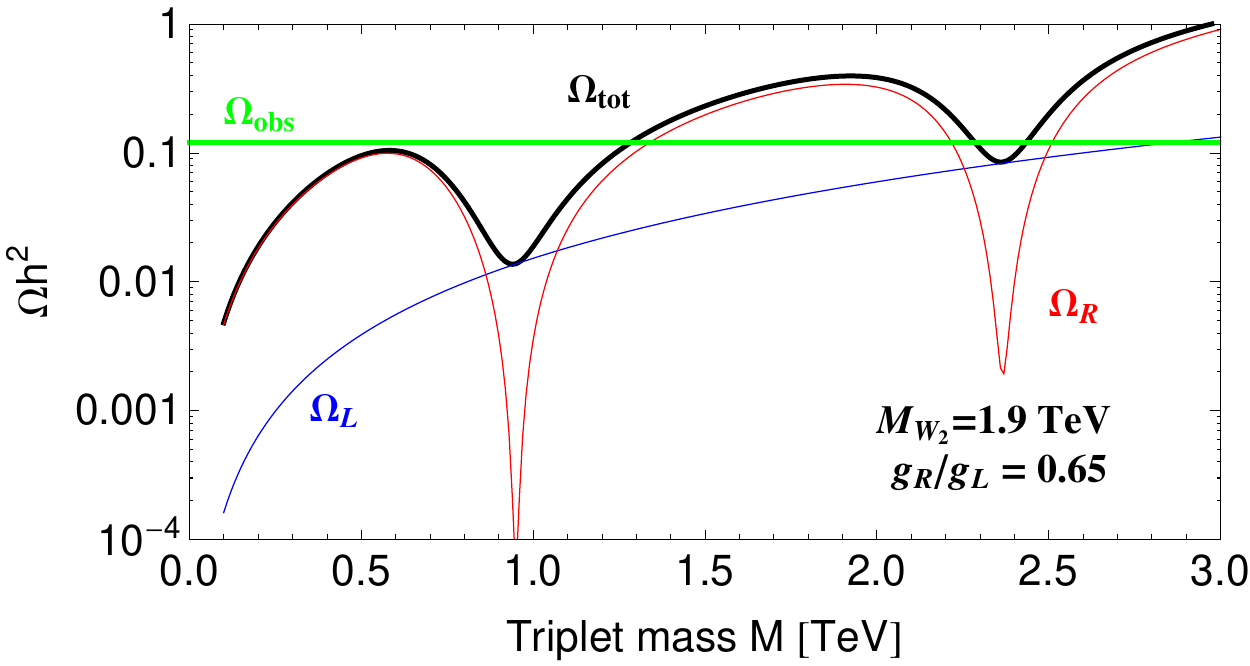}\hspace{2ex}
\includegraphics[width=0.48\textwidth]{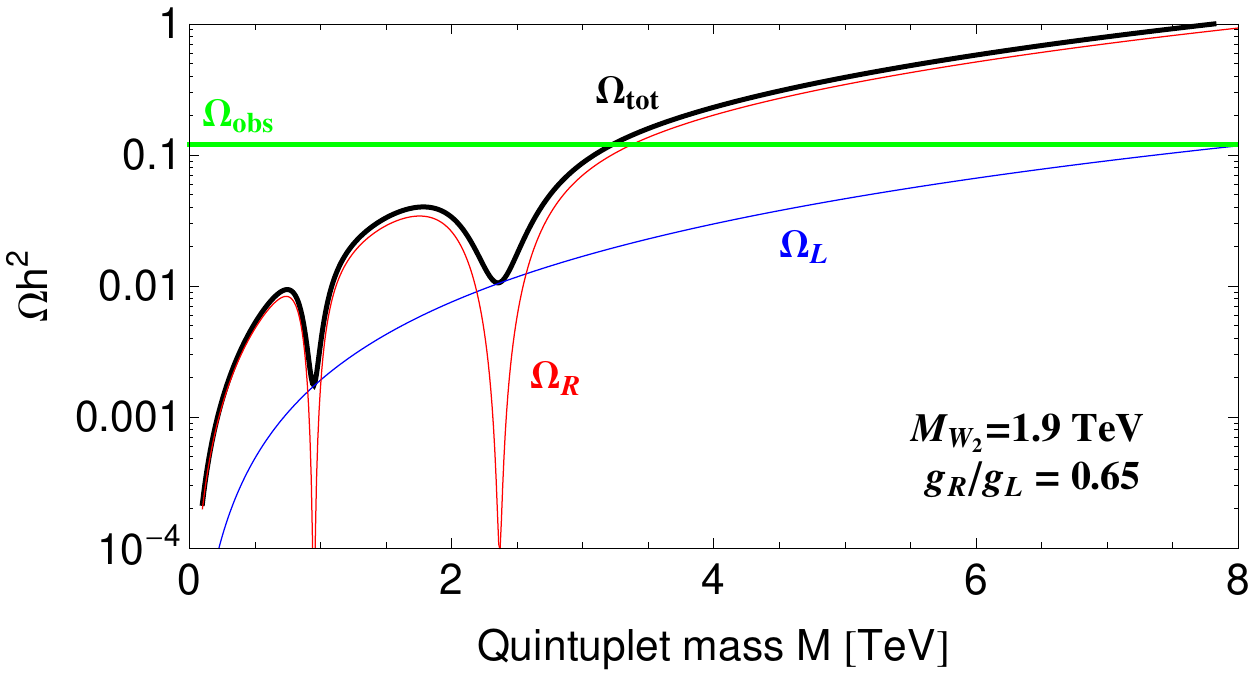}
\caption{Relic density (black) for the LR fermion triplet (left) and quintuplet (right) for $M_{W_2} = \unit[1.9]{TeV}$ and $g_R/g_L = 0.65$.
}
\label{fig:relic_densities_gR}
\end{figure}

The relic density for triplet and quintuplet is shown in Fig.~\ref{fig:relic_densities_gR}, calculated using our $SU(2)_L$-symmetric Sommerfeld formulae. 
The abundance $\Omega_L$ of the left-handed multiplet remains unaffected by the change of $g_R$, and will in particular give rise to the same indirect-detection constraints as before (see Fig.~\ref{HESS:limits}).
Since the mass splitting of the multiplet components no longer excludes the region $M>M_{W_2}$, we can easily obtain the correct relic density. For the quintuplet, this yields $M = \unit[3.2]{TeV}$, slightly disfavored by indirect detection for the Einasto DM profile.
The smaller $g_R$ suppresses the annihilation cross sections and thus increases the relic abundance for a given mass $M$, which opens up more than one solution for the triplet (see also Table~\ref{tab:dibosonDM}). In particular, the correct relic density can be obtained for \emph{three} different triplet DM masses (with an additional region around $M\simeq 550$--$\unit[600]{GeV}$ that requires a more exact relic density calculation to evaluate it). 
The two triplet solutions around the $Z_2$ resonance are, however, robustly excluded by $\gamma$-line searches in the Milky Way and continuum searches in dwarf galaxies, see Fig.~\ref{HESS:limits}.

\begin{figure}[t]
\includegraphics[width=0.48\textwidth]{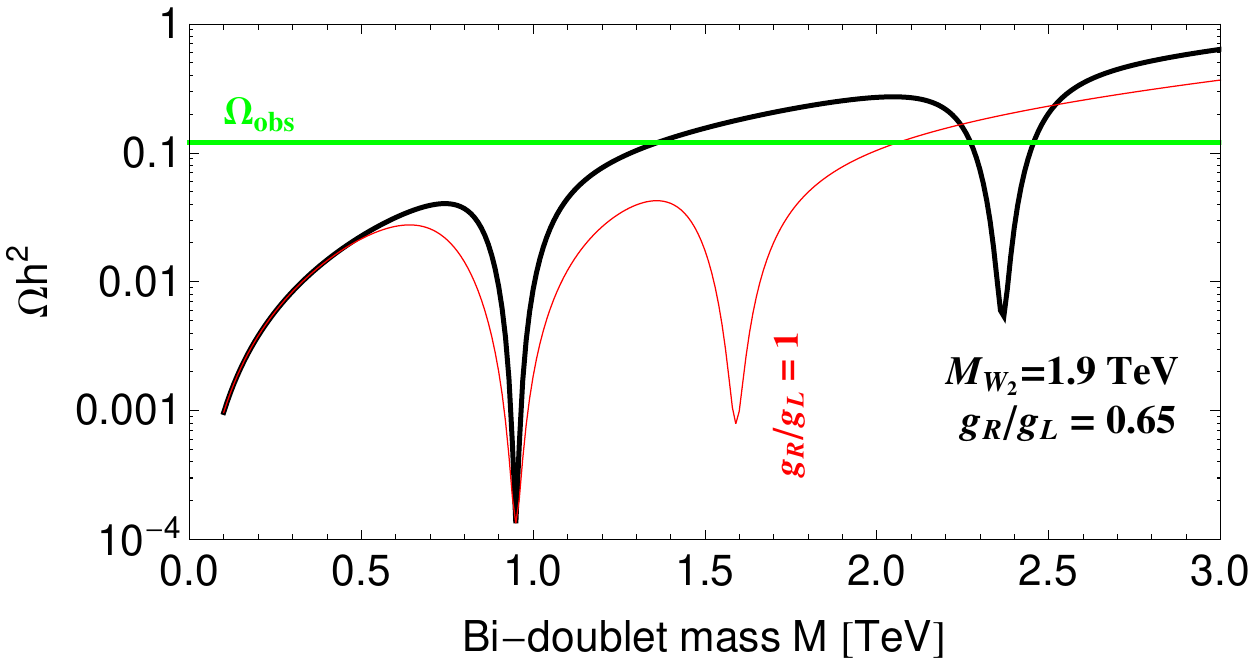}\hspace{2ex}
\includegraphics[width=0.48\textwidth]{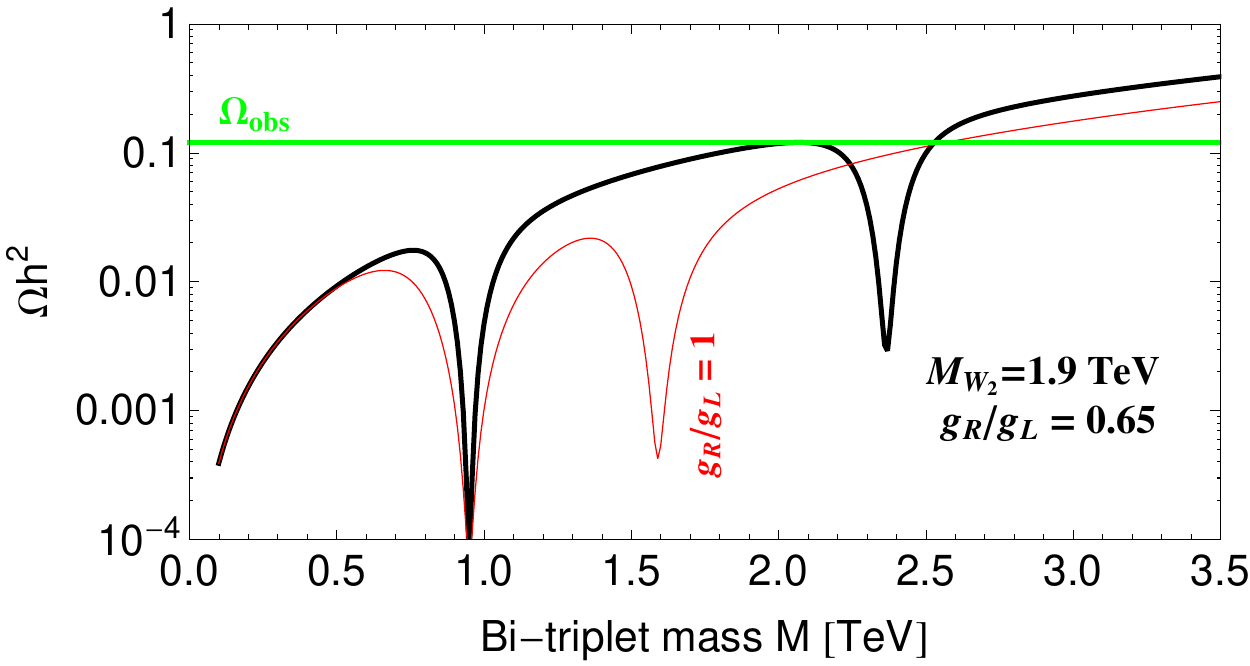}
\caption{Relic density for the LR fermion bi-doublet (left) and bi-triplet (right) for $M_{W_2} = \unit[1.9]{TeV}$ and $g_R/g_L = 0.65$ ($1$) in black (red).
}
\label{fig:relic_densities_bimultiplets_gR}
\end{figure}

The fermion bi-doublet can easily provide DM for the $\unit[2]{TeV}$ $W_2$ diboson solution, as can be seen from Fig.~\ref{fig:relic_densities_bimultiplets_gR} (left). The two solutions around the $Z_2$ resonance are obviously somewhat sensitive to additional model details, such as the spectrum of the non-SM-like scalars, which we have neglected in our discussion. Since Sommerfeld enhancement remains small for the bi-doublet, the indirect detection prospects of are rather dim.
For the bi-\emph{triplet}, the valid masses are around the $Z_2$ resonance (Fig.~\ref{fig:relic_densities_bimultiplets_gR} (right)). The broad region around $M\simeq \unit[2]{TeV}$ requires a more exact calculation of the relic density in order to be sure of its validity. In any way, all bi-triplet solutions are robustly excluded by indirect detection, as shown in Fig.~\ref{HESS:limits}.

The fermionic DM candidates for a LR model with $M_{W_2}=\unit[1.9]{TeV}$ and $g_R/g_L = 0.65$ are collected in Table~\ref{tab:dibosonDM}. Indirect detection already excludes many possibilities, leaving us with only a handful of possible masses. Only marginal better sensitivity is necessary to probe the remaining candidates via indirect detection, certainly of interest should the diboson anomaly survive the next LHC run. Notice that all candidates of Table~\ref{tab:dibosonDM} around the $Z_2$ co-annihilation resonance are sensitive to $g_R$, because the $Z_2$ mass depends strongly on it: $M_{Z_2}\simeq \sqrt{2} M_{W_2}/c_M$, with $c_M^2 = 1- (g_L/g_R)^2 \tan^2\theta_W$.
A reevaluation of LR DM is thus worthwhile once the data provides more precise values for both $g_R$ and $M_{W_2}$. Our present analysis should illustrate the predictive power of this framework.

We close this section with the observation that some of the DM masses of Table~\ref{tab:dibosonDM} allow for new $Z_2$ decay modes, which can have a huge impact on future searches for this neutral vector boson. ($M_{W_2}=\unit[1.9]{TeV}$ and $g_R/g_L = 0.65$ imply $M_{Z_2} = \unit[4.7]{TeV}$.)
For fermion triplet and quintuplets, the partial widths are given by
\begin{align}
\Gamma (Z_2\to \overline{\Psi}_R^Q \Psi_R^Q ) \simeq Q^2 \frac{g_R^2 c_M^2}{12\pi}M_{Z_2} \left(1+ \frac{2M^2}{M_{Z_2}^2}\right) \sqrt{1-\frac{4M^2}{M_{Z_2}^2}}\,, 
\end{align}
while the decay into the fermion bi-doublet states $\Psi^+$ or $\chi_{1,2}$ is given by
\begin{align}
\Gamma (Z_2\to \overline{\Psi}^+ \Psi^+ ) \simeq 2\times\Gamma (Z_2\to \chi_j \chi_j )  \simeq  \frac{g_R^2 c_M^2}{48\pi}M_{Z_2} \left(1+ \frac{2M^2}{M_{Z_2}^2}\right) \sqrt{1-\frac{4M^2}{M_{Z_2}^2}}\,,
\end{align}
where we neglected all mass splittings among the fermions.
Needless to say, a precise determination of the $Z_2$ (and $W_2$) decay modes will help to pin down the DM representation realized in nature.
The charged components will, of course, decay into the DM state plus pions, leading to testable signatures (see Sec.~\ref{sec:bidoublet_decays} and Refs.~\cite{Cirelli:2005uq,Cirelli:2014dsa}). This will be studied elsewhere.

\begin{table*}[t]
	\centering
	\renewcommand{\arraystretch}{1.5}
		\begin{tabular}{ccc}
			Fermion representation &  & DM mass $M/\unit{TeV}$ \\
			\hline
			\hline
			$(\vec{3},\vec{1}, 0)\oplus (\vec{1},\vec{3},0)$ & & 1.3, 2.3*, 2.4*\\
			$(\vec{5},\vec{1}, 0)\oplus (\vec{1},\vec{5},0)$ & & 3.2**\\
			$(\vec{2},\vec{2}, 0)$ & & 1.4, 2.3, 2.5\\
			$(\vec{3},\vec{3}, 0)$ & & 2.0*--2.1*, 2.5*\\
		\end{tabular}
\caption{DM candidates that yield the observed abundance for $M_{W_2}=\unit[1.9]{TeV}$ and $g_R/g_L = 0.65$.
Solutions with one asterisk are robustly excluded by indirect detection, while those with two asterisks are only excluded for the Einasto profile.
\label{tab:dibosonDM}}
\end{table*}

\subsection{Diboson excess with \texorpdfstring{$g_R=g_L$}{gR=gL}}

The diboson excess can also be fitted in LR models with $g_R = g_L$ if a new decay channel (e.g.~into DM) for $W_2$ is introduced, thus lowering the branching ratios into dijets and dibosons~\cite{Dev:2015pga,Dobrescu:2015qna}.  
The cross sections for all these excesses in the narrow-width approximation are given by 
\begin{align}
\sigma_X \equiv\sigma (pp\to W_2)\times \BR(W_2\to X) \propto g_R^2\times \BR(W_2\to X) \,.
\end{align} 
Instead of lowering $\sigma (pp\to W_2)$ by $g_R^2/g_L^2\simeq (0.65)^2$, we can lower $\BR(W_2\to X)$ by this factor by introducing a new -- sufficiently invisible -- channel $\BR (W_2\to \text{new})\simeq 55$--$60\%$, while keeping $g_R = g_L$.
Note that this kind of solution to the $\unit[2]{TeV}$ excesses is in conflict with the $\Delta F = 2$ meson constraint $M_{W_2} \gtrsim \unit[3]{TeV}\,(g_R/g_L)$~\cite{Bertolini:2014sua}, which arises from an off-shell $W_2$ and is hence insensitive to additional invisible decay channels. Since these bounds are based on loop processes, we can however speculate about additional new-physics contributions that cancel those of the $W_2$, allowing for $M_{W_2}\simeq\unit[2]{TeV}$. (Underestimated hadronic uncertainties are a possibility as well.)
Let us see if the LR DM candidates discussed in this article can provide the new $W_2$ decay channels necessary for this solution to the diboson excess.

With $g_R = g_L$, we can consider our analysis from the main text (Sec.~\ref{sec:majorana}) and look for possible DM candidates with $M<M_{W_2}/2\simeq \unit[1]{TeV}$ (abandoning the lower bound on $W_2$ from meson data). 
The $W_2$ decay widths into triplets and quintuplets are
\begin{align}
\Gamma (W_2\to\text{triplet}) &\simeq\frac{g_R^2}{12\pi}M_{W_2} \left(1+ \frac{2M^2}{M_{W_2}^2}\right) \sqrt{1-\frac{4M^2}{M_{W_2}^2}}\,, 
\label{eq:W2totriplet}\\
\Gamma (W_2\to\text{quintuplet}) &\simeq 5 \times\frac{g_R^2}{12\pi}M_{W_2} \left(1+ \frac{2M^2}{M_{W_2}^2}\right) \sqrt{1-\frac{4M^2}{M_{W_2}^2}}\,,
\end{align}
leading to $W_2$ branching ratios of about $30\%$ and $70\%$ in the limit $M\ll M_{W_2}<M_N$ for the triplet and quintuplet, respectively. 
One quintuplet with some phase-space suppression -- say $M\simeq \unit[800]{GeV}$ -- can thus easily give the required $W_2$ branching ratio for the diboson excess, but can only account for less than $5\%$ of the DM abundance (see Fig.~\ref{fig:relic_densities} (right)). This is a possible, but admittedly somewhat unsatisfactory, explanation of the diboson excess. 
Indirect detection bounds disfavor this solution for the Einasto DM profile.

\emph{One} triplet cannot account for the $W_2$ branching ratio, nor the full DM abundance; \emph{three} copies of the triplet can however give the required $W_2$ branching ratios for the diboson excess \emph{and} account for the entire DM abundance. Assuming the three triplets to be degenerate, this requires a common mass of either $\unit[300]{GeV}$ or $\unit[625]{GeV}$. The left-handed fraction of the abundance is then $\Omega_L/\Omega_\text{obs} \simeq 0.03$ or $0.15$, respectively. 
Three triplets with common mass $M\simeq \unit[300]{GeV}$ are, however, excluded by LHC searches~\cite{Aad:2013yna}, seeing as they increase the wino production cross section by a factor of 3 and hence increase the wino bound of $\unit[270]{GeV}$.
Three triplets around $\unit[625]{GeV}$ are in experimental reach and provide a simple explanation for the diboson excess with $g_R = g_L$.

The fermion bi-doublet does not suffer from the mass-splitting constraint and can thus give $100\%$ of DM for $M_{W_2}=\unit[1.9]{TeV}$ and $M\simeq \unit[2]{TeV}$ (see Fig.~\ref{fig:MvsMWR_bidoublet}). However, with these values the decay $W_2\to\text{DM}$ is kinematically forbidden and the $W_2$ branching ratios can not fit the diboson excess. So, even in the bi-doublet case one has to postulate an additional source -- maybe simply more bi-doublet generations -- of DM to account for observations. One bi-doublet gives $\Gamma (W_2\to\text{bi-doublet})\simeq\Gamma (W_2\to\text{triplet})/2$ for the same mass, and hence a branching ratio $\BR (W_2\to \text{DM}) \simeq 18\%$. Similar to the triplet, multiple copies of the bi-doublet are consequently needed to lower the $W_2$ branching ratios sufficiently for the diboson excess.

More than \emph{one} of our new multiplets are therefore always necessary to explain the $\unit[2]{TeV}$ anomalies in an LR model with $g_R=g_L$ \emph{and} make up all of the DM density of our universe. 
One possible solution is a quintuplet with mass $M\simeq\unit[800]{GeV}$ to lower the $W_2$ branching ratios plus one bi-doublet with mass $M \simeq \unit[2.4]{TeV}$ to set the DM abundance.\footnote{Note that both multiplets are separately stable, but the additional annihilation rate of bi-doublets into quintuplets requires a larger bi-doublet mass to obtain $\Omega_\text{obs}$ compared to the case without a quintuplet.}
The quintuplet component  is again constrained by indirect detection data.
The dominant $W_2$ decay mode $\Gamma (W_2\to\text{multiplet})$ is of course not invisible but rather leads to a (slightly) displaced vertex signature, e.g.~$W_2^+\to \chi \Psi_R^+\to \chi\chi \pi^+$~\cite{Cirelli:2005uq,Cirelli:2014dsa}. Note that the mass splitting of charged and neutral components of $\Psi_R$ is about $\unit[1]{GeV}$ in this case, much larger than the corresponding mass splitting of $\Psi_L$ (see Fig.~\ref{fig:mass-splitting}).
Other combinations of multiplets are possible as well.
A study of the LHC prospects will be presented elsewhere.

\section{Conclusion}
\label{sec:conclusion}

Left--right symmetric extensions of the Standard Model based on the gauge group $SU(2)_L\times SU(2)_R\times U(1)_{B-L}$ are theoretically appealing because they shed light on the maximal parity violation of weak interactions and give rise to seesaw-suppressed neutrino masses. Here we studied in detail how dark matter can be introduced into the LR framework, using only the gauge group to make the new particle stable, without requiring an ad hoc stabilizing symmetry. The new fermion/scalar multiplet can be stable either because of its representation under the remaining matter parity $\mathbb{Z}_2^{B-L}$ symmetry of the vacuum, or because decay-operators only arise at mass-dimension $>4$, in the Minimal Dark Matter spirit.
For fermions, the only new parameter of these new multiplets is its common mass, split by radiative corrections and fixed to obtain the observed DM abundance once $M_{W_2}$ is given. Scalar multiplets behave similarly, but bring additional couplings in the scalar potential. We have surveyed fermion multiplets that lead to Majorana DM and scalar multiplets with real DM; some cases have multicomponent DM (see Table~\ref{tab:results} for an overview). 

In our phenomenological study we accounted for the Sommerfeld effect, which arises in Center of the Galaxy --  where indirect detection signatures are expected -- and in the early Universe.
The dominant constraints come from indirect detection, in particular from DM annihilating into monochromatic gamma rays, which already exclude large regions of parameter space of some of the candidates (depending on the DM profile employed), see Figs.~\ref{fig:MvsMWR} (triplet and quintuplet), \ref{fig:MvsMWR_bidoublet} (bi-doublet), and \ref{fig:bitriplet} (bi-triplet).
Future observations, most notably with the CTA, will further improve the limits on the dark matter fraction by a factor of $1.2$--$3$~\cite{Garcia-Cely:2015dda, Ibarra:2015tya} (assuming line-like spectral feature searches and $\unit[112]{h}$ of observation time). Moreover, CTA will be able to probe candidates with dark matter masses above $\unit[20]{TeV}$, which are not constrained by searches with current gamma-ray telescopes.

To illustrate the predictive power of our framework we have taken the recent hints for a $\unit[2]{TeV}$ gauge boson $W_2$ as a benchmark value. We have shown that our new fermion multiplets can easily provide DM for models with $g_R < g_L$, predicting the masses of Table~\ref{tab:dibosonDM}. For $g_R = g_L$, our new particles can open invisible decay channels for $W_2$ that suppress the branching ratios into SM particles, thus resolving the diboson excess in a qualitatively different way. More data is required to pin down the exact model realization behind the diboson excess -- or confirm it to be a statistical fluctuation -- but the framework presented here should serve as a useful tool for a wide range of theories.

All of the DM particles considered here (see Table~\ref{tab:results}) are reminiscent of other WIMPs or even contain components of them. The triplet and bi-triplet behave like a wino in supersymmetric theories, the bi-doublet like a Higgsino, the scalar doublet like a sneutrino or inert doublet, and the quintuplet and scalar 7-plet contain Minimal Dark Matter. With the exception of the latter, these candidates typically require the ad hoc introduction of a stabilizing symmetry, which in our case is provided automatically by the LR gauge group.
LR models hence provide the perfect environment for WIMPs, with few (relevant) parameters and hence highly testable DM phenomenology.
In a grander scheme, LR models are often viewed as a stepping stone towards grand unification, so we list possible $SO(10)$ representations for our DM candidates in Table~\ref{tab:results}. As is well known, the accidental matter parity survives this embedding, and the $SO(10)$ representations of interest for DM have long been identified~\cite{Martin:1992mq}.
Grand unified models with DM along these lines have been studied already, e.g.~in Refs.~\cite{Kadastik:2009dj,Kadastik:2009cu,Frigerio:2009wf,Mambrini:2015vna,Dev:2015pga,Nagata:2015dma}, and certainly deserve attention. 

\begin{table*}
	\centering
	\renewcommand{\arraystretch}{1.5}
		\begin{tabular}{ccccccc}
			Representation & DM & Decays & $M/\unit{TeV}$ & $M_{W_2}/\unit{TeV}$ & $\sigma_\mathrm{SI}^p/\unit{cm^2}$ & $SO(10)$ embedding\\
			\hline
			\hline
			$(\vec{3},\vec{1}, 0)\oplus (\vec{1},\vec{3},0)$ & 2 MF & -- & $ 0.5$--$2.8$ & $\gtrsim 2.6$ & $\lesssim 2\times 10^{-47}$ &  $\vec{45}$, $\vec{210}$, $\vec{770}$\\
			$(\vec{5},\vec{1}, 0)\oplus (\vec{1},\vec{5},0)$ & 2 MF & -- & $ 2.4$--$8.0$ & $\gtrsim 7.8$  & $\lesssim 2\times 10^{-46}$ & $\vec{770}$, $\vec{5940}$\\
			$(\vec{2},\vec{2}, 0)$ & 1 MF & -- &  $ 1.2$--$30$ & $\lesssim 75$ & $\ll 10^{-48}$ & $\vec{10}$, $\vec{120}$, $\vec{126}$, $\vec{210'}$, $\vec{320}$\\
			$(\vec{3},\vec{3}, 0)$ & 1 MF & -- &  $ 1.8$--$40$ & $\gtrsim 4.5$ & $2\times 10^{-47}$ & $\vec{54}$, $\vec{660}$, $\vec{770}$\\
			\hline
			$(\vec{7},\vec{1}, 0)\oplus (\vec{1},\vec{7},0)$ & 2 RS & dim-5 & $1.7$--$22.7$ & -- & $\lesssim 3\times 10^{-44}$ & $\vec{7644}$\\
			$(\vec{2},\vec{1}, -1)\oplus (\vec{1},\vec{2},-1)$ & 1 RS & -- & $\gtrsim 0.15$ & -- & $\lesssim  2\times 10^{-46}$ & $\vec{16}$, $\vec{144}$, $\vec{560}$
		\end{tabular}
\caption{Left--right dark matter candidates discussed in this work, indicating the $SU(2)_L\times SU(2)_R\times U(1)_{B-L}$ 
representation, the number of DM components (MF $=$ Majorana fermion, RS $=$ real scalar), the mass-dimension of possible decay 
operators. Restrictions on the DM mass $M$ and the $W_2$ mass $M_{W_2}$ arise from obtaining the correct relic abundance (for $g_R = g_L$); indirect detection further constrains the allowed values, see Figs.~\ref{fig:MvsMWR} (triplet and quintuplet), \ref{fig:MvsMWR_bidoublet} (bi-doublet), and \ref{fig:bitriplet} (bi-triplet). 
We also give the spin-independent cross sections $\sigma_\mathrm{SI}^p$ for direct detection~\cite{Cirelli:2005uq,Hisano:2015rsa,Klasen:2013btp}, with an inequality for the cases where only the left-handed component gives limits.
For scalars, the mass ranges and cross sections change if scalar--scalar interactions are taken into account. 
In the last column we also give some $SO(10)$ representations that contain our LR DM candidates.
\label{tab:results}}
\end{table*}

\section*{Acknowledgements}
 We thank Sudhanwa Patra for collaboration in the early stages of this project and Michel Tytgat and Laura Lopez Honorez for discussions.
CGC is supported by the IISN and the Belgian Federal Science Policy through the Interuniversity Attraction Pole P7/37 ``Fundamental Interactions''.
JH is a postdoctoral researcher of the F.R.S.-FNRS.
We acknowledge the use of \texttt{Package-X}~\cite{Patel:2015tea} for calculating loop integrals and \texttt{JaxoDraw}~\cite{Binosi:2003yf} for drawing Feynman diagrams.

{\texttt{Note added:} After finalizing this work, a preliminary analysis of the $\sqrt{s}=\unit[13]{TeV}$ run-2 data by ATLAS and CMS puts slight pressure on the significance of the diboson excess, but a proper evaluation requires more data~\cite{Dias:2015mhm}.}

\appendix

\section{Gauge boson masses and decay rates for \texorpdfstring{$g_L \neq g_R$}{gL != gR}}
\label{app:gauge_boson_mixing}

Here we present some useful formulae for LR models with $g_L\neq g_R$, as potentially relevant for the diboson excess (see Sec.~\ref{sec:diboson}).
In this general case, the spontaneous symmetry breaking results in different mass relations for charged and neutral gauge bosons than the ones given in the main text.

\subsection{Gauge boson masses and mixing}
\label{sec:gauge_mixing}

In the following, we assume $\kappa_{1,2}$ to be real and set $\kappa^2 \equiv \kappa_1^2+\kappa_2^2$ and $\tan\beta = \kappa_1/\kappa_2$. Approximations will be made in the limit $\kappa \ll v_R$.
For $g_L\neq g_R$, the charged-boson mass (squared) matrix in the basis $(W^+_L, W^+_R)$ takes the form
\begin{align}
	M_{\rm charged}^2 = \frac{1}{4}
	\matrixx{
			g^2_L \left(\kappa^2 +2 v_L^2\right) & - 2g_L g_R \kappa^*_1 \kappa_2\\
			- 2g_L g_R \kappa_1 \kappa^*_2 & g^2_R \left(\kappa^2 +2 v_R^2\right) 
	} ,
\end{align}
diagonalized by the rotation
\begin{align}
\matrixx{W_L^+\\W_R^+} = \matrixx{\cos\xi & \sin\xi\\-\sin\xi & \cos\xi} \matrixx{W_1^+\\W_2^+} .
\end{align}
The masses for the physical charged gauge bosons $W_{1,2}$ are as follows (neglecting $v_L$)
\begin{align}
M_{W_1}^2 \simeq \frac{ g^2_L}{4} \kappa^2 , &&
M_{W_2}^2 \simeq \frac{ g^2_R}{4} \left(2 v^2_R+\kappa^2 \right)  ,
\end{align}
while the mixing angle $\xi$ is given by
\begin{align}
\begin{split}
	\tan 2\xi &= -\frac{2 g_L g_R \kappa^2 \sin 2\beta}{2 g_R^2 v_R^2+(g_R^2-g_L^2) \kappa^2}\\
	&	\simeq -2 \, \frac{g_R}{g_L} \frac{M_{W_1}^2}{M_{W_2}^2} \sin 2\beta \,.
\end{split}
\label{eq:xi}
\end{align}
Similarly, the neutral gauge boson mass (squared) matrix, in the basis $(W_L^3, W_R^3, B)$, is given by
\begin{align}
M_{\rm neutral}^2 = \frac{1}{4}
\matrixx{
 g^2_L \kappa^2     & -g_L g_R \kappa^2               & -4 g_L g_{BL}  v_L^2   \\
 -g_R g_L \kappa^2&g^2_R\left( \kappa^2+4 v_R^2\right)  &-4 g_R g_{BL}  v_R^2\\
 -4 g_R g_{BL}  v_L^2  &-4 g_R g_{BL} v_R^2           &4 g_{BL}^2 v_R^2
},
\end{align}
with $v_L\simeq 0$. 
This real symmetric neutral gauge boson mass matrix can be diagonalized 
by an orthogonal mixing matrix. For $g_L\neq g_R$, the physical mass eigenstates are 
related to the flavor eigenstates as follows
\begin{align}
\matrixx{W_{L}^3 \\ W_{R}^3 \\ B} = \matrixx{c_W c_\phi & c_W s_\phi & s_W \\ -s_W s_M c_\phi  - 
c_M s_\phi & -s_W s_M s_\phi +c_M c_\phi &  c_W s_M\\ -s_W c_M c_\phi + s_M s_\phi & -s_W c_M s_\phi  - s_M c_\phi & c_W c_M} \matrixx{Z_1\\Z_2\\A} ,
\end{align}
with the standard abbreviations $s_x\equiv \sin x$ and $c_x\equiv \cos x$.
The masses of the neutral gauge bosons $Z_{1,2}$ are given by
\begin{align}
M_{Z_1}^2 \simeq \frac{ g_L^2}{4 c_W^2} \kappa^2 , &&
M_{Z_2}^2 \simeq \frac{g^2_R}{c_M^2} v_R^2 +\frac{g_R^2}{4} c_M^2 \kappa^2  \,,
\label{eq:neutral_masses}
\end{align}
implying the ratios $M_{W_1}/M_{Z_1} \simeq c_W$ (as in the SM) and $M_{W_2}/M_{Z_2} \simeq c_M/\sqrt{2}$, and
the mixing angles are
\begin{align}
\sin \theta_W &= \frac{g_{BL} g_R}{\sqrt{g_R^2 g_{BL}^2 + g_L^2 g_{BL}^2 +
g_L^2 g_R^2}} \,, \\
\sin \theta_M &= \frac{g_{BL}}{\sqrt{g_{BL}^2 + g_R^2}} = \frac{g_L}{g_R} \tan \theta_W \,,
\label{eq:thetaM}\\
\begin{split}
\tan 2\phi &= \frac{2 g_L g_R \kappa^2 c_M^3/c_W}{g_L^2 \kappa^2 c_M^2+g_R^2 \left(-\kappa^2-4 v_R^2+\kappa^2 (2+\cos(2 \theta_M)) s_M^2\right)}\\
& \simeq - 2 c_W c_M \frac{g_R}{g_L} \frac{M_{Z_1}^2}{M_{Z_2}^2}\,.
\end{split}
\end{align}
We note that the mass ratio $M_{Z_2}/M_{W_2}\simeq \sqrt{2}/c_M$ depends on $g_R/g_L$ via $\theta_M$, which in particular implies $M_{Z_2} \geq \sqrt{2} M_{W_2}$. $M_{Z_2}/M_{W_2}$ obviously becomes infinitely large for $c_M\to 0$; from the definition of $\theta_M$ (Eq.~\eqref{eq:thetaM}) we find the consistency relation $g_R/g_L \geq \tan\theta_W\simeq 0.55$.
Most importantly, we have the definition of the electric charge as
\begin{align}
e = g_L s_W = g_R c_W s_M = g_{BL} c_W c_M \,.
\end{align}
For $g_L=g_R$, these relations give back $s_M = \tan\theta_W$, $g_L=g_R=e/\sin\theta_W$, and $g_{BL}=e/\sqrt{\cos 2\theta_W}$, as used in the main text~\cite{Duka:1999uc}.

\subsection{Gauge boson decay rates}
\label{sec:gauge_boson_decays}

\begin{figure}[t]
\includegraphics[width=0.55\textwidth]{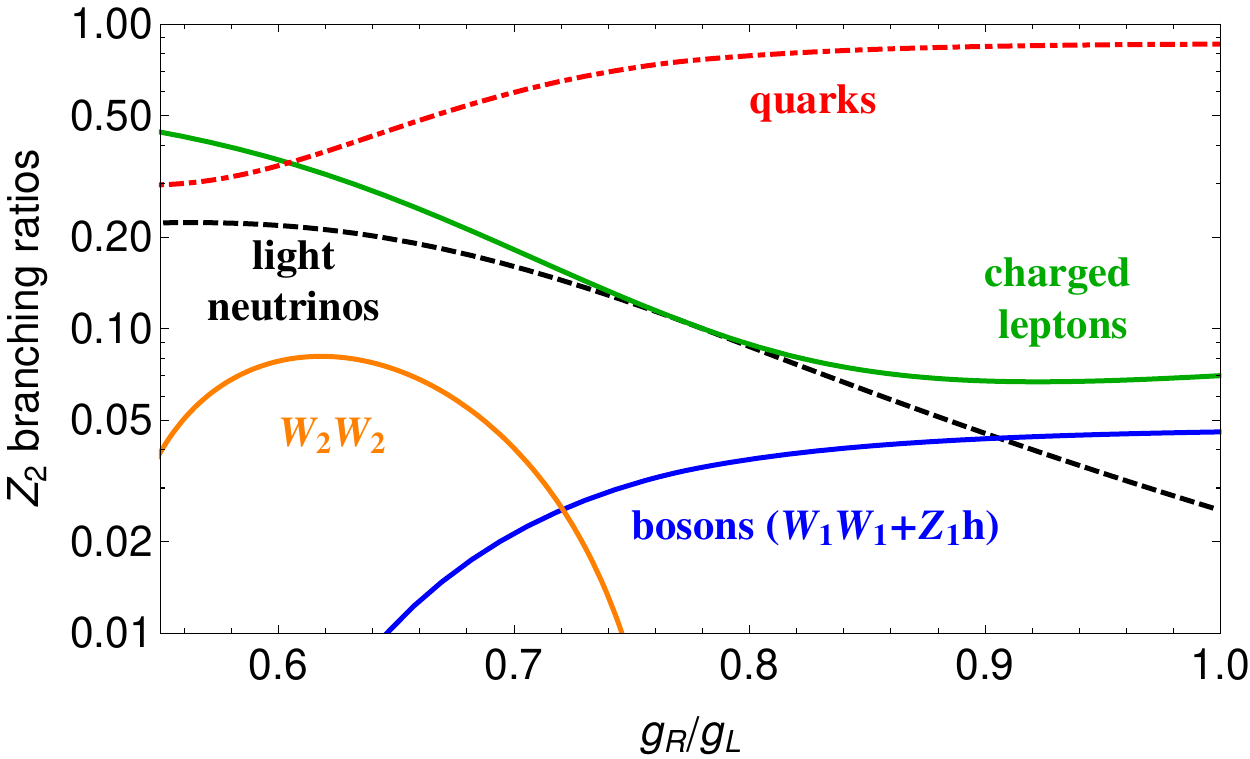} 
\caption{Branching ratio of $Z_2$ to light neutrinos (black, dashed), charged leptons (green), quarks (red, dot-dashed), and bosons (blue) in the limit $2 M_N > M_{Z_2}\gg M_t$.}
\label{fig:branching_ratios_Z2}
\end{figure}

For reference, we give the decay widths of the new gauge bosons into fermions 
and gauge bosons, assuming the non-SM-like scalars are too heavy to contribute. 
For the neutral gauge boson $Z_2$, the expressions are somewhat unwieldy due to the non-negligible mixing angle $\theta_M$. We have, per fermion generation and ignoring mixing~\cite{Deshpande:1988qq,Deandrea:1992ag},
\begin{align}
\begin{split}
\Gamma (Z_2 \to f\bar f) &\simeq \frac{N_c^f}{48\pi}  \left[ (G_L^f + G_R^f)^2 \left(1+\frac{2 M_f^2}{M_{Z_2}^2}\right) + (G_L^f - G_R^f)^2 \left(1-\frac{4 M_f^2}{M_{Z_2}^2}\right)\right] \\
&\qquad\times \left(1-\frac{4 M_f^2}{M_{Z_2}^2}\right)^{1/2}  M_{Z_2}\,,\quad \text{ with } f \in \{e,u,d\} \,,
\end{split}\\
\Gamma (Z_2 \to \nu \nu ) &\simeq \frac{1}{96\pi} \frac{g_R^2 s_M^4}{c_M^2}  M_{Z_2}\,,\\
\Gamma (Z_2 \to N N) &\simeq \frac{1}{96\pi} \frac{g_R^2}{c_M^2}  \left(1 - \frac{4 M_N^2}{ M_{Z_2}^2}\right)^{3/2} M_{Z_2} \,,\\
\Gamma (Z_2 \to W_2^+ W_2^-) &\simeq \frac{g_R^2 }{192\pi}\frac{1}{c_M^2} \left( 1- \frac{4 M_{W_2}^2}{M_{Z_2}^2}\right)^{3/2} \left(1 + 20 \frac{M_{W_2}^2}{M_{Z_2}^2} + 12 \frac{M_{W_2}^4}{M_{Z_2}^4} \right)  M_{Z_2} \,,\\
\Gamma (Z_2 \to W_1^+ W_1^-) &\simeq \frac{g_R^2 }{192\pi}c_M^2 \left( 1- \frac{4 M_{W_1}^2}{M_{Z_2}^2}\right)^{3/2} \left(1 + 20 \frac{M_{W_1}^2}{M_{Z_2}^2} + 12 \frac{M_{W_1}^4}{M_{Z_2}^4} \right)  M_{Z_2} \,,\\
\begin{split}
\Gamma (Z_2 \to Z_1 h) &\simeq \frac{g_R^2}{192\pi} c_M^2 \sin^2 (\alpha -\beta) \left(1-2 \frac{M_{Z_1}^2+ M_{h}^2}{M_{Z_2}^2} + \frac{(M_{Z_1}^2- M_{h}^2)^2}{M_{Z_2}^4} \right)^{1/2}  \\
	&\qquad \times \left(1+\frac{ 10 M_{Z_1}^2- 2 M_{h}^2}{M_{Z_2}^2} + \frac{(M_{Z_1}^2- M_{h}^2)^2}{M_{Z_2}^4} \right) M_{Z_2} \,,
\end{split}
\end{align}
where $\nu$ ($N$) denotes the light (heavy) Majorana neutrinos.
The couplings $G_{L,R}^f$ of the charged-fermion $f \in \{e,u,d\}$ are 
\begin{align}
G_L^f &= -\frac{g_R s_M^2}{2 c_M} (B-L)(f) \,, & G_R^f = g_R c_M Q(f) - \frac{g_R}{2 c_M} (B-L)(f) \,,
\end{align}
with $(B-L)\{e,u,d\} = \{-1,\tfrac13,\tfrac13\}$, $Q\{e,u,d\} = \{-1,\tfrac23,-\tfrac13\}$, and the color factor $N_c^{\{e,u,d\}} = \{1,3,3\}$.
We work exclusively in the alignment limit, where the scalar mixing angle $\alpha$ is given by $\alpha = \beta - \pi/2$, which gives $\Gamma (Z_2 \to W_1^+ W_1^-)\simeq\Gamma (Z_2 \to Z_1 h)$ up to final state masses.
For $2 M_N > M_{Z_2}\gg M_t$, the full width into SM particles is then given by
\begin{align}
\Gamma_{Z_2} \simeq \frac{g_R^2}{96\pi c_M^2} \left(22 - 50 s_M^2 +41 s_M^4 \right)  M_{Z_2} \,,
\label{eq:totalSMwidthZ2}
\end{align}
which numerically equates to $\Gamma_{Z_2}/M_{Z_2} \simeq 2.2\%$ ($1.5\%$) for $g_R /g_L=1$ ($0.65$). In Fig.~\ref{fig:branching_ratios_Z2} we show the branching ratios into SM particles as a function of $g_R/g_L$.
For $g_R = g_L$, the dominant decays are into down-quarks, up-quarks, and right-handed neutrinos (if sufficiently light), but for $g_R < g_L$, leptons become increasingly dominant. In particular, the rate into the three light neutrinos increases by almost an order of magnitude from $\BR (Z_2\to \nu\nu)\simeq 2.5\%$ ($g_R=g_L$) to $\BR (Z_2\to \nu\nu)\simeq 20\%$ ($g_R/g_L = 0.65$).
The channel $Z_2\to W_2 W_2$ opens up for $g_R/g_L \lesssim 0.77$ because this implies $M_{W_2}/M_{Z_2} \simeq c_M/\sqrt{2} \lesssim 1/2$ (see Eq.~\eqref{eq:thetaM}), but this channel is never dominant.

For the charged gauge boson $W_2$, we obtain much simpler expressions (neglecting any fermion mixing and masses of down-type fermions)~\cite{Deshpande:1988qq,Deandrea:1992ag,Brehmer:2015cia,Dev:2015pga}
\begin{align}
	\Gamma(W_2^+ \to u \overline{d}) &\simeq \frac{g_R^2}{16\pi} M_{W_2} \left(1+\frac{M_u^2}{2 M_{W_2}^2}\right)  \left(1-\frac{M_u^2}{M_{W_2}^2}\right)^2 ,  \\
	\Gamma(W_2^+ \to  e^+ N) & \simeq
		\frac{g_R^2}{48\pi} M_{W_2}  \left(1+\frac{M_N^2}{2 M_{W_2}^2}\right) \left(1-\frac{M_N^2}{M_{W_2}^2}\right)^2 ,   \\
\begin{split}
	\Gamma(W_2^+ \to W_1^+ Z_1) &\simeq \frac{g_R^2 }{192\pi} M_{W_2}  \sin^2(2\beta) \left(1-2 \frac{M_{W_1}^2+ M_{Z_1}^2}{M_{W_2}^2} + \frac{(M_{W_1}^2- M_{Z_1}^2)^2}{M_{W_2}^4} \right)^{3/2}  \\
	&\qquad \times \left(1+10 \frac{M_{W_1}^2+ M_{Z_1}^2}{M_{W_2}^2} + \frac{M_{W_1}^4+ 10 M_{W_1}^2 M_{Z_1}^2 + M_{Z_1}^4}{M_{W_2}^4} \right) ,
\end{split}\\
\begin{split}
	\Gamma(W_2^+ \to W_1^+ h) &\simeq \frac{g_R^2}{192\pi} M_{W_2} \cos^2(\alpha + \beta)  \left(1-2 \frac{M_{W_1}^2+ M_{h}^2}{M_{W_2}^2} + \frac{(M_{W_1}^2- M_{h}^2)^2}{M_{W_2}^4} \right)^{1/2}  \\
	&\qquad \times \left(1+\frac{ 10 M_{W_1}^2- 2 M_{h}^2}{M_{W_2}^2} + \frac{(M_{W_1}^2- M_{h}^2)^2}{M_{W_2}^4} \right) .
\end{split}
\end{align}
The color factor for the quarks has already been taken into account and makes them the dominant $W_2$ decay channel.
In the alignment limit, we have $\Gamma(W_2^+ \to W_1^+ Z_1)\simeq \Gamma(W_2^+ \to W_1^+ h)$ for $M_{W_2}\gg M_h$, in accordance with the Goldstone boson equivalence theorem ($h$ being the SM-like scalar with mass $M_h =\unit[125]{GeV}$).
The limits on $M_{W_2}$ depend strongly on the masses of the right-handed neutrinos $N$,
which decide whether the channels $W_2\to \ell N$ are kinematically allowed. 
For $M_N > M_{W_2}\gg M_t$, the full width into SM particles is given by
\begin{align}
\Gamma_{W_2} \simeq \frac{g_R^2}{96\pi } \left(18+\sin^2 2\beta \right)  M_{W_2} \,.
\label{eq:totalSMwidthW2}
\end{align}

If the mass of any of our newly introduced DM multiplets is below $M_{W_2}/2$ or $M_{Z_2}/2$, new decay channels open up. Since most of our new particles reside in rather large representations, the branching ratio into DM can be significant, barring phase space suppression. The calculation of the partial widths is straightforward with the explicit interaction Lagrangians provided in the main text for each candidate, so we will omit them here.

\section{Real representations of \texorpdfstring{$SU(2)$}{SU(2)}}
\label{app:real_reps}

For reference purposes we provide here a pedagogical introduction to real $SU(2)$ representations.
The Lie group $SU(2)$ is defined by the well-known Lie algebra commutator relations of the three generators $T_{1,2,3}$:
\begin{align}
[T_a, T_b] = i \epsilon_{abc} T_c \,.
\end{align}
The $(2j+1)$ dimensional representation -- where $j=0,\frac12,1,\frac32,2,\dots$ -- has components $\ket{j,m}$, where $m=-j,-j+1,\dots,+j$. Working in the eigenbasis of $T_3$, the generators act as
\begin{align}
T_3 \ket{j,m} &= m \ket{j,m} ,\\
T_\pm \ket{j,m} &= \sqrt{j(j+1)-m(m\pm 1)} \ket{j,m\pm 1} ,\\
T^2\ket{j,m} &= j(j+1) \ket{j,m} .
\end{align}
Here, we introduced the raising and lowering operators $T_\pm \equiv T_1\pm i T_2$, as well as the Casimir operator $T^2 \equiv T_1^2 + T_2^2+T_3^2$.

With the above formulae one can immediately write down the $(2j+1)\times (2j+1)$ generator matrices relevant for e.g.~the covariant derivative of a field in the representation $\vec{2j+1}$ of $SU(2)$. For a \emph{real} representation -- those with $j=1,2,3,\dots$ -- one can however find a more convenient basis, in which the generators $i T_a^r$ are real, i.e.~$T_a^r = - (T_a^r)^*$, so the $T_a^r$ are generators of $SO(3)$. These generator matrices can be obtained from the complex ones from above by a basis rotation via $T_a^r = U^\dagger T_a U$, $U$ being a unitary matrix. For the lowest cases of interest, $j=1,2,3,4$, possible $U$ matrices are
\begin{align}
U_\vec{3} &= \frac{1}{\sqrt{2}} \left(
\begin{array}{ccc}
 i & 0 & 1 \\
 0 & \sqrt{2} & 0 \\
 i & 0 & -1 \\
\end{array}
\right), &
U_\vec{5} &= \frac{1}{\sqrt{2}} \left(
\begin{array}{ccccc}
 i & 0 & 0 & 0 & 1 \\
 0 & i & 0 & 1 & 0 \\
 0 & 0 & \sqrt{2} & 0 & 0 \\
 0 & i & 0 & -1 & 0 \\
 -i & 0 & 0 & 0 & 1 \\
\end{array}
\right),\\
U_\vec{7} &= \frac{1}{\sqrt{2}} \left(
\begin{array}{ccccccc}
 i & 0 & 0 & 0 & 0 & 0 & 1 \\
 0 & i & 0 & 0 & 0 & 1 & 0 \\
 0 & 0 & i & 0 & 1 & 0 & 0 \\
 0 & 0 & 0 & \sqrt{2} & 0 & 0 & 0 \\
 0 & 0 & i & 0 & -1 & 0 & 0 \\
 0 & -i & 0 & 0 & 0 & 1 & 0 \\
 i & 0 & 0 & 0 & 0 & 0 & -1 \\
\end{array}
\right), & 
U_\vec{9} &= \frac{1}{\sqrt{2}} \left(
\begin{array}{ccccccccc}
 i & 0 & 0 & 0 & 0 & 0 & 0 & 0 & 1 \\
 0 & i & 0 & 0 & 0 & 0 & 0 & 1 & 0 \\
 0 & 0 & i & 0 & 0 & 0 & 1 & 0 & 0 \\
 0 & 0 & 0 & i & 0 & 1 & 0 & 0 & 0 \\
 0 & 0 & 0 & 0 & \sqrt{2} & 0 & 0 & 0 & 0 \\
 0 & 0 & 0 & i & 0 & -1 & 0 & 0 & 0 \\
 0 & 0 & -i & 0 & 0 & 0 & 1 & 0 & 0 \\
 0 & i & 0 & 0 & 0 & 0 & 0 & -1 & 0 \\
 -i & 0 & 0 & 0 & 0 & 0 & 0 & 0 & 1 \\
\end{array}
\right) ,
\end{align}
readily generalized to arbitrary dimension by noting the recurring pattern.
With the so defined generator matrices one can then define e.g.~real scalar fields $\phi_r\sim \vec{3},\vec{5},\vec{7},\dots$ that have the Lagrangian 
\begin{align}
\L = \tfrac12 (D_\mu\phi_r)^T D^\mu \phi_r -\tfrac12 M^2 \phi_r^T\phi_r \,,
\end{align}
with $D_\mu \equiv \del_\mu - i g_2 T_a^r W^a_\mu$ being real. This obviously only makes sense if $\phi_r$ transforms in a real representation of the \emph{full} gauge group, so $\phi_r$ is in particular uncharged under $U(1)$ factors. While conceptually simple, the components of $\phi_r$ are not charge (or $T_3$) eigenstates, so the Lagrangian is not explicitly gauge invariant. The charge eigenstates can be obtained by a simple rotation 
\begin{align}
\phi = U_\vec{2j+1} \phi_r = \left( \dots, \phi^{+++}, \phi^{++},\phi^{+}, \phi^0, -\phi^{-}, +\phi^{--}, - \phi^{---},\dots\right)^T ,
\end{align}
where $\phi^0$ is still a real field, but $\phi^Q$, $Q\neq 0$, are complex scalars with $(\phi^Q)^* = \phi^{-Q}$. It is easy to see that $\phi$ fulfills a self-conjugacy condition,
\begin{align}
\phi = \epsilon \phi^* \,, 
\end{align}
where $\epsilon \equiv U_\vec{2j+1} U_\vec{2j+1}^T $ is an anti-diagonal symmetric matrix with alternating entries $\pm 1$ (here normalized so that $+1$ is in the center).
Substituting $\phi_r = U_\vec{2j+1}^\dagger \phi$ in $\L$ finally gives the Lagrangian for the charge (and mass) eigenstates $\phi^Q$, properly normalized. The above procedure has a straightforward extension to bi-multiplet fields, relevant to the discussion in the main text.

In more generality the matrix $\epsilon$ can be written as $\epsilon = (-1)^j \exp (i \pi T_2)$, acting on a field $\phi\sim \vec{2j+1}$, now also including the $j= 1/2, 3/2,\dots$ representations. This allows for the definition of a field $\tilde \phi \equiv \epsilon \phi^*$, which transforms again as $\vec{2j+1}$, but is not necessarily identical to $\phi$ (e.g.~if $\phi$ is charged under other (gauge) groups).

Let us now turn to a \emph{chiral} fermion $\psi$ that transforms in a real representation $\vec{2j+1}$, $j\in \mathbb{N}_0$, of $SU(2)$. Here, we can define a field $\tilde \psi \equiv \epsilon \psi^c$, which again transforms as $\vec{2j+1}$, but has opposite chirality. If $\psi$ does not carry other quantum numbers -- except maybe for $\mathbb{Z}_2$ charges -- the Lagrangian allows for a mass term 
\begin{align}
\L = i\overline{\psi} \slashed{D} \psi - \tfrac12 M\left( \overline{\psi} \tilde \psi + \hc \right) .
\end{align}
The charge and mass eigenstate field 
\begin{align}
\Psi \equiv \psi + \tilde \psi = \left( \dots, \Psi^{+++}, \Psi^{++},\Psi^{+}, \Psi^0, -\Psi^{-}, +\Psi^{--}, - \Psi^{---},\dots\right)^T
\end{align}
then fulfills the self-conjugacy condition $\Psi = \tilde \Psi$ and consists of one Majorana fermion $\Psi^0$ and $j$ Dirac fermions $\Psi^Q = (\Psi^{-Q})^c$ for $Q=1,\dots, j$.

\section{Mass splitting}
\label{app:mass_splitting}

\begin{figure}[t]
\includegraphics[width=0.8\textwidth]{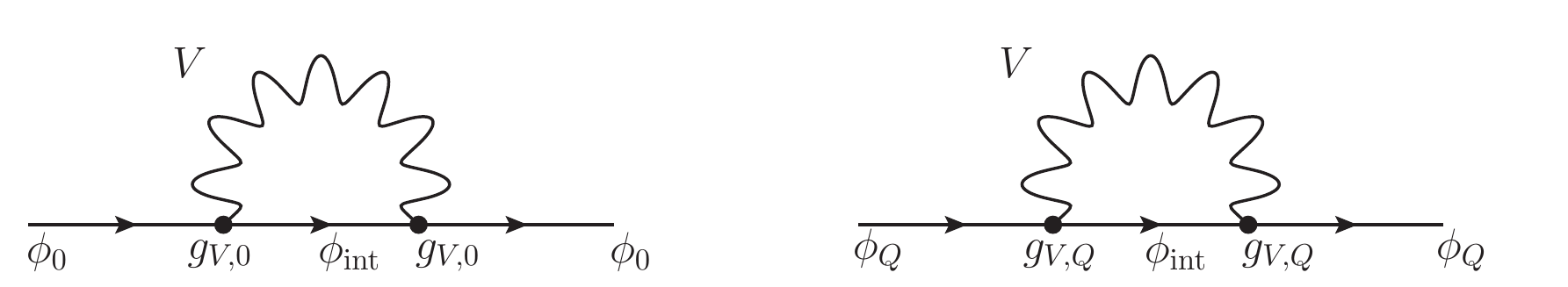}
\caption{Feynman diagrams for the fermion self energies relevant for the mass splitting.}
\label{fig:self-energy}
\end{figure}

The radiative mass splitting of a degenerate $SU(2)_L\times SU(2)_R$ multiplet via vector-current interactions can be readily calculated from the self-energies (see Fig.~\ref{fig:self-energy})~\cite{Cheng:1998hc,Feng:1999fu}. At one loop, the mass splitting between the neutral Majorana component $\phi_0$ (as relevant to our discussion) and a Dirac component $\phi_Q$ of charge $Q\in \mathbb{Z}$ takes the simple form
\begin{align}
M_Q - M_0 = \frac{M}{16\pi^2} \left[ \sum_V \sum_{\phi_\text{int}}c_{\phi_\text{int}} g_{V,0}^2 f(r_V) - \sum_V \sum_{\phi_\text{int}} g_{V,Q}^2 f(r_V)\right] ,
\end{align}
$M$ being the degenerate multiplet mass, $g_{V,X}$ the vector coupling of the vector boson $V$ to $\phi_X$ and the internal fermion $\phi_\text{int}$ (see Fig.~\ref{fig:self-energy}), $r_V \equiv M_V/M$, and the loop function
\begin{align}
\begin{split}
f(r)&\equiv 2\int_0^1 \dd x \, (1+x) \log \left[ x^2 + (1-x) r^2\right]\\
&= -5-r^2+r^4 \log r +\frac{r}{2}  \sqrt{r^2-4} \left(2+r^2\right) \log \left[ \frac{r^2-2-r \sqrt{r^2-4}}{2}\right] .
\end{split}
\label{eq:loop_function_f}
\end{align}
The parameter $c_{\phi_\text{int}}$ is equal to $1$ ($2$) if $\phi_\text{int}$ is a Majorana (Dirac) fermion.
For small $r$, we have $f(r)\simeq -5 + 2\pi r$. For large $r$, one can show that $f(r) \to -7/2+6\log r$.

The same discussion applies to the finite scalar mass splitting, replacing the self-energy function~$f(r)$ by the function
\begin{align}
\begin{split}
g(r)\equiv -5-\frac{r}{4}  \left(2 r^3 \log r+\left(r^2-4\right)^{3/2} \log\left[ \frac{r^2-2-r \sqrt{r^2-4}}{2}\right]\right) .
\label{eq:loop_function_g}
\end{split}
\end{align} 
For small $r$, one has again $g(r)\simeq -5+ 2\pi r$, while for $r\to\infty$:
\begin{align}
g(r)\to -\frac{1}{4} \left(11+2 r^2\right)-3 \left(r^2-1\right) \log r \,.
\end{align}
One important difference arises in the scalar case compared to the fermion case: 
the mass splitting is finite \emph{only if} no tree-level mass splitting due to scalar couplings exists. In this case the mass splitting is invariant under $g(r)\to g(r) + c_1 + c_2 r^2$ (essentially due to custodial symmetry), and we have already used this invariance above to remove a UV-divergent $r^2$ term.\footnote{Due to the invariance, different functions $g$ have been proposed in the literature that give the same mass splitting~\cite{Cirelli:2005uq,Aoki:2015nza}.}

Purely radiative mass splitting occurs for the MDM scalars without hypercharge, because the potentially dangerous coupling of the form $(H H)_\vec{3} (\phi  \phi)_\vec{3}$ vanishes for a self-conjugate $\phi$~\cite{Cirelli:2005uq}. For MDM scalars \emph{with} hypercharge, the coupling does not vanish and it leads to a tree-level mass splitting (which provides a counterterm for the divergent $r^2$ terms in $g(r)$).
In our LR, model custodial symmetry is broken by the triplet's VEVs, so one unavoidably generates a tree-level mass splitting for the new scalar multiplets (see Sec.~\ref{sec:real}).

\section{Sommerfeld effect in the center of the galaxy}
\label{sec:SEinGC}

The exchange of gauge bosons $V$ among non-relativistic dark matter particles that are much heavier than $V$ gives rise to a long-range force. This significantly distorts the DM wave function and leads to an enhancement of the corresponding annihilation cross section. This is called Sommerfeld effect and takes place, for instance, in the Center of the Galaxy, where the DM velocity is of the order of $10^{-3}c$. 

In order to calculate this effect, one must consider the corresponding Schr\"odinger equation and solve for the distortion of the wave function.  Moreover, the exchange of $W$ bosons not only leads to a long-range force but also to transitions of pairs of DM particles into pairs of charged particles belonging to the same $SU(2)_L$ representation. Because of this, the state vector entering the Schr\"odinger equation is a $n\times n$ matrix, where $n$ is the number of particles in the DM $SU(2)_L$ representation. Since we consider  velocities of the order of $10^{-3}c$, it is a very good approximation to take only the $s$-wave part of the annihilation cross section into account and  consequently to only solve the Schr\"odinger equation for the $L=0$ states.   
Then, if $v$ is the relative velocity of the initial state particles, the Schr\"odinger equation is  
\begin{align}
-\frac{1}{M} g''(r) +\left(-\dfrac{1}{4} M v^2 {1\!\!1} +2 \delta m + V(r) \right) g(r) =0 \,.
\label{SoDE}
\end{align}
In this equation, $\delta m $ is a diagonal matrix whose entries are the mass splittings between the corresponding  $SU(2)_L$ components  and the DM particle, as discussed in Appendix~\ref{app:mass_splitting}. In addition, $V(r)$ is the Yukawa potential induced by the gauge boson exchange. We list these potentials for the representations that are considered in this work in Table~\ref{table:potentials}. They agree with the expressions reported in Refs.~\cite{Hisano:2004ds,Cirelli:2007xd}. (The scalar doublet representation is not included here because its indirect detection signals crucially depend on the mass splitting, which can not be fixed without a precise knowledge of the scalar potential.) 
The expressions of Table~\ref{table:potentials} correspond to annihilating initial states with  total charge $Q=0$ and total spin $S=0$. The latter condition stems from the fact that our DM candidates  are  scalar or Majorana particles  and that the $s$-wave annihilation ($L=0$) precludes the $S=1$ configuration for pairs of particles.

\begin{table}
\centering
\begin{tabular}{|c|c|c|}\hline
 Representation & Basis  & $V(r)$ \\\hline
&&\\
\multicolumn{1}{|c|}{
\multirow{2}{*}{ $\begin{array}{c} \text{Triplet} \\ (\vec{3},\vec{1},0)\oplus(\vec{1},\vec{3},0) \end{array}$}
} 
& $ \left(
\begin{array}{c}
 \Psi_L^+\,\Psi_L^-\\
 \Psi_L^0\,\Psi_L^0\\
\end{array}
\right)$
&$ \frac{\alpha_2}{r} \begin{pmatrix}  - U & -\,\sqrt2\,e^{-M_{W_1} r}  \\ -\,\sqrt2\,e^{-M_{W_1} r}   & 0\end{pmatrix}$\\
&&\\\hline
&&\\
\multicolumn{1}{|c|}{
\multirow{2}{*}{$\begin{array}{c} \text{Quintuplet} \\ (\vec{5},\vec{1},0)\oplus(\vec{1},\vec{5},0) \end{array}$}
} & 
$
\left(
\begin{array}{c}
 \Psi_L^{++}\,\Psi_L^{--}\\
 \Psi_L^{+}\,\Psi_L^{-}\\
 \Psi^0\,\Psi^0 \\
\end{array}
\right)
$
&
$ \frac{\alpha_2}{r} \begin{pmatrix} -4\, U  & -2\,e^{-M_{W_1} r}  & 0 \\ -2\, e^{-M_{W_1} r}  & - U & -3\,\sqrt2\,e^{-M_{W_1} r}  \\ 0 & -3\,\sqrt2\,e^{-M_{W_1} r}   & 0\end{pmatrix}$\\
&&
\\\hline
&&\\
\multicolumn{1}{|c|}{
\multirow{2}{*}{$\begin{array}{c} \text{Bi-doublet} \\ (\vec{2},\vec{2},0) \end{array}$}
} & 
$
\left(
\begin{array}{c}
 \Psi^{+}\,\Psi^{-}\\
 \chi_2\,\chi_2\\
 \chi_1\,\chi_1 \\
\end{array}
\right)
$

&$
\frac{\alpha_2}{r}\left(
\begin{array}{ccc}
 -\frac{ \cos ^2(2 \theta_W) e^{-M_{Z_1} r} +\sin ^2(2 \theta_W) }{4 \,c_W^2} &
   \frac{e^{-M_{W_1} r} }{2 \sqrt{2} } & -\frac{e^{-M_{W_1} r}}{2 \sqrt{2} } \\
 \frac{e^{-M_{W_1} r} }{2 \sqrt{2} } & 0 & \frac{e^{-M_{Z_1} r}  }{4\, c_W^2}
   \\
 -\frac{e^{-M_{W_1} r} }{2 \sqrt{2} } & \frac{e^{-M_{Z_1} r}  }{4\, c_W^2} & 0
   \\
\end{array}

\right)$\\
&&
\\\hline
&&\\
\multicolumn{1}{|c|}{
\multirow{1}{*}{$\begin{array}{c} \text{Bi-triplet} \\ (\vec{3},\vec{3},0) \end{array}$}
} & 
$ \left(
\begin{array}{c}
 \Psi_1^+\,\Psi_1^-\\
 \chi\,\chi\\
\end{array}
\right)$
&$\begin{array}{c} \text{Same as} \\ \text{for the triplet.} \end{array} $ \\
&&
 \\\hline
&&\\
\multicolumn{1}{|c|}{
\multirow{2}{*}{ $\begin{array}{c} \text{Scalar 7-plet} \\ (\vec{7},\vec{1},0)\oplus(\vec{1},\vec{7},0) \end{array}$}
} & 
$\begin{pmatrix}  \phi_L^{+++}\, \phi_L^{---}\\ \phi_L^{++}\, \phi_L^{--}\\ \phi_L^{+}\, \phi_L^-\\ \phi_L^{0}\, \phi_L^0 \end{pmatrix}$

&$ \frac{\alpha_2}{r} \begin{pmatrix} -9\, U&  -3\ e^{-M_{W_1} r} & 0 & 0\\ -3\, e^{-M_{W_1} r} & -4 \,U  & -5\, e^{-M_{W_1} r} & 0 \\ 0 & -5\, e^{-M_{W_1} r}  & - U & -6\,\sqrt2\,e^{-M_{W_1} r}   \\ 0 &0  &  -6\,\sqrt2\,e^{-M_{W_1} r}   & 0\end{pmatrix}$\\
&&
\\\hline
\end{tabular}
\caption{ Potential matrices describing the long-range interaction that gives rise to the Sommerfeld effect today. Notice that they correspond to $Q=0$ and $S=0$ states. Here $U\equiv c_W^2 e^{-M_{Z_1} r} +s_W^2$.}
\label{table:potentials}
\end{table}

Eq.~\eqref{SoDE} requires two boundary conditions. On the one hand,   at the origin one demands $ g(0)={1\!\!1 }$. On the other hand, for large values of $r$, the solution describes pairs according to the mass splitting  $\delta m_{ii',ii'}$. If the latter is smaller than the kinetic energy $\tfrac{1}{4}M v^2$, then  on-shell states of the corresponding pair are kinematically allowed and hence the matrix elements $g_{ii',jj'}(r)$ at infinity behave as an out-going wave, whose  momentum is $ p_{ii'}= \sqrt{M \left(\tfrac{1}{4} M v^2 - 2\,  \delta m_{ii',ii'}\right)}$, as inferred from Eq.~(\ref{SoDE}). The corresponding boundary condition at infinity is 
\begin{align}
\frac{\dd g_{ii',jj' }(\infty)}{\dd r} = i \, p_{ii'} \, g_{ii',jj' } (\infty)\,, \quad \text{if}  \quad\delta m_{ii',ii'} < \tfrac{1}{4}M v^2\,.
\end{align}
When on-shell states are not allowed at infinity because $\delta m_{ii', ii'}> \tfrac{1}{4}M v^2$, the matrix elements $g_{ii',jj'}(r)$ describe a damping wave and therefore
\begin{align}
g_{ii',jj' }(\infty) = 0 \,, \quad \text{if}  \quad \delta m_{ii',ii'} > \tfrac{1}{4}M v^2\,.
\end{align}
As a consequence of the boundary conditions, $g(r)$ at large values of $r$ can be recast as
\begin{align}
g(r) \to e^{ ir\sqrt{M \left(\tfrac{1}{4} M v^2 {1\!\!1} - 2\,  \delta m\right)}} d\,.
\end{align}
Notice that $d={1\!\!1}$ when the potential in Eq.~(\ref{SoDE}) is negligible. Because of this, the matrix $d$ encodes the distortion of the wave function. In fact, the $s$-wave cross section for the annihilation of the pair $(i,i')$ into a final state $f$ is calculated by means of the equation
\begin{align}
\sigma v \left(i i' \to f\right) \Big|_\text{$s$ wave} = c_{ii'} (d\,\Gamma^{(f)} \, d^\dagger)_{ii',ii'} \,,
\label{SEsigmav}
\end{align}
where $c_{ii'}=2$ if the particles of the pair are identical and $c_{ii'}=1$ otherwise. $\Gamma^{(f)}$ is a matrix that describes the annihilation process, given by  
\begin{align}
\Gamma^{(f)}_{ii',jj'} = \frac{N_{ii'} N_{jj'} }{4M^2} \int
{\cal M} \left(ii'  \to f \right) {\cal M}^*\left(jj' \to f \right) (2\pi)^4 \delta^{(4)}\left(P_{ii'}-P_f \right)
\left(\prod_{a \in f } \frac{\dd^3 q_a}{(2\pi)^3 2 E_a} \right) 
\label{SEsigmav2}
\end{align}
in terms of the annihilation amplitudes ${\cal M} (ii' \to f)$.
Here, $N_{ii'}=1/\sqrt{2}$ if the particles of the pair are identical and $N_{ii'}=1$ otherwise. Again, notice that when the potential is negligible, $d={1\!\!1}$ and Eqs.~\eqref{SEsigmav} and \eqref{SEsigmav2} reduce to the usual formulae for calculating the $s$-wave piece of the annihilation cross sections.  

In this work we are interested in calculating the annihilation cross sections for processes that generate gamma rays. On the one hand, annihilations into $WW$, $ZZ$ and $Z\gamma$ produce a soft continuum of photons from the decay and fragmentation of the massive gauge bosons. On the other hand, annihilations into $Z\gamma$ and $\gamma\gamma$ produce monochromatic photons with an energy roughly equal to the DM mass. Employing Eq.~\eqref{SEsigmav2} we calculate the annihilation matrices for those final states. For the triplet and the bi-triplet, these are  
\begin{align}
\Gamma^{WW}=  \frac {\alpha_2^2 \pi }{2 \,M^2} \left(
 \begin{array}{cc}
  1   & \sqrt{2} \\
 \sqrt{2} & 2 \\
\end{array}\right) , &&
\Gamma^{\gamma\gamma}= \tfrac{s_W^2}{2 c_W^2} \Gamma^{\gamma Z} =  \tfrac{s_W^4}{ c_W^4}  \Gamma^{ZZ} =  \frac {\alpha_2^2 s_W^4 \pi }{M^2} 
  \left(
 \begin{array}{cc}
  1   & 0\\
  0 & 0 \\
\end{array}\right) .
\end{align}
Likewise, for the quintuplet 
\begin{align}
\Gamma^{WW}=  \frac {\alpha_2^2 \pi }{2 \,M^2} 
\left(
\begin{array}{ccc}
 4 & 10 & 6 \sqrt{2} \\
 10 & 25 & 15 \sqrt{2} \\
 6 \sqrt{2} & 15 \sqrt{2} & 18 \\
\end{array}
\right) , &&
\Gamma^{\gamma\gamma}= \tfrac{s_W^2}{2 c_W^2} \Gamma^{\gamma Z} =  \tfrac{s_W^4}{ c_W^4}  \Gamma^{ZZ} =  \tfrac {\alpha_2^2 s_W^4 \pi }{M^2} 
 \left(
\begin{array}{ccc}
 16 & 4 & 0 \\
 4 & 1 & 0 \\
 0 & 0 & 0 \\
\end{array}
\right) \label{GammaQuin}.
\end{align}
Finally, for the bi-doublet,
\begin{align}
\Gamma^{WW} &=  \frac {\alpha_2^2 \pi }{16 \,M^2}
\left(
\begin{array}{ccc}
 2 & -\sqrt{2} & \sqrt{2} \\
 -\sqrt{2} & 1 & -1 \\
 \sqrt{2} & -1 & 1 \\
\end{array}
\right)
, \quad
\Gamma^{\gamma\gamma}= \tfrac{\tan^2(2\theta_W)}{2} \Gamma^{\gamma Z} =  \frac {\alpha_2^2 s_W^4 \pi }{M^2} 
  \left(
 \begin{array}{ccc}
  1  &0  & 0\\
  0  &0  & 0 \\
  0  &0  & 0 \\
\end{array}\right),\\
\Gamma^{ZZ}&=  \frac {\alpha_2^2 \pi }{32 \, c_W^4 \,M^2}
\left(
\begin{array}{ccc}
 2 \cos^4(2\theta_W) & -\sqrt{2} \cos^2(2\theta_W) & \sqrt{2} \cos^2(2\theta_W) \\
 -\sqrt{2} \cos^2(2\theta_W) & 1 & -1 \\
 \sqrt{2} \cos^2(2\theta_W) & -1 & 1 \\
\end{array}
\right).
\end{align}
The previous matrices are written in the basis stated in Table~\ref{table:potentials} and they again correspond to $Q=0$ and $S=0$ states. Subsequently, in order to solve the Schr\"odinger Eq.~\eqref{SoDE}, we use the method introduced in Ref.~\cite{Garcia-Cely:2015dda} and find the matrix $d$. Then, we use Eq.~\eqref{SEsigmav} to obtain the annihilation cross sections.

\section{Relic density in the \texorpdfstring{$SU(2)_L$}{SU(2)L} symmetric limit}
\label{sec:SU2LL}

\input{AppendicesTables.tex}

The relic-density calculation for left--right symmetric dark matter requires us to account for the Sommerfeld effect because the masses of the DM candidates are typically at the TeV scale. Although the effect on the relic density is much smaller than in the Center of the Galaxy, it is still significant and affects the DM phenomenology. In order to calculate it, we consider the so-called instantaneous freeze-out approximation for solving the Boltzmann equation associated to the abundance of DM. This leads to  
\begin{align}
\Omega h^2 =  \frac{\unit[1.03\times 10^9]{GeV^{-1}}}{g_\star^{1/2} M_\mathrm{Pl} } \left(\int^\infty_{x_f} \frac{\langle \sigma v \rangle_\mathrm{eff}}{x^2} \dd x \right)^{-1} ,
\label{Omega}
\end{align}
where $x_f \equiv M/T_f$ is the inverse freeze-out temperature normalized to the DM mass, $g_\star$ is the relativistic number of degrees of freedom at the freeze-out, and $\langle \sigma v \rangle_\mathrm{eff}$ is the thermal average of the effective total annihilation cross section accounting for all the co-annihilating species (discussed at length below). We find $x_f$ by calculating the temperature at which  the DM annihilation rate drops below the Hubble expansion rate. By considering this, one finds 
\begin{align}
x_f = \log \left( \frac{0.038 \,M_\mathrm{Pl}\, M\, g \,x_f^{1/2}  \langle \sigma v \rangle_\mathrm{eff} }{g_\star^{1/2}}\right) .
\end{align}
Here $g$ is the number of degrees of freedom associated to the DM particle. For (real) scalar, Majorana, and Dirac particles, $g=1$, $2$, and $4$, respectively.  Typically, one finds $x_f \simeq 20$--$30$, which corresponds to  temperatures greater than the electroweak scale for TeV DM.
Then, we can reasonably ignore the mass splittings for the co-annihilating species in the relic density calculation  since they are orders of magnitude below the freeze-out temperature, as follows from Appendix~\ref{app:mass_splitting}. 
Under these circumstances, the effective annihilation cross section is 
\begin{align}
\langle \sigma v\rangle_\mathrm{eff} = \sum_{ij} \frac{g_i g_j}{g_\mathrm{tot}^2} \langle\sigma_{ij} v \rangle ,
\end{align} 
where the sum runs over all the co-annihilating pairs, $g_\mathrm{tot} = \sum_i g_i$ is the total number of degrees of freedom, and the cross sections are $\sigma_{ij} v = \sum_f \sigma v (ij\to f)$, using the  notation of Appendix~\ref{sec:SEinGC}.  

In general, in order to calculate $\langle \sigma v\rangle_\mathrm{eff}$, one has to determine the $\Gamma$ and $V$ matrices for all the co-annihilating pairs and solve the corresponding Schr\"odinger equation  for every velocity, as explained in Appendix~\ref{sec:SEinGC},  and then calculate the thermal average. This procedure can be greatly simplified if one neglects the mass splittings among co-annihilating particles as well as the gauge boson masses~\cite{Cirelli:2009uv}. In that case, the potential matrix $V(r)$ can be written in the basis of total isospin, in which it is diagonal. Furthermore, in this basis the $\Gamma$ matrices are block-diagonal, with each block corresponding to the same isospin $I$, spin $S$ and total hypercharge  $Y$. This, in a very convenient notation, reads as
\begin{align}
V \,|I_3, I\rangle^S_{Y} = V^S_{I,Y} |I_3, I\rangle^S_{Y}\,, &&
\Gamma \,|I_3, I\rangle^S_{Y} = \Gamma^S_{I,Y} |I_3, I\rangle^S_{Y}\,.
\end{align}
Notice that the eigenvalues do not depend on $I_3$ because of isospin invariance.  With this, it is possible to estimate the Sommerfeld enhancement in each block with definite $I$, $S$ and $Y$  because in each of them $V^S_{I,Y}$ is just a Coulomb potential
\begin{align}
V^S_{I,Y} =\frac{\alpha^S_{I,Y}}{r}\,.
\end{align}
In that case,  the total annihilation cross-sections reads
\begin{align}
(\sigma v )_\mathrm{eff} = \frac{2}{k^2} \text{Tr} \sum_{I, Y, S}  (2I+1)\, (2S+1)\, {\cal S}\, (  \pi\alpha^S_{I,Y}/v) \, \Gamma^S_{I,Y} \,.
\label{master}
\end{align}
The factor of $2$ in front is analogous to the one in Eq.~\eqref{SEsigmav} and stems from the fact that all our DM candidates are self-conjugated. The trace here is taken over possible states that are not determined only by specifying $I$, $Y$ and $S$. The factor $(2I+1)(2S+1)$ is the multiplicity associated to the  isospin and spin. Here, $k=2n+1$ for the left and right components of the $(\vec{2n+1},\vec{1},0)\oplus(\vec{1},\vec{2n+1},0)$ representations,  whereas  $k=n^2$ for the chiral bi-multiplets $(\vec{n},\vec{n},0)$. Also,  ${\cal S} (t) = t/(e^{t}-1)$ is the Sommerfeld enhancement factor associated to a Coulomb potential~\cite{Cirelli:2007xd}.  
By taking the thermal average of $(\sigma v)_\mathrm{eff}$ and then using the instantaneous freeze-out approximation, one can estimate the DM relic density using Eq.~\eqref{Omega}.

In summary, neglecting the masses of the electroweak gauge bosons, we first determine the annihilation and potential matrices in the basis of co-annihilating pairs, as described in Appendix~\ref{sec:SEinGC}. For this we also assume that $Z_2$, $W_2$, $N_R$, as well as all the scalars except for the SM-like Higgs particle, are much heavier than the DM, so that they do not enter as final states in the annihilation process.  As usual, the co-annihilation pairs are conveniently classified according to their electric charge $Q$ and their spin $S$. Then, using the appropriate Clebsch--Gordon coefficients, we rewrite these pairs in the basis of isospin $I$ and hypercharge $Y$.  
We make sure that the annihilation  and the potential matrices are simultaneously block diagonalized with those unitary transformations. The resulting matrices in the blocks are reported in tables~\ref{table:GVmultiplet},~\ref{table:GVbidoublet},~\ref{table:GVbitriplet}. As shown there, the annihilation matrices describing the $s$-channel $Z_2$ and $W_2$ exchange always correspond to singlet isospin states $I=0$ with $S=1$ and are proportional to 
\begin{align}
A_{Z_2} &\equiv \frac{\pi  \alpha_2^2}{3M^2 \,\left(\gamma_{Z_2}^2
   r_{Z_2}^4+\left(r_{Z_2}^2-4\right)^2\right)} \left(\frac{g_R}{g_L}\right)^4\left(22-50\left(\frac{r_{Z_2}}{2}\right)^2 s_M^2+41 \left(\frac{r_{Z_2}}{2}\right)^4 s_M^4  \right) ,
\label{eq:AZ2}
\\
A_{W_2} &\equiv \frac{ \pi  \alpha_2^2 }{3M^2 \left(\gamma_{W_2}^2 r_{W_2}^4+\left(r_{W_2}^2-4\right)^2\right)}\left(\frac{g_R}{g_L}\right)^4 \left(18+\sin^2 2\beta\right) ,
\label{eq:AW2}
\end{align}
with $\gamma_{V}\equiv \Gamma_V/M_V$.
Notice that exactly at the resonance ($r_{W_2/Z_2} = 2$) these expressions are proportional to the $Z_2$ and $W_2$ widths of Eqs.~\eqref{eq:totalSMwidthZ2} and~\eqref{eq:totalSMwidthW2}, as expected from the Breit--Wigner formula.
Finally, with block-diagonal matrices we estimate the  relic density 
by means of Eq.~\eqref{master}. 

For the sake of clarity, we now discuss the details of each DM candidate separately. The case of the scalar doublet is not discussed because the Sommerfeld effect for that DM candidate is negligible in the early Universe~\cite{Garcia-Cely:2015khw}.

\subsection{Scalar and fermionic multiplets \texorpdfstring{$(\vec{2 n +1},\vec{1},0)\oplus (\vec{1},\vec{2 n +1},0)$}{Multiplets (2n+1,1,0)+(1,2n+1,0)}}

In this case, we work under the assumption that the densities corresponding to the left- and right-handed components evolve independently of each other. The total abundance is then $ \Omega h^2 = \Omega_L h^2 + \Omega_R h^2$.

For the right handed sector, applying the procedure described above  is trivial because the corresponding co-annihilating pairs are obviously $SU(2)_L$ singlets  and are thus already in the isospin basis. They correspond to $I=0$ and their hypercharge equals their electric charge $Y=Q$. The annihilation and potential matrices associated to these states are reported in table~\ref{table:GVmultiplet}.  Notice that specifying the quantum numbers $I$, $Y$ and $S$ does not suffice to specify each right-handed co-annihilating pair, and  accordingly in each subspace $\Gamma$ and $V$ are matrices -- not numbers -- even though the latter is diagonal because transitions by means of a $W$ boson exchange can not take place.    

For the left-handed sector, the procedure is more involved. However, it is strictly the same as in the Minimal Dark Matter scenarios \cite{Cirelli:2007xd,Cirelli:2009uv}. First of all, notice that $Y=0$ always. Second, the co-annihilating pairs break into different isospin components according to 
\begin{align}
(\vec{2n+1})\otimes (\vec{2n+1}) = \vec{1}\oplus\vec{3}\oplus\ldots\oplus(\vec{4n+1})\,.
\label{product}
\end{align}
The terms of the right-hand side correspond to $I=0$, $1$, \ldots, and $2n$. When $I$ is even (odd), the isospin state is (anti-)symmetric.  Accordingly, for fermions even (odd) isospin corresponds to  $S=0$ ($S=1$).  For scalars, since the co-annihilating pairs have no spin, the states with odd isospin are precluded. Using these selection rules, it is straightforward to calculate the unitary transformations from the isospin basis to the co-annihilating pairs by means of the appropriated Clebsch--Gordan coefficients. The results are shown in tables~\ref{table:triplet},~\ref{table:quintuplet},~\ref{table:sevenplet}. The corresponding annihilation and potential matrices are in table~\ref{table:GVmultiplet}.
\begin{table}[H]
  \begin{center}
    \begin{tabular}{|l|}\hline
        \usebox{\tripletL}\\\hline
    \end{tabular}
  \end{center}
  \caption{Transformation matrices for the fermionic triplet $(\vec{3},\vec{1},0)\oplus(\vec{1},\vec{3},0)$. Only the left-handed pairs are shown.  }
  \label{table:triplet}
\end{table}
\begin{table}[H]
  \begin{center}
    \begin{tabular}{|l|}\hline
       \usebox{\quintupletL}\\\hline
    \end{tabular}
  \end{center}
\caption{Transformation matrices for the fermionic quintuplet $(\vec{5},\vec{1},0)\oplus(\vec{1},\vec{5},0)$. Only the left-handed pairs are shown.  }
  \label{table:quintuplet}
\end{table}

\begin{table}[H]
  \begin{center}
    \begin{tabular}{|l|}\hline
       \usebox{\sevenpletL}\\\hline
    \end{tabular}
  \end{center}
\caption{Transformation matrices for the scalar 7-plet $(\vec{7},\vec{1},0)\oplus(\vec{1},\vec{7},0)$. Only the left-handed pairs are shown.  }
  \label{table:sevenplet}
\end{table}

\begin{table}[H]
\centering
\usebox{\GVmultiplet}
\caption{\small Annihilation and potential matrices for states of definite spin, isospin and hypercharge in the case of the multiplets $(\vec{2n+1},\vec{1},0) \oplus (\vec{1},\vec{2n+1},0)$. The states not listed  do not contribute to the annihilation process. These matrices apply even for $g_R \neq g_L$.}
\label{table:GVmultiplet}
\end{table}

For the left-handed quintuplet we would like to comment on a difference that we find with respect to the matrices that have been reported in the literature. As a by-product, we show a rather pedestrian example of how to go from the isospin basis to the co-annihilating pairs basis. Concretely, taking the eigenvalues $V^S_{I,Y}$ and using the transformation matrices of table~\ref{table:quintuplet}, we find 
\begin{align}
V^{S=0}_{Q=2}&= \frac{\alpha_2}{r}
    \left(
    \begin{array}{cc}
     \sqrt{\frac{3}{7}} & \frac{2}{\sqrt{7}} \\
 \frac{2}{\sqrt{7}} & -\sqrt{\frac{3}{7}} \\
\end{array}
    \right)
    \left(
    \begin{array}{cc}
     4 & 0 \\
 0 & -3 \\
\end{array}
    \right)
    \left(
    \begin{array}{cc}
     \sqrt{\frac{3}{7}} & \frac{2}{\sqrt{7}} \\
 \frac{2}{\sqrt{7}} & -\sqrt{\frac{3}{7}} \\
\end{array}
    \right)^\dagger = \frac{\alpha_2}{r}\left(
  \begin{array}{cc}
   0 & 2 \sqrt{3} \\
 2 \sqrt{3} & 1 \\
\end{array}
  \right) ,\\
%
%
V^{S=0}_{Q=1}&=\frac{\alpha_2}{r}
    \left(
    \begin{array}{cc}
     -\frac{1}{\sqrt{7}} & -\sqrt{\frac{6}{7}} \\
 \sqrt{\frac{6}{7}} & -\frac{1}{\sqrt{7}} \\
\end{array}
    \right)
    \left(
    \begin{array}{cc}
     4 & 0 \\
 0 & -3 \\
\end{array}
    \right)
    \left(
    \begin{array}{cc}
     -\frac{1}{\sqrt{7}} & -\sqrt{\frac{6}{7}} \\
 \sqrt{\frac{6}{7}} & -\frac{1}{\sqrt{7}} \\
\end{array}
    \right)^\dagger = \frac{\alpha_2}{r}\left(
  \begin{array}{cc}
   -2 & -\sqrt{6} \\
 -\sqrt{6} & 3 \\
\end{array}
  \right),\\
%
%
V^{S=1}_{Q=1}&=\frac{\alpha_2}{r}
    \left(
    \begin{array}{cc}
     -\sqrt{\frac{3}{5}} & -\sqrt{\frac{2}{5}} \\
 \sqrt{\frac{2}{5}} & -\sqrt{\frac{3}{5}} \\
\end{array}
    \right)
    \left(
    \begin{array}{cc}
     0 & 0 \\
 0 & -5 \\
\end{array}
    \right)
    \left(
    \begin{array}{cc}
     -\sqrt{\frac{3}{5}} & -\sqrt{\frac{2}{5}} \\
 \sqrt{\frac{2}{5}} & -\sqrt{\frac{3}{5}} \\
\end{array}
    \right)^\dagger = \frac{\alpha_2}{r}\left(
  \begin{array}{cc}
   -2 & -\sqrt{6} \\
 -\sqrt{6} & -3 \\
\end{array}
  \right),\\
%
%
V^{S=0}_{Q=0}&=\frac{\alpha_2}{r}
    \left(
    \begin{array}{ccc}
     \frac{1}{\sqrt{35}} & \frac{2}{\sqrt{7}} & \sqrt{\frac{2}{5}} \\
 -\frac{4}{\sqrt{35}} & -\frac{1}{\sqrt{7}} & \sqrt{\frac{2}{5}} \\
 3 \sqrt{\frac{2}{35}} & -\sqrt{\frac{2}{7}} & \frac{1}{\sqrt{5}} \\
\end{array}
    \right)
    \left(
    \begin{array}{ccc}
     4 & 0 & 0 \\
 0 & -3 & 0 \\
 0 & 0 & -6 \\
\end{array}
    \right)
    \left(
    \begin{array}{ccc}
     \frac{1}{\sqrt{35}} & \frac{2}{\sqrt{7}} & \sqrt{\frac{2}{5}} \\
 -\frac{4}{\sqrt{35}} & -\frac{1}{\sqrt{7}} & \sqrt{\frac{2}{5}} \\
 3 \sqrt{\frac{2}{35}} & -\sqrt{\frac{2}{7}} & \frac{1}{\sqrt{5}} \\
\end{array}
    \right)^\dagger\nonumber \\
		&= \frac{\alpha_2}{r}\left(
  \begin{array}{ccc}
   -4 & -2 & 0 \\
 -2 & -1 & -3 \sqrt{2} \\
 0 & -3 \sqrt{2} & 0 \\
\end{array}
  \right),\\
%
%
V^{S=1}_{Q=0}&=\frac{\alpha_2}{r}
    \left(
    \begin{array}{cc}
     \frac{1}{\sqrt{5}} & \frac{2}{\sqrt{5}} \\
 -\frac{2}{\sqrt{5}} & \frac{1}{\sqrt{5}} \\
\end{array}
    \right)
    \left(
    \begin{array}{cc}
     0 & 0 \\
 0 & -5 \\
\end{array}
    \right)
    \left(
    \begin{array}{cc}
     \frac{1}{\sqrt{5}} & \frac{2}{\sqrt{5}} \\
 -\frac{2}{\sqrt{5}} & \frac{1}{\sqrt{5}} \\
\end{array}
    \right)^\dagger = \frac{\alpha_2}{r}\left(
  \begin{array}{cc}
   -4 & -2 \\
 -2 & -1 \\
\end{array}
  \right).
\end{align}
Notice that $V^{S=0}_{Q=1}$ and $V^{S=1}_{Q=1}$ differ by a sign in their $(2,2)$ component in contrast to Ref.~\cite{Cirelli:2007xd}. Furthermore, $V^{S=0}_{Q=2}$ has an extra sign on the off-diagonal terms. Since our matrices are derived from group theoretical arguments, they must be used in the $SU(2)_L$ symmetric limit, otherwise the diagonalization of the potential and the annihilation matrices would not be possible. In fact, if the potentials for the states $S=0$, $Q=1$  and $S=1$, $Q=1$ equaled, they would have the same isospin eigenvalues in contradiction with the fact that those states have different isospin, namely, $I=2,4$ and $I=1,3$, respectively. 
Following the same procedure for the annihilation matrices, we find  
\begin{align}
\Gamma^{S=0}_{Q=2} &=
\dfrac{\pi \alpha_2^2}{2 M^2}
\left(
\begin{array}{cc}
 12 & -6 \sqrt{3} \\
 -6 \sqrt{3} & 9 \\
\end{array}
\right) ,\\
\Gamma^{S=0}_{Q=1} &=
\dfrac{\pi \alpha_2^2}{2M^2}
\left(
\begin{array}{cc}
 18 & 3 \sqrt{6} \\
 3 \sqrt{6} & 3 \\
\end{array}
\right) , \quad
\Gamma^{S=1}_{Q=1} =
\dfrac{25\pi \alpha_2^2}{24 M^2}
\left(
\begin{array}{cc}
 2 & \sqrt{6} \\
 \sqrt{6} & 3 \\
\end{array}
\right) ,\\
\Gamma^{S=0}_{Q=0} &=
\dfrac{\pi \alpha_2^2}{2M^2}
\left(
\begin{array}{ccc}
 36 & 18 & 6 \sqrt{2} \\
 18 & 27 & 15 \sqrt{2} \\
 6 \sqrt{2} & 15 \sqrt{2} & 18 \\
\end{array}
\right)= \Gamma^{WW} + \Gamma^{ZZ} + \Gamma^{Z\gamma} +\Gamma^{\gamma\gamma} \,,\\
\Gamma^{S=1}_{Q=0} &=
\dfrac{25\pi \alpha_2^2}{24M^2}
\left(
\begin{array}{cc}
 4 & 2 \\
 2 & 1 \\
\end{array}
\right) .
\end{align}
Here, $\Gamma^{WW}$, $\Gamma^{ZZ}$, $\Gamma^{Z\gamma}$, and  $\Gamma^{\gamma\gamma}$ are the annihilation matrices introduced in Eq.~\eqref{GammaQuin}. The matrices $\Gamma^S_Q$ coincide with the ones reported in Ref.~\cite{Cirelli:2007xd}, if the factor $(2/k^2) (2I+1)$ of Eq.~\eqref{master} is absorbed in them. However, we do not have an agreement with the corresponding $S=1$ matrices of Ref.~\cite{Cirelli:2015bda}.
In spite of these discrepancies, none of this turns out to be numerically significant, since the relic density calculation employing the $SU(2)_L$ symmetric limit gives us approximately the same result for the thermal quintuplet mass as in Ref.~\cite{Cirelli:2009uv}, namely $M\simeq \unit[8]{TeV}$.  
Similarly, we find an agreement with the corresponding matrices that have been reported for our left-handed triplet, known as wino in the literature~\cite{Hisano:2006nn, Cirelli:2007xd}.

We would like to comment on the validity of our approach to calculate the relic density for the scalar representations. In order to produce the resonances associated with the $Z_2$ and $W_2$ bosons, the co-annihilating pairs must constitute a state with total angular momentum $J=1$. In the $s$-wave channels, that is, when the orbital angular momentum vanishes, this can only happen if the total spin is $S=1$. Accordingly, if annihilations are approximated by their $s$ wave, as we do here, we can only probe the resonances with the \emph{fermionic} representations. This is clear from Table~\ref{table:GVmultiplet}, because the scalar 7-plet does not annihilate via resonances, in contrast with the triplet and the quintuplet fermions. In spite of this, the scalars \emph{do} annihilate via $W_2$ and $Z_2$ resonances by means of the $p$-wave part of the annihilation process, which in that case corresponds to $J=L=1$. One faces then a situation in which the $p$-wave cross section is velocity suppressed but resonantly enhanced and thus not negligible a priori. As a consequence, the formalism discussed here is expected to give a fairly poor approximation to relic density in the case of scalars, in particular around the resonances.   

Finally, notice that in the case of  the $(\vec{2n+1},\vec{1},0)\oplus (\vec{1},\vec{2n+1},0)$ representations, for very high masses,  $M\gg M_{W_2}$, one recovers one more symmetric limit, the one associated to the $SU(2)_L\times SU(2)_R$ symmetry. In that case, the relic density $\Omega_R$ approaches $\Omega_L$ for $g_R=g_L$ due to the enhanced symmetry. For this it is crucial to include $W_2$ and $Z_2$ in the annihilation final states, which requires a small modification of our formulae.

\subsection{Chiral bi-multiplets \texorpdfstring{$(\vec{n},\vec{n},0)$}{Chiral bi-multiplets (n,n,0)}}

Every chiral bi-multiplet $(\vec{n},\vec{n},0)$ can be thought of as a collection of $n$ $SU(2)_L$ $n$-plets with different hypercharge. Then one can decompose the co-annihilating pairs as we did above for the other cases. The only crucial difference now is that the hypercharge plays and important role and we must (anti)-symmetrize with respect to it in order to obtain the different states in the basis of definite isospin. 
 
For instance, for the fermionic bi-doublet  the co-annihilating pairs consist of $\Psi^0,(\Psi^0)^c,\Psi^+,\Psi^-$.  Since they belong to $SU(2)_L$ doublets, the subspace generated by such pairs can be decomposed into singlets and triplets of $SU(2)_L$. Concretely, one self-conjugate isospin singlet $|I_3=0,I=0\rangle_{Y=0}$, one self-conjugate triplet $|I_3,I=1\rangle_{Y=0}$ and one isospin triplet $|I_3,I=1\rangle_{Y=1}$ and its corresponding complex conjugate $|I_3,I=1\rangle_{Y=-1}$. In terms of these states, we show the co-annihilating pairs with charge $Q=0$, $Q=1$ and $Q=2$ in table~\ref{table:bidoublet}. There, we find it convenient to introduce the notation 
\begin{align}
|...Y_1=\tfrac{1}{2},Y_2=-\tfrac{1}{2}\rangle_{Sym} &= \tfrac{1}{\sqrt{2}} \left(|...,Y_1=\tfrac{1}{2},Y_2=-\tfrac{1}{2}\rangle+|...,Y_1=-\tfrac{1}{2},Y_2=\tfrac{1}{2}\rangle\right) ,\\
|...Y_1=\tfrac{1}{2},Y_2=-\tfrac{1}{2}\rangle_{Ant\;} &= \tfrac{1}{\sqrt{2}} \left(|...,Y_1=\tfrac{1}{2},Y_2=-\tfrac{1}{2}\rangle-|...,Y_1=-\tfrac{1}{2},Y_2=\tfrac{1}{2}\rangle\right) ,
\end{align}
where the subscript $Sym$ or $Ant$ denotes  symmetrization or antisymmetrization on the hypercharge {\it only}.
A similar decomposition can be done for the bi-triplet. The corresponding transformation matrices are shown in table~\ref{table:bitriplet}.

As in the previous section, if the factor $(2/k^2) (2I+1)$ of Eq.~\eqref{master} is absorbed in the annihilation matrices of the bi-doublet,  we find an agreement with the corresponding matrices that were reported for Higgsino DM in Ref.~\cite{Cirelli:2007xd}.
\begin{table}[H]
  \begin{center}
    \begin{tabular}{|l|}\hline
       \usebox{\bidoublet}\\\hline
    \end{tabular}
  \end{center}
  \caption{Transformation matrices for the bi-doublet $(\vec{2},\vec{2},0)$. }
  \label{table:bidoublet}
\end{table}

\begin{table}[H]
  \begin{center}
    \begin{tabular}{|l|}\hline
       \usebox{\bitriplet}\\\hline
    \end{tabular}
  \end{center}
\caption{Transformation matrices for the bi-triplet $(\vec{3},\vec{3},0)$. }
  \label{table:bitriplet}
\end{table}

\begin{table}[H]
\centering
\usebox{\GVbidoublet}
\caption{ \small Annihilation and potential matrices for states of definite spin, isospin and hypercharge in the case of the bi-doublet $(\vec{2},\vec{2},0)$. The states not listed  do not contribute to the annihilation process. These matrices apply even for $g_R \neq g_L$.}
\label{table:GVbidoublet}
\end{table}

\begin{table}[H]
\centering
\usebox{\GVbitriplet}
\caption{ \small Annihilation and potential matrices for states of definite spin, isospin and hypercharge in the case of the bi-triplet $(\vec{3},\vec{3},0)$. The states not listed  do not contribute to the annihilation process. These matrices apply even for $g_R \neq g_L$.}
\label{table:GVbitriplet}
\end{table}

\bibliographystyle{utcaps_mod}
\bibliography{BIB}

\end{document}

%% file: AppendicesTables.tex
\newsavebox\tripletL
\newsavebox\tripletR
\newsavebox\quintupletL
\newsavebox\quintupletR
\newsavebox\sevenpletL
\newsavebox\bidoublet
\newsavebox\bitriplet

\newsavebox\GVmultiplet
\newsavebox\GVbidoublet
\newsavebox\GVbitriplet


\begin{lrbox}{\bidoublet}
\begin{minipage}{\textwidth}
\begin{eqnarray}
\begin{array}{c}
Q=2\\
S=0\\
\end{array}: \hspace{20pt}
 \Psi^+\,\Psi^+ \hspace{15pt} &=& |I_3=1, I=1, Y_1 =\tfrac{1}{2}, Y_2 =\tfrac{1}{2}\rangle^{S=0}\nonumber \\
\hline
\begin{array}{c}
Q=1\\
S=0\\
\end{array} : 
\left(
\begin{array}{c}
 \Psi^+\,\Psi^0 \\
 \Psi^+\,(\Psi^0)^c\\
\end{array}
\right)
&=&
\left(
\begin{array}{c}
 |I_3=1, I=1, Y_1 =\tfrac{1}{2}, Y_2 =-\tfrac{1}{2}\rangle^{S=0}_{Sym}\\
 -|I_3=0, I=1, Y_1 =\tfrac{1}{2}, Y_2 =\tfrac{1}{2}\rangle^{S=0}\\
\end{array}
\right)\nonumber\\
\hline
\begin{array}{c}
Q=1\\
S=1\\
\end{array} :
\left(
\begin{array}{c}
 \Psi^+\,\Psi^0 \\
 \Psi^+\,(\Psi^0)^c\\
\end{array}
\right)
&=&
\left(
\begin{array}{c}
 |I_3=1, I=1, Y_1 =\tfrac{1}{2}, Y_2 =-\tfrac{1}{2}\rangle^{S=1}_{Ant}\\
 -|I_3=0, I=0, Y_1 =\tfrac{1}{2}, Y_2 =\tfrac{1}{2}\rangle^{S=1}_{Sym}\\
\end{array}
\right)\nonumber\\
\hline
\begin{array}{c}
Q=0\\
S=0\\
\end{array}: 
\left(
\begin{array}{c}
 \Psi^0\,(\Psi^0)^c\\
 \Psi^+\,\Psi^-\\
\end{array}
\right)
&=&
\left(
\begin{array}{ccc}
  -\frac{1}{\sqrt{2}} & \frac{1}{\sqrt{2}} \\
  \frac{1}{\sqrt{2}} & \frac{1}{\sqrt{2}}   
\end{array}
\right)
\left(
\begin{array}{c}
 |I_3=0, I=1,Y_1=\tfrac{1}{2},Y_2=-\tfrac{1}{2}\rangle^{S=0}_{Sym}\\
 |I_3=0, I=0,Y_1=\tfrac{1}{2},Y_2=-\tfrac{1}{2}\rangle^{S=0}_{Ant}\\
\end{array}
\right)\nonumber \nonumber\\
\hspace{40pt}(\Psi^0)^c\,(\Psi^0)^c \hspace{5pt}&=&
 |I_3=-1, I=1,Y_1=\tfrac{1}{2},Y_2=\tfrac{1}{2}\rangle^{S=0}\nonumber
\label{BiDQ0S0}\\
\hline
\begin{array}{c}
Q=0\\
S=1\\
\end{array} :
\left(
\begin{array}{c}
 \Psi^0\,(\Psi^0)^c\\
 \Psi^+\,\Psi^-\\
\end{array}
\right)
&=&
\left(
\begin{array}{cc}
  \frac{1}{\sqrt{2}} & -\frac{1}{\sqrt{2}} \\
  \frac{1}{\sqrt{2}} & \frac{1}{\sqrt{2}} 
\end{array}
\right)
\left(
\begin{array}{c}
 |I_3=0, I=1,Y_1=\tfrac{1}{2},Y_2=-\tfrac{1}{2}\rangle^{S=1}_{Ant}\\
 |I_3=0, I=0,Y_1=\tfrac{1}{2},Y_2=-\tfrac{1}{2}\rangle^{S=1}_{Sym}\\
\end{array}
\right)\nonumber
\end{eqnarray}
\end{minipage}
\end{lrbox}

\begin{lrbox}{\bitriplet}
\begin{minipage}{\textwidth}
\begin{eqnarray}
%
%
%
\begin{array}{c}
Q=2\\
S=0\\
\end{array}
: \hspace{40pt} 
\Psi^{++}\Psi^0 \hspace{10pt}&=&
|I_3=2, I=2,  Y_1=1, Y_2=-1\rangle^{S=0}_{Sym}\nonumber\\
 \hspace{40pt} 
\Psi_1^+\Psi_1^+ \hspace{15pt}&=&
|I_3=2, I=2,  Y_1=0, Y_2=0\rangle^{S=0}\nonumber\\
\hline
\begin{array}{c}
Q=1\\
S=0\\
\end{array}
: \hspace{40pt} 
\Psi_1^+\chi \hspace{20pt}&=&
|I_3=1, I=2,  Y_1=0, Y_2=0\rangle^{S=0}\nonumber\\
\left(
\begin{array}{c}
 \Psi^{++}\,\Psi_2^-\\
 \Psi_2^+\,\chi\\
 \Psi_1^+\,(\Psi^0)^c \\
\end{array}
\right)
&=&
\left(
\begin{array}{ccc}
 -\frac{1}{\sqrt{6}} & -\frac{1}{\sqrt{2}} & -\frac{1}{\sqrt{3}} \\
 \sqrt{\frac{2}{3}} & 0 & -\frac{1}{\sqrt{3}} \\
 \frac{1}{\sqrt{6}} & -\frac{1}{\sqrt{2}} & \frac{1}{\sqrt{3}} \\
\end{array}
\right)
\left(
\begin{array}{c}
 |I_3=0, I=2,  Y_1=1, Y_2=0\rangle^{S=0}_{Sym}\\
 |I_3=0, I=1,  Y_1=1, Y_2=0\rangle^{S=0}_{Ant}\\
 |I_3=0, I=0, Y_1=1, Y_2=0\rangle^{S=0}_{Sym}\\
\end{array}
\right)\nonumber \nonumber\\
\hspace{20pt}
\left(
\begin{array}{c}
 \Psi^{++}\,\Psi_2^-\\
 \Psi_2^+\Psi^0\\
\end{array}
\right)
&=&
\left(
\begin{array}{cc}
 -\frac{1}{\sqrt{2}} & -\frac{1}{\sqrt{2}} \\
 -\frac{1}{\sqrt{2}} & \frac{1}{\sqrt{2}} \\
\end{array}
\right)\left(
\begin{array}{c}
 |I_3=1, I=1, Y_1=1, Y_2=-1\rangle^{S=0}_{Ant}\\
 |I_3=1, I=2, Y_1=1, Y_2=-1\rangle^{S=0}_{Sym}\\
\end{array}
\right)\nonumber\\
\hline
\begin{array}{c}
Q=1\\
S=1\\
\end{array}
: \hspace{40pt} 
\Psi_1^+\chi \hspace{20pt} &=&
|I_3=1, I=1,  Y_1=0, Y_2=0\rangle^{S=1} \nonumber\\
\left(
\begin{array}{c}
 \Psi^{++}\,\Psi_2^-\\
 \Psi_2^+\,\chi\\
 \Psi_1^+\,(\Psi^0)^c \\
\end{array}
\right)
&=&
\left(
\begin{array}{ccc}
 -\frac{1}{\sqrt{6}} & -\frac{1}{\sqrt{2}} &
   -\frac{1}{\sqrt{3}} \\
 \sqrt{\frac{2}{3}} & 0 & -\frac{1}{\sqrt{3}} \\
 -\frac{1}{\sqrt{6}} & \frac{1}{\sqrt{2}} &
   -\frac{1}{\sqrt{3}} \\
\end{array}
\right)
\left(
\begin{array}{c}
 |I_3=0, I=2,  Y_1=1, Y_2=0\rangle^{S=1}_{Ant}\\
 |I_3=0, I=1,  Y_1=1, Y_2=0\rangle^{S=1}_{Sym}\\
 |I_3=0, I=0, Y_1=1, Y_2=0\rangle^{S=1}_{Ant}\\
\end{array}
\right)\nonumber \nonumber\\
\hspace{20pt}
\left(
\begin{array}{c}
 \Psi^{++}\,\Psi_2^-\\
 \Psi_2^+\Psi^0\\
\end{array}
\right)
&=&
\left(
\begin{array}{cc}
 -\frac{1}{\sqrt{2}} & -\frac{1}{\sqrt{2}} \\
 \frac{1}{\sqrt{2}} & -\frac{1}{\sqrt{2}} \\
\end{array}
\right)
\left(
\begin{array}{c}
 |I_3=1, I=2, Y_1=1, Y_2=-1\rangle^{S=1}_{Ant}\\
 |I_3=1, I=1, Y_1=1, Y_2=-1\rangle^{S=1}_{Sym}\\
\end{array}
\right)\nonumber\\
\hline
\begin{array}{c}
Q=0\\
S=0\\
\end{array}
 : 
\left(
\begin{array}{c}
 \Psi_2^+\,\Psi_2^-\\
 (\Psi^0)^c\,\Psi^0\\
 \Psi^{++}\,\Psi^{--} \\
\end{array}
\right)
&=&
\left(
\begin{array}{ccc}
 \frac{1}{\sqrt{3}} & 0 & -\sqrt{\frac{2}{3}} \\
 \frac{1}{\sqrt{3}} & -\frac{1}{\sqrt{2}} & \frac{1}{\sqrt{6}} \\
 \frac{1}{\sqrt{3}} & \frac{1}{\sqrt{2}} & \frac{1}{\sqrt{6}} \\
\end{array}
\right)
\left(
\begin{array}{c}
 |I_3=0, I=0,  Y_1=1, Y_2=-1\rangle^{S=0}_{Sym}\\
 |I_3=0, I=1,  Y_1=1, Y_2=-1\rangle^{S=0}_{Ant}\\
 |I_3=0, I=2, Y_1=1, Y_2=-1\rangle^{S=0}_{Sym}\\
\end{array}
\right)\nonumber \nonumber\\
\hspace{20pt}
\left(
\begin{array}{c}
 \Psi_1^+\,\Psi_1^-\\
 \chi\chi\\
\end{array}
\right)
&=&
\left(
\begin{array}{cc}
 -\sqrt{\frac{2}{3}} & -\frac{1}{\sqrt{3}} \\
 -\frac{1}{\sqrt{3}} & \sqrt{\frac{2}{3}} \\
\end{array}
\right)
\left(
\begin{array}{c}
 |I_3=0, I=0, Y_1=0, Y_2=0\rangle^{S=0}\\
 |I_3=0, I=2, Y_1=0, Y_2=0\rangle^{S=0}\\
\end{array}
\right)\nonumber\\
\hline
\begin{array}{c}
Q=0\\
S=1\\
\end{array}
 :
\left(
\begin{array}{c}
 \Psi_2^+\,\Psi_2^-\\
 (\Psi^0)^c\,\Psi^0\\
 \Psi^{++}\,\Psi^{--} \\
\end{array}
\right)
&=& 
\left(
\begin{array}{ccc}
 \frac{1}{\sqrt{3}} & 0 & -\sqrt{\frac{2}{3}} \\
 \frac{1}{\sqrt{3}} & -\frac{1}{\sqrt{2}} & \frac{1}{\sqrt{6}} \\
 \frac{1}{\sqrt{3}} & \frac{1}{\sqrt{2}} & \frac{1}{\sqrt{6}} \\
\end{array}
\right)
\left(
\begin{array}{c}
 |I_3=0, I=0,  Y_1=1, Y_2=-1\rangle^{S=1}_{Ant}\\
 |I_3=0, I=1,  Y_1=1, Y_2=-1\rangle^{S=1}_{Sym}\\
 |I_3=0, I=2, Y_1=1, Y_2=-1\rangle^{S=1}_{Ant}\\
\end{array}
\right)\nonumber \nonumber\\
\Psi_1^+\Psi_1^- \hspace{20pt}&=& |I_3=0, I=1, Y_1=0, Y_2=0\rangle^{S=1}\nonumber
\end{eqnarray}

\end{minipage}
\end{lrbox}

\begin{lrbox}{\tripletL}
\begin{minipage}{\textwidth}
\begin{eqnarray}
\begin{array}{c}
Q=2\\
S=0\\
\end{array}: \hspace{20pt}
 \Psi_L^+\,\Psi_L^+ \hspace{15pt} &=& |I_3=2, I=2,  Y=0\rangle^{S=0}\nonumber \\
\hline
\begin{array}{c}
Q=1\\
S=0\\
\end{array}: \hspace{20pt}
 \Psi_L^0\,\Psi_L^+ \hspace{15pt} &=& |I_3=1, I=2, Y=0\rangle^{S=0}\nonumber \\
\hline
\begin{array}{c}
Q=1\\
S=1\\
\end{array}: \hspace{20pt}
 \Psi_L^0\,\Psi_L^+ \hspace{15pt} &=& -|I_3=1, I=1, Y=0\rangle^{S=1}\nonumber \\
\hline
\begin{array}{c}
Q=0\\
S=0\\
\end{array} :\hspace{10pt}
\left(
\begin{array}{c}
 \Psi_L^+\,\Psi_L^-\\
 \Psi_L^0\,\Psi_L^0\\
\end{array}
\right)
&=&
\left(
\begin{array}{cc}
 -\frac{1}{\sqrt{3}} & -\sqrt{\frac{2}{3}} \\
 \sqrt{\frac{2}{3}} & -\frac{1}{\sqrt{3}} \\
\end{array}
\right)
\left(
\begin{array}{c}
 |I_3=0, I=2,Y=0\rangle^{S=0}\\
 |I_3=0, I=0,Y=0\rangle^{S=0}\\
\end{array}
\right)\nonumber\nonumber\\
\hline
\begin{array}{c}
Q=0\\
S=1\\
\end{array} :
\hspace{20pt}
\Psi_L^+\,\Psi_L^- \hspace{15pt}&=&
 |I_3=0, I=0, Y=0\rangle^{S=1}\nonumber \nonumber%
\end{eqnarray}
\end{minipage}
\end{lrbox}

\begin{lrbox}{\tripletR}
\begin{minipage}{\textwidth}
\begin{eqnarray}
\begin{array}{c}
Q=1\\
S=0\\
\end{array}: \hspace{20pt}
 \Psi_R^0\,\Psi_R^+ \hspace{15pt} &=& |I_3=0, I=0, Y=1\rangle^{S=0}\nonumber\\
\hline
\begin{array}{c}
Q=1\\
S=1\\
\end{array}: \hspace{20pt}
 \Psi_R^0\,\Psi_R^+ \hspace{15pt} &=& |I_3=0, I=0, Y=1\rangle^{S=1}\nonumber\\
\hline
\begin{array}{c}
Q=0\\
S=1\\
\end{array} :
\hspace{20pt}
\Psi_R^+\,\Psi_R^- \hspace{15pt}&=& |I_3=0, I=0, Y=0\rangle^{S=1}\nonumber
\end{eqnarray}

\end{minipage}
\end{lrbox}

\begin{lrbox}{\quintupletL}
\begin{minipage}{\textwidth}
\begin{eqnarray}
\begin{array}{c}
Q=2\\
S=0\\
\end{array} :\hspace{10pt}
\left(
\begin{array}{c}
 \Psi_L^{++}\,\Psi_L^0\\
 \Psi_L^{+}\,\Psi_L^{+}\\
\end{array}
\right)
&=&
\left(
\begin{array}{cc}
 \sqrt{\frac{3}{7}} & \frac{2}{\sqrt{7}} \\
 \frac{2}{\sqrt{7}} & -\sqrt{\frac{3}{7}} \\
\end{array}
\right)
\left(
\begin{array}{c}
 |I_3=2, I=4,Y=0\rangle^{S=0}\\
 |I_3=2, I=2,Y=0\rangle^{S=0}\\
\end{array}
\right)\nonumber\\
\hline
\begin{array}{c}
Q=1\\
S=0\\
\end{array} :\hspace{10pt}
\left(
\begin{array}{c}
 \Psi_L^{++}\,\Psi_L^-\\
 \Psi_L^{+}\,\Psi_L^0\\
\end{array}
\right)
&=&
\left(
\begin{array}{cc}
 -\frac{1}{\sqrt{7}} & -\sqrt{\frac{6}{7}} \\
 \sqrt{\frac{6}{7}} & -\frac{1}{\sqrt{7}} \\
\end{array}
\right)
\left(
\begin{array}{c}
 |I_3=1, I=4,Y=0\rangle^{S=0}\\
 |I_3=1, I=2,Y=0\rangle^{S=0}\\
\end{array}
\right)\nonumber\\
\hline
\begin{array}{c}
Q=1\\
S=1\\
\end{array} :\hspace{10pt}
\left(
\begin{array}{c}
 \Psi_L^{++}\,\Psi_L^-\\
 \Psi_L^{+}\,\Psi_L^0\\
\end{array}
\right)
&=&
\left(
\begin{array}{cc}
 -\sqrt{\frac{3}{5}} & -\sqrt{\frac{2}{5}} \\
 \sqrt{\frac{2}{5}} & -\sqrt{\frac{3}{5}} \\
\end{array}
\right)
\left(
\begin{array}{c}
 |I_3=1, I=3,Y=0\rangle^{S=1}\\
 |I_3=1, I=1,Y=0\rangle^{S=1}\\
\end{array}
\right)\nonumber\nonumber\\
\hline
\begin{array}{c}
Q=0\\
S=0\\
\end{array}
 :
\left(
\begin{array}{c}
 \Psi_L^{++}\,\Psi_L^{--}\\
 \Psi_L^{+}\,\Psi_L^{-}\\
 \Psi_L^0\,\Psi_L^0 \\
\end{array}
\right)
&=& 
\left(
\begin{array}{ccc}
 \frac{1}{\sqrt{35}} & \frac{2}{\sqrt{7}} & \sqrt{\frac{2}{5}} \\
 -\frac{4}{\sqrt{35}} & -\frac{1}{\sqrt{7}} & \sqrt{\frac{2}{5}} \\
 3 \sqrt{\frac{2}{35}} & -\sqrt{\frac{2}{7}} & \frac{1}{\sqrt{5}} \\
\end{array}
\right)
\left(
\begin{array}{c}
 |I_3=0, I=4, Y=0\rangle^{S=0}\\
 |I_3=0, I=2, Y=0\rangle^{S=0}\\
 |I_3=0, I=0, Y=0\rangle^{S=0}\\
\end{array}
\right)\nonumber\\
\hline
\begin{array}{c}
Q=0\\
S=1\\
\end{array}
 :
\left(
\begin{array}{c}
 \Psi_L^{++}\,\Psi_L^{--}\\
 \Psi_L^{+}\,\Psi_L^{-}\\
\end{array}
\right)
&=& 
\left(
\begin{array}{cc}
 \frac{1}{\sqrt{5}} & \frac{2}{\sqrt{5}} \\
 -\frac{2}{\sqrt{5}} & \frac{1}{\sqrt{5}} \\
\end{array}
\right)
\left(
\begin{array}{c}
 |I_3=0, I=3, Y=0\rangle^{S=1}\\
 |I_3=0, I=1, Y=0\rangle^{S=1}\\
\end{array}
\right)\nonumber 
\end{eqnarray}

\end{minipage}
\end{lrbox}

\begin{lrbox}{\sevenpletL}
\begin{minipage}{\textwidth}
\begin{eqnarray}
\begin{array}{c}
Q=2\\
S=0\\
\end{array} :\hspace{10pt}
\left(
\begin{array}{c}
 \phi_L^{+++}\,\phi_L^{-}\\
 \phi_L^{++}\,\phi_L^{0}\\
 \phi_L^{+}\,\phi_L^{+}\\
\end{array}
\right)
&=&
\left(
\begin{array}{ccc}
 -\sqrt{\frac{5}{21}} & 3 \sqrt{\frac{6}{77}} &
   -\sqrt{\frac{2}{33}} \\
 -\sqrt{\frac{10}{21}} & -\sqrt{\frac{3}{77}} &
   \frac{4}{\sqrt{33}} \\
 \sqrt{\frac{2}{7}} & 2 \sqrt{\frac{5}{77}} &
   \sqrt{\frac{5}{11}} \\
\end{array}
\right)
\left(
\begin{array}{c}
 |I_3=2, I=6,Y=0\rangle^{S=0}\\
 |I_3=2, I=4,Y=0\rangle^{S=0}\\
 |I_3=2, I=2,Y=0\rangle^{S=0}\\
\end{array}
\right)\nonumber\\
\hline
\begin{array}{c}
Q=1\\
S=0\\
\end{array} :\hspace{10pt}
\left(
\begin{array}{c}
 \phi_L^{+++}\,\phi_L^{--}\\
 \phi_L^{++}\,\phi_L^-\\
 \phi_L^{+}\,\phi_L^0\\
\end{array}
\right)
&=&
\left(
\begin{array}{ccc}
 \frac{5}{\sqrt{42}} & -\sqrt{\frac{30}{77}} &
   \frac{1}{\sqrt{66}} \\
 \sqrt{\frac{5}{14}} & 4 \sqrt{\frac{2}{77}} &
   -\sqrt{\frac{5}{22}} \\
 \frac{1}{\sqrt{21}} & \sqrt{\frac{15}{77}} &
   \frac{5}{\sqrt{33}} \\
\end{array}
\right)
\left(
\begin{array}{c}
 |I_3=1, I=6,Y=0\rangle^{S=0}\\
 |I_3=1, I=4,Y=0\rangle^{S=0}\\
 |I_3=1, I=2,Y=0\rangle^{S=0}\\
\end{array}
\right)\nonumber\\
\hline
\begin{array}{c}
Q=0\\
S=0\\
\end{array}
 :
\left(
\begin{array}{c}
 \phi_L^{+++}\,\phi_L^{---}\\
 \phi_L^{++}\,\phi_L^{--}\\
 \phi_L^{++}\,\phi_L^{--}\\
 \phi_L^0\,\phi_L^0 \\
\end{array}
\right)
&=& 
\left(
\begin{array}{cccc}
 \sqrt{\frac{2}{7}} & -\frac{5}{\sqrt{42}} &
   \frac{3}{\sqrt{77}} & -\frac{1}{\sqrt{462}} \\
 \sqrt{\frac{2}{7}} & 0 & -\sqrt{\frac{7}{11}} &
   \sqrt{\frac{6}{77}} \\
 \sqrt{\frac{2}{7}} & \sqrt{\frac{3}{14}} &
   \frac{1}{\sqrt{77}} & -5 \sqrt{\frac{3}{154}} \\
 \frac{1}{\sqrt{7}} & \frac{2}{\sqrt{21}} & 3
   \sqrt{\frac{2}{77}} & \frac{10}{\sqrt{231}} \\
\end{array}
\right)
\left(
\begin{array}{c}
 |I_3=0, I=6, Y=0\rangle^{S=0}\\
 |I_3=0, I=4, Y=0\rangle^{S=0}\\
 |I_3=0, I=2, Y=0\rangle^{S=0}\\
 |I_3=0, I=0, Y=0\rangle^{S=0}\\
\end{array}
\right)\nonumber
\end{eqnarray}

\end{minipage}
\end{lrbox}


\begin{lrbox}{\GVmultiplet}
\begin{tabular}{|c|c|c|c|c|c|c|}\hline
Rep.   & Basis  & $S$ & $I$  & $Y$ & $\Gamma^{S}_{I,Y}$ & $\alpha^S_{I,Y}/\alpha_2$ \\\hline\hline
\multirow{3}{*}{$\begin{array}{c}  \text{Fermionic} \\ \text{L-triplet} \\ (\vec{3},\vec{1},0)\end{array}$} & \multirow{3}{*}{$|I_3,I,Y\rangle^S$}  & \multirow{2}{*}{0} & 2 &0  &
$
\frac{\pi  \alpha_2^2}{2\, M^2} 
$
&
$
 1
$
\\\cline{4-7}
& & & 0 & 0 
&$
 \frac{2 \pi  \alpha_2^2 }{ M^2} 
$
&
$
 -2
$
\\\cline{3-7}
 && 1 & 1 & 0 &
$
 \frac{25\pi  \alpha_2^2 }{24 M^2} $ 
&
$
-1
$
\\\cline{1-7}
\multirow{3}{*}{$\begin{array}{c}  \text{Fermionic} \\ \text{R-triplet} \\ (\vec{1},\vec{3},0)\end{array}$}
 & $\Psi_R^+ \Psi_R^- $ & 0 & 0  & 0  &
$
 \frac{\pi  \alpha_2^2 \tan ^4\theta_W }{ M^2} 
$
&
$
 -\tan^2\theta_W
$\\\cline{2-7}
& $\Psi_R^+ \Psi_R^- $ &1& 0 & 0 &
$
2 A_{Z_2}
$ 
&
$
-\tan^2\theta_W 
$
\\\cline{2-7} 
&$\Psi_R^0 \Psi_R^+ $ &1 & 0 & 1 & 
$
2A_{W_2}
$
&
$
0
$
\\\hline\hline
%


 \multirow{5}{*}{ $\begin{array}{c}  \text{Fermionic} \\  \text{L-quintuplet} \\ (\vec{5},\vec{1},0) \end{array}$} &  \multirow{5}{*}{$|I_3,I,Y\rangle^S$} & \multirow{3}{*}{0}  & 4 & 0  & 0 & 4 \\\cline{4-7} 
 && & 2 & 0  
&
$
 \frac{21 \pi  \alpha_2^2 }{2 M^2} 
$ 
& $-3$ \\\cline{4-7}
&&& 0 & 0  
& 
$
 \frac{30 \pi  \alpha_2^2 }{ M^2} 
$ & $-6$ \\\cline{3-7}
 && \multirow{2}{*}{1} & 3 & 0 & 0 & 0 \\\cline{4-7} 
& & & 1 & 0  & 
$
 \frac{125 \pi  \alpha_2^2 }{24 M^2} 
$
 & $-5$\\\cline{1-7}
\multirow{5}{*}{ $\begin{array}{c}  \text{Fermionic} \\  \text{R-quintuplet} \\ (\vec{1},\vec{5},0) \end{array}$}  & $\begin{pmatrix} \Psi_R^{++}\, \Psi_R^{--}\\ \Psi_R^{+}\, \Psi_R^-\\ \end{pmatrix}$ & 0 & 0 & 0 & 
$\frac{ \pi  \alpha_2^2 \tan ^4\theta_W }{ M^2}
\left(
\begin{array}{cc}
 16  & 4\\
 4 & 1 \\
\end{array}
\right)
$ 
&
$\tan^2\theta_W
\left(
\begin{array}{cc}
 -4  & 0\\
 0 & -1\\
\end{array}
\right)
$
\\\cline{2-7}
& $\begin{pmatrix} \Psi_R^{++}\, \Psi_R^{--}\\ \Psi_R^{+}\, \Psi_R^-\\ \end{pmatrix}$  & 1 & 0 & 0 &
$2 A_{Z_2}
\left(
\begin{array}{cc}
 4  & 2\\
 2 & 1 \\
\end{array}
\right)
$ 
&
$\tan^2\theta_W
\left(
\begin{array}{cc}
 -4  & 0\\
 0 & -1\\
\end{array}
\right)
$
\\\cline{2-7}
 &  $\begin{pmatrix} \Psi_R^{++}\, \Psi_R^{-}\\ \Psi_R^{+}\, \Psi_R^0\\ \end{pmatrix}$ 
& 1 & 0 & 1 & 
$A_{W_2}
\left(
\begin{array}{cc}
 4  & 2\sqrt{6}\\
 2\sqrt{6} & 6 \\
\end{array}
\right)
$
& 
$\tan^2\theta_W
\left(
\begin{array}{cc}
 -2  & 0\\
 0 & 0 \\
\end{array}
\right)
$
\\\hline\hline
%


 \multirow{4}{*}{$\begin{array}{c}  \text{Scalar} \\  \text{L-7-plet} \\ (\vec{7},\vec{1},0) \end{array}$} &  \multirow{4}{*}{$|I_3,I,Y\rangle^S$} & \multirow{4}{*}{0}  & 6 & 0  & 0 & 9 \\\cline{4-7} 
 && & 4 & 0  
&0  & $-2$ \\\cline{4-7}
&&& 2 & 0  & 
$
 \frac{126 \pi  \alpha_2^2 }{ M^2} 
$ & $-9$ \\\cline{4-7}
 &&  & 0 & 0 &
$
 \frac{336 \pi  \alpha_2^2 }{ M^2} 
$ 
 & $-12$ \\\cline{1-7} 
$\begin{array}{c}  \text{Scalar} \\  \text{R-7-plet} \\ (\vec{1},\vec{7},0) \end{array}$ & $\begin{pmatrix}  \phi_R^{+++}\, \phi_R^{---}\\ \phi_R^{++}\, \phi_R^{--}\\ \phi_R^{+}\, \phi_R^- \end{pmatrix}$ & 0 & 0 & 0 & 
$\frac{162 \pi  \alpha_2^2 \tan ^4\theta_W }{ M^2}
\left(
\begin{array}{ccc}
 1 & \frac{4}{9} & \frac{1}{9} \\
 \frac{4}{9} & \frac{16}{81} & \frac{4}{81} \\
 \frac{1}{9} & \frac{4}{81} & \frac{1}{81} \\
\end{array}
\right)
$ 
&
$\tan^2\theta_W
\left(
\begin{array}{ccc}
 -9 & 0 & 0 \\
 0 & -4 & 0 \\
 0 & 0 & -1 \\
\end{array}
\right)$
\\\hline
\end{tabular}
\end{lrbox}


\begin{lrbox}{\GVbidoublet}
\begin{tabular}{|c|c|c|c|c|}\hline
 $S$  & $I$  & $Y$ & $\Gamma^{S}_{I,Y}$ & $\alpha^S_{I,Y}/\alpha_2$ \\\hline\hline
\multirow{3}{*}{0} & 0 & 0  & 
$
\frac{\pi  \alpha_2^2 (3+\tan^4\theta_W)}{8\, M^2} 
$
&
$
 -\frac{1+2 c_W^2}{4 c_W^2}
$
\\\cline{2-5}
  & 1 & 0 & 
$
 \frac{\pi  \alpha_2^2 \tan ^2\theta_W }{4 M^2} 
$
&
$
 -\frac{1-2 c_W^2}{4 c_W^2}
$ 
\\\cline{2-5}
  & 1 & 1 & 0 &
$
 \frac{1}{4 c_W^2}
$ 
\\\hline
 \multirow{3}{*}{1} & 0 & 0 & 
$
A_{Z_2}
$ 
&
$
 -\frac{1+2 c_W^2}{4 c_W^2}
$
\\\cline{2-5}
  & 1 & 0 & 
$
\frac{25 \pi  \alpha_2^2}{48 M^2} 
$
&
$
-\frac{1-2 c_W^2}{4 c_W^2}
$
\\\cline{2-5}
  & 0 & 1 &  
$
A_{W_2}
$
&
$
-1+\frac{1}{4 c_W^2}
$
\\\hline
\end{tabular}
\end{lrbox}


\begin{lrbox}{\GVbitriplet}
\begin{tabular}{|c|c|c|c|c|c|}\hline
 $S$  & $I$  & $Y$ & Basis & $\Gamma^{S}_{I,Y}$ & $\alpha^S_{I,Y}/\alpha_2$ \\\hline\hline
\multirow{3}{*}{0} & 0 & 0  & 
$
\begin{pmatrix}
|Y_1=1,Y_2=-1 \rangle_{Sym}\\
|Y_1=0,Y_2=0 \rangle
\end{pmatrix}
$
&
$
\frac{2 \, \pi  \alpha_2^2 }{ M^2} 
\left(
\begin{array}{cc}
2+\frac{3}{2} \tan^4\theta_W   & - \sqrt{2} \\
 - \sqrt{2} & 1 \\
\end{array}
\right)
$
&
$
\left(
\begin{array}{cc}
 -\frac{\cos 2 \theta_W +3}{2 c_W^2}   & 0\\
 0 & -2 \\
\end{array}
\right)
$
\\\cline{2-6}
& 1 & 0 & 
$
|Y_1=1,Y_2=-1 \rangle_{Ant}
$
&
$
 \frac{4\pi  \alpha_2^2 \tan ^2\theta_W }{ M^2} 
$
&
$
- \frac{1 }{c_W^2}
$
\\\cline{2-6}
  & 2 & 0 &  
$
\begin{pmatrix}
|Y_1=1,Y_2=-1 \rangle_{Sym}\\
|Y_1=0,Y_2=0 \rangle
\end{pmatrix}
$
&
 $
\frac{\, \pi  \alpha_2^2 }{2\, M^2} 
\left(
\begin{array}{cc}
2  & - \sqrt{2} \\
 - \sqrt{2} & 1 \\
\end{array}
\right)
$
&
$
\left(
\begin{array}{cc}
 \frac{\cos2\theta_W}{c_W^2}   & 0\\
 0 & 1 \\
\end{array}
\right)
$
\\\hline
 \multirow{3}{*}{1} & 0 & 0 & 
$
|Y_1=1,Y_2=-1 \rangle_{Ant}
$
&
$
6 A_{Z_2}
$ 
&
$
 -\frac{\cos 2 \theta_W +3}{2 c_W^2}
$
\\\cline{2-6}
  & 0 & 1 &
$
|Y_1=1,Y_2=0 \rangle_{Ant}
$
&  
$
 6 A_{W_2}
$
&
$
 -2
$
\\\cline{2-6}
  & 1 & 0 & 
$
\begin{pmatrix}
|Y_1=1,Y_2=-1 \rangle_{Sym}\\
|Y_1=0,Y_2=0 \rangle
\end{pmatrix}
$
&
 $
\frac{25 \, \pi  \alpha_2^2 }{24\, M^2} 
\left(
\begin{array}{cc}
2  & - \sqrt{2} \\
 - \sqrt{2} & 1 \\
\end{array}
\right)
$
&
$
\left(
\begin{array}{cc}
 -\frac{1}{ c_W^2}   & 0\\
 0 & -1 \\
\end{array}
\right)
$
\\\hline
\end{tabular}
\end{lrbox}